%% file: preprint.tex
\documentclass[11pt]{article}

\usepackage[english]{babel}

\usepackage[letterpaper,top=2cm,bottom=2cm,left=2cm,right=2cm,marginparwidth=1.75cm]{geometry}

\usepackage{xcolor}
\definecolor{pumpkin}{HTML}{E67E22}
\definecolor{belize}{HTML}{2980b9}

\usepackage{amsmath}
\usepackage{amsfonts}
\usepackage{amssymb}
\usepackage{graphicx}
\usepackage{booktabs}
\usepackage[colorlinks=true, allcolors=belize]{hyperref}
\usepackage[capitalize,noabbrev]{cleveref}
\usepackage[suppress]{color-edits}
\usepackage{paralist}
\usepackage[font=footnotesize,labelfont=bf]{caption}
\usepackage{algorithm}
\usepackage{algorithmic}
\usepackage{subfigure}
\usepackage{amsthm}
\usepackage{tcolorbox}
\DeclareMathOperator*{\argmax}{arg\,max}

\theoremstyle{definition}

\usepackage{array}
\newcolumntype{H}{>{\setbox0=\hbox\bgroup}c<{\egroup}@{}}

\addauthor{mp}{red}
\addauthor{yy}{magenta}

\addauthor{r}{blue}

\usepackage{listings}
\crefname{appendix}{SI Appendix}{SI Appendices}

\title{Network Formation and Dynamics among Multi-LLMs\thanks{Accepted at PNAS Nexus. Corresponding author: Marios Papachristou (\texttt{mpapachr@asu.edu}).}}

\author{Marios Papachristou\footnote{Department of Information Systems, W.P. Carey School of Business, Arizona State University, Tempe, AZ, USA and Department of Computer Science, Cornell University, Ithaca, NY, USA. \\ Supported by a scholarship from the Onassis Foundation (Scholarship ID: F ZT 056-1/2023-2024), and in part by a Simons Investigator Award, a Vannevar Bush Faculty Fellowship, AFOSR grant FA9550-19-1-0183, a Simons Collaboration grant, and a grant from the MacArthur Foundation.} \and Yuan Yuan\footnote{Graduate School of Management, University of California Davis, Davis, CA, USA.}}
\date{}

\begin{document}
\maketitle

\begin{abstract}
\redit{Social networks profoundly influence how humans form opinions, exchange information, and organize collectively. As large language models (LLMs) are increasingly embedded into social and professional environments, it is critical to understand whether their interactions approximate human-like network dynamics. We develop a framework to study the network formation behaviors of multiple LLM agents and benchmark them against human decisions. Across synthetic and real-world settings, including friendship, telecommunication, and employment networks, we find that LLMs consistently reproduce fundamental micro-level principles such as preferential attachment, triadic closure, and homophily, as well as macro-level properties including community structure and small-world effects. Importantly, the relative emphasis of these principles adapts to context: for example, LLMs favor homophily in friendship networks but heterophily in organizational settings, mirroring patterns of social mobility. A controlled human-subject survey confirms strong alignment between LLMs and human participants in link-formation decisions. These results establish that LLMs can serve as powerful tools for social simulation and synthetic data generation, while also raising critical questions about bias, fairness, and the design of AI systems that participate in human networks.}
\medskip


\medskip

\noindent \emph{Code:} \url{https://doi.org/10.5281/zenodo.16969696}

\noindent \emph{Data:} \url{https://doi.org/10.5281/zenodo.17196412}

\end{abstract}

% \subsection*{Significance Statement}

% \input{significance}

\newpage

\section*{Introduction}

Recent progress in large language models (LLMs),
such as GPT~\cite{openai2023gpt}, Claude~\cite{anthropic2024claude}, and Llama~\cite{touvron2023llama2}, have shown promising developments in AI techniques and their integration into real-life applications. It is thus crucial to comprehend AI actions to ensure they align with human expectations, mitigate potential risks, and maximize their benefits. Misaligned AI actions may lead to unintended consequences, such as biased decision-making, fairness issues, and the miscoordinative or non-cooperative behavior~\cite{rahwan2019machine}. Recently, researchers have started to apply social science methodologies, such as methods analogous to laboratory experiments \cite{horton2023large,aher2023using,veselovsky2023artificial,manning2024automated}, agent-based modeling \cite{park2018strength,gao2023s,he2023homophily,de2023emergence,fatemi2023talk,perozzi2024let}, and qualitative methods~\cite{chew2023llm}, to study LLMs. These methods not only reveal the capabilities and interpretability of LLMs but also suggest their potential for applications in social science~\cite{horton2023large,park2023generative,chen2023emergence,leng2023interpretable}.

In human societies, social networks play a crucial role in shaping individual behaviors, preferences, and connections, as well as influencing the diffusion of information and norms across communities~\cite{diffusionRogers,bakshy2012role,fowler2008dynamic,banerjee2013diffusion,yuan2018interpretable}.
LLMs have shown great potential in social contexts, notably as intelligent personal assistants that facilitate social and prosocial interactions (see, e.g., \cite{papachristou2023leveraging,chopra2023conducting,veselovsky2023artificial}). However, less is known about how LLMs' behaviors and preferences align with human network formation principles~\cite{jackson2008social,zhou2023sotopia,park2023generative}.
This is particularly crucial, as it sheds light on the potential of these models to shape and be shaped by the networks of human relationships, which is a fundamental aspect of social systems.

\redit{Recently, there has been an emerging body of work regarding LLM-based agent-based models, where the research community has devised frameworks for social simulation \cite{zhou2023sotopia,park2018strength,rossetti2024social,ferraro2024agent}, and have shown that collectives of LLMs exhibit linguistic collective biases \cite{ashery2025emergent,ferraro2024agent}, and the friendship paradox \cite{orlando2025can}, showing that LLMs are promising at simulating real-world social networks.}

\redit{Our study complements these works, and explores LLMs' behaviors and preferences in the context of network formation with both synthetic and real-world social networks, and answers the question \emph{``Which complex phenomena emerge by interactions between multiple LLMs?''} beyond merely focusing on the modeling and simulation aspect.  By analyzing such interactions, we aim to understand the implications of LLMs representing humans in social and professional settings.}

Specifically, we examine micro-level social network properties including preferential attachment \cite{barabasi1999emergence}, triadic closure \cite{granovetter1973strength}, and homophily~\cite{mcpherson2001birds}, as well as macro-level properties including community structure \cite{newman2004finding}, and the small-world phenomenon \cite{kleinberg2000small,watts1998collective}. 

\redit{By analyzing LLM agents interacting dynamically in several different contexts and environments, with a variety of models (both closed and open-weight), and variations of the prompts, we find that in synthetic network simulations, LLMs displayed preferential attachment, homophily, and triadic closure, resulting in the formation of community structures and small-world dynamics.}
More notably, in real-world social network simulations, we find that LLMs prioritize triadic closure and homophily over preferential attachment when forming new links, indicating a strong preference for connecting with similar nodes or shared acquaintances. Additionally, in a telecommunication network, LLMs tended to prioritize homophily and preferential attachment over triadic closure, and in a company network, the agents who corresponded to employees formed links frequently with managers, which showcases behavior that is consistent with human social mobility principles. 

Generally, LLMs not only exhibit fundamental social network formation principles in synthetic simulations but also adapt their strategies based on the context of real-world networks, mirroring human social behaviors specific to each setting.

As LLM technology continues to evolve, our study serves as an early exploration of their potential in social network studies, with several significant implications for future studies.
First, our study demonstrates the potential of LLMs for agent-based modeling. By simulating decision-making processes that approximate human-like behavior across various network settings, LLMs can provide valuable insights into the emergence of social phenomena. Although these models are still in early stages of development, they offer an intriguing framework for studying and designing systems that can mimic key aspects of real-world dynamics. This opens up possibilities for applying LLMs to explore and understand complex behaviors in social, professional, and collaborative environments.
Second, our work highlights the potential of LLMs for synthetic dataset generation, a critical area in network science. Although the accuracy of LLM-based predictions is not yet perfect, like all other link prediction models, this approach is particularly valuable in scenarios where privacy concerns limit access to real-world data. By simulating realistic datasets that capture important network properties, LLMs can facilitate research and experimentation without compromising sensitive information. 

\begin{figure*}[!ht]
    \centering
    % \subfigure[]{\includegraphics[width=0.45\linewidth]{figures/probabilitytopk_all_.png}\label{subfig:principle_1_models_environments_topk_all}}
    % \subfigure[]{\includegraphics[width=0.45\linewidth]{figures/probabilitytopk.png}\label{subfig:principle_1_models_environments_topk}}
    % \subfigure[]{\includegraphics[width=0.45\linewidth]{figures/exponents.png}\label{subfig:principle_1_models_environments}}
    % \subfigure[]{\includegraphics[width=0.9\textwidth]{figures/principle_1_final_graphs_neighbors.png}\label{subfig:principle_1_final_graphs}}
    \includegraphics[width=0.9\textwidth]{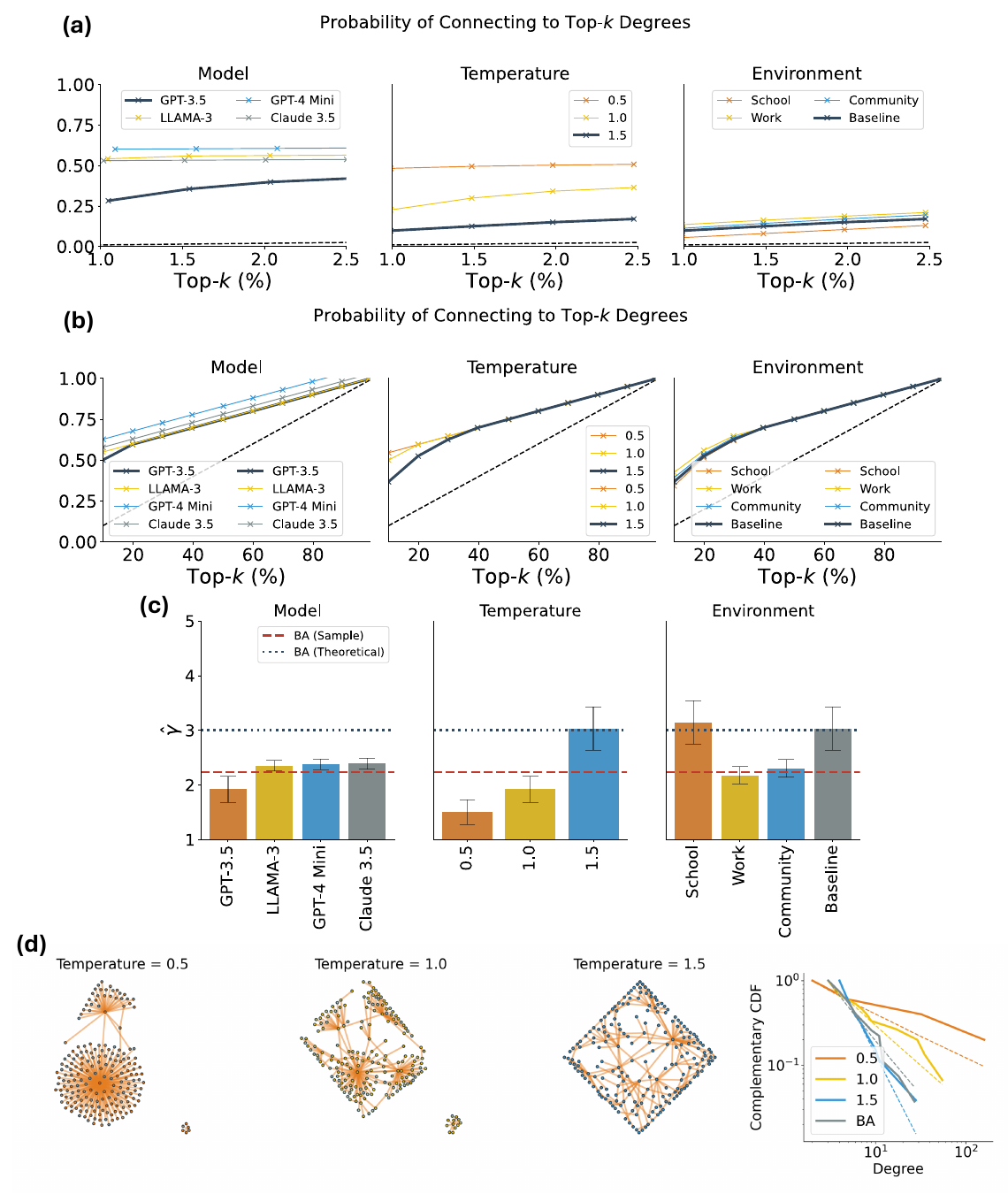}
    \caption{\textbf{Results for Principle 1 (preferential attachment)} The multi-LLM setup was given neighborhood information $\{ N_{j, t} : j \in V_t \}$. \textbf{(a, b):} Probability of connecting to top-$k$-degree nodes for varying model (temperature is fixed to 1.0 and environment to baseline), temperature (model fixed to GPT-3.5 and environment to baseline) and environment (model fixed to GPT-3.5 and environment temperature to 1.5) for networks generated according to Principle 1 with $n = 200$ nodes. \redit{(a) shows the whole range of $k$, and (b) shows the top $1-2.5\%$ nodes.} \textbf{(c):} Power Law exponents and standard errors for varying model, temperature, and environment. 
    \textbf{(d):} Simulated networks. Power-law degree distributions are evident ($P > 0.5$, K-S test), with the networks at a temperature of 1.5 closely resembling the Barab\'asi-Albert model ($P > 0.1$, K-S test) for GPT-3.5 agents.}
    \label{fig:principle_1}
\end{figure*}

\section*{Results}

In this study, we investigated whether LLMs exhibit fundamental principles of network formation observed in human social networks. By simulating multiple LLM agents acting independently within separate conversational threads, we examined their behaviors in decision-making scenarios involving network connections.
We focused on three micro-level network principles -- preferential attachment, triadic closure, and homophily -- and two macro-level phenomena--community structure and the small-world effect. To assess the robustness of our findings, we varied the temperature settings of different LLM models, including GPT-3.5-turbo, GPT-4o Mini, Llama 3 (70b-instruct), and Claude 3.5 Sonnet. We also experimented with different environmental prompts (e.g., friendship, collaboration, community) to test prompt sensitivity.
Additionally, we employed an interview-like method to probe the LLMs' decision-making rationale and conducted experiments using Chain-of-Thought (CoT) reasoning~\cite{wei2022chain} (the experiments are deferred to \cref{app:cot}). Finally, we extended our analysis to real-world networks, including a social media friendship network, a telecommunication network, and a company collaboration network, to compare the network formation preferences between LLMs and humans.

More information about the experimental procedure, methods, and materials can be found in the \cref{app:methods}. 

\subsection*{Micro-Level Properties} \label{sec:micro_level}
\subsubsection*{Principle 1: Preferential Attachment} \label{sec:principle_1}

Preferential attachment is a fundamental concept in network science, illustrating how nodes in a network gain connections over time, leading to a scale-free degree distribution characterized by a few highly connected nodes~\cite{barabasi1999emergence,bianconi2001competition}.

To test if LLM agents exhibit preferential attachment, we simulated network growth by sequentially adding nodes to an initially empty network. Each new node was prompted with information about existing nodes, and the person to connect with was decided. We generated networks with $n = 200$ nodes to observe meaningful degree distributions\footnote{Note that we provide the full network structure in the prompt, so models are not inherently biased toward forming links with the highest-degree nodes.}.

{On a micro-scale, \cref{fig:principle_1} illustrates the probability of connecting to a top-$k$ node as a function of its degree percentile ($k / n$). To demonstrate the tendency toward preferential attachment, we compare these probabilities to a null model assuming random connections (represented by dashed lines), where the likelihood of connecting to a top-$k$ node is simply $k / n$. 
Our findings reveal that all models prefer connecting to higher-degree nodes. Notably, GPT-3.5 exhibits a weaker preference, while other, arguably more capable models, show an even stronger inclination toward preferential attachment. Using GPT-3.5 as an example, we examine the effect of temperature -- a parameter controlling the variability of model output -- on this tendency. At lower temperatures, the model makes fewer stochastic choices and, as a result, is more likely to connect to high-degree nodes.
We also vary the prompt to explore the influence of ``environment''-contextual settings such as school, work, or community. The results show slight variations compared to the baseline (GPT-3.5 with temperature = 1.5), yet the tendency for preferential attachment persists across environments.
In all cases, the observed curves lie above the null model, underscoring the presence of preferential attachment.}

Next, we investigate the degree distribution of the resulting graphs. As shown in \cref{fig:principle_1}, the resulting networks display a pattern where a few nodes have many connections while most have few, indicative of a scale-free distribution, with form:
\begin{align}
\pi(d) \propto d^{-\gamma}, \quad \text{where} \quad \gamma > 1.\label{eq:pi}
\end{align}
\noindent We estimated the exponent $\gamma$ for different models and temperatures. Our analysis reveals several notable patterns in the networks generated by LLM agents under different conditions. First, models newer than GPT-3.5 exhibit a slightly larger $\hat \gamma$ than GPT-3.5. This implies that these models display a stronger tendency toward preferential attachment and the formation of hubs. 
% \yuan{Is gamma in BA model 3?} \mpcomment{$\gamma = 3$ as $n \to \infty$}
Second, as the temperature increases the power-law exponent $\hat{\gamma}$ generally becomes larger. This indicates that higher temperatures introduce more variance in node connectivity, leading to degree distributions with heavier tails. Third, the environmental context significantly affects the value of $\hat{\gamma}$. For example, when the network is framed within a ``school'' environment, the exponent increases, suggesting a more uniform distribution of connections and fewer highly central nodes.  

% {Surprisingly, we find that when simulating networks with GPT-3.5 and temperature 1.5, the resulting networks have identical degree distribution with a Barabasi-Albert model graph with $n = 200$ nodes ($P > 0.5$, K-S test between the degree distributions of the LLM-generated network).}

Finally, while the prompts we have utilized thus far have provided the model with the complete existing network structure, we also explore an alternative scenario: what happens if agents are supplied solely with the degree of other alternatives, without access to the network's full structure? As detailed in \cref{app:omitted}, our findings reveal that limiting agents to degree information alone also leads to notable structural differences in the networks that emerge (cf \cref{fig:app_principle_1_final_graphs}). Thus, degree information alone yields more restrictive structures than providing the agents with the full topological information (i.e., the neighbors). 

The findings highlight the practical potential of LLMs in modeling complex networks, such as social, economic, or biological systems, by leveraging their ability to simulate preferential attachment and scale-free distributions. These models can be used to study real-world phenomena like information diffusion, hub formation, or connectivity patterns under varying conditions. Additionally, the sensitivity of network structures to parameters like temperature and context underscores the importance of prompt design in steering outcomes, making LLMs versatile tools for tailored simulations.

\subsubsection*{Principle 2: Triadic Closure} \label{sec:principle_2}

 \begin{figure*}[!ht]
    \centering
    
    \includegraphics[width=0.7\textwidth]{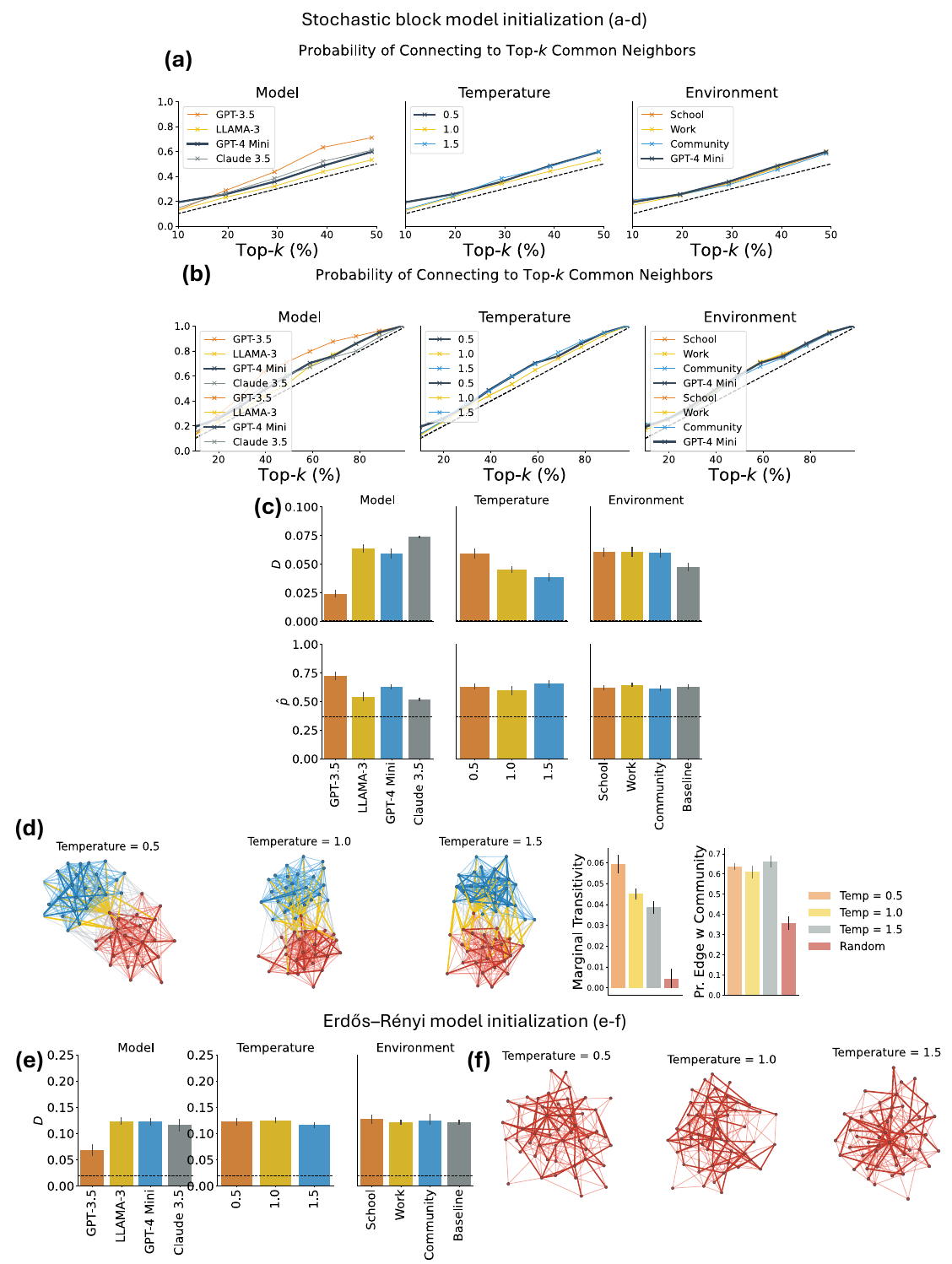}
    
    \caption{\textbf{Results for Principle 2 (triadic closure).}  \textbf{(a, b):} Probability of connecting to top-$k$ nodes (in terms of common neighbors) for varying model (temperature is fixed to 1.0 and environment to baseline), temperature (model fixed to GPT-4 Mini and environment to baseline) and environment (model fixed to GPT-4 Mini and environment temperature to 0.5) for networks generated according to Principle 2 ($n = 50$, 10 simulations for each model, environment and temperature). \redit{The dotted diagonal line corresponds to the null model, where connections are made at random.} \redit{Panel (a) shows top-$k$ for $k$ for $k$ in the range $10-50\%$, and Panel (b) shows top-$k$ for $k$ in the range of $10-100\%$.} \textbf{(c):} Marginal transitivity ($D$) and probability of an edge within a community ($\hat p$) for networks generated according to Principle 2 in different models, temperatures, and environments. \redit{The dotted line corresponds to the random null model.} \textbf{(d):} The figure shows the resulting networks created by GPT-4 Mini, according to Principle 2 when the intersection of the neighborhoods of the query node and each alternative is provided \redit{and comparison of the metrics $D$ and $\hat p$ with the random null model.} The node colors correspond to the groups to which each node belongs. The bold edges (red or blue) correspond to the newly created inter-cluster edges, and the orange edges correspond to the new intra-cluster edges. \redit{\textbf{(e, f):} Marginal transitivity ($D$) and network instances when the initial network is an Erd\"os-R\'enyi graph with $n = 50$ and $p = 0.1$.}}
    \label{fig:principle_2}
\end{figure*}

The second micro-level principle we examine is triadic closure, which posits that individuals are more likely to form connections with friends of friends, thus creating closed triads in the network. This process strengthens network structure and cohesion, grounded in the idea that two nodes are more likely to connect if they share a common neighbor~\cite{granovetter1973strength,mosleh2024tendencies}.

To investigate triadic closure, we employ an assortative stochastic block model (SBM) \cite{newman2003mixing} to create an initial network $G_1$ with $n$ nodes divided into two equal-sized clusters $A$ and $B$. Connections within each cluster are formed with a probability of $0.5$, while inter-cluster connections occur with a probability of $0.1$. This setup mirrors our assumption that nodes within the same cluster are more inclined to connect due to a higher number of shared neighbors. In subsequent time steps, we then examine each node $i$, considering the intersection of neighborhoods of $i$'s non-neighbors\footnote{Similar outcomes arise when providing neighbors instead of common neighbors.}.

We conducted ten simulations with $n = 50$ nodes to facilitate clear visualization and ensure statistical significance\footnote{Choosing $n = 50$ instead of a larger number like $n = 200$ aids in visualization and maintains statistical significance.}. 

{On a micro-scale, \cref{fig:principle_2} illustrates the probability of connecting to a top-$k$-percentile node as a function of the number of common neighbors. The dashed lines represent the results of null models, where connections are chosen randomly; which corresponds to the probability of connecting to a top-$k$ percentile node in terms of the common neighbors being $k / n$. Our findings reveal that, across all models, there is a consistently higher probability of forming links with nodes that share more common neighbors. Unlike the behavior observed in preferential attachment, temperature does not appear to severely impact this probability. This tendency to form links with nodes that have more common neighbors is consistent across various contexts, including school, work, and community environments. These results suggest that the triadic closure tendency is a robust phenomenon, persisting across different model families, configurations, and environments.}

Then, for evaluating triadic closure on the network (macroscopic) level, we utilize two metrics: \textit{marginal transitivity} and \textit{probability of edge formation within the same community}. Marginal transitivity ($D$) represents the change in the ratio of closed triangles to all triads, \redit{transitioning from the SBM-generated network $G_1$ to the final network $G_T$ after $T = 50$ iterations:}
\begin{align*}
    D = 3 \times \frac {\text{\# triangles}(G_T)} {\text{\# triads}(G_T)} - 3 \times \frac {\text{\# triangles}(G_1)} {\text{\# triads}(G_1)}.
\end{align*}
where a large positive $D$ indicates a strong triadic closure tendency.

As we investigate under SBM, the same community membership indicates more open triads being closed.

Marginal transitivity ($D$), presented in \cref{fig:principle_2}, demonstrates a statistically significant increase across all models, temperatures, and environments, underscoring the robust nature of triadic closure.

In \cref{fig:principle_2}, sample networks from GPT-3.5 are displayed, with the upper panel showing networks where the entire structure is provided and the lower panel showing those with only common neighbor numbers provided. Nodes are color-coded to indicate their cluster memberships in the SBM, with red and blue edges within clusters and orange edges between clusters. Newly formed edges are highlighted with thicker lines.

\redit{Finally, to eliminate the possibility that the results are due to structural bias from the initial structure (SBM), we note that we can obtain the same results when we start from a more ``neutral'' initial topology. Specifically, we get the same results and an even stronger effect for the marginal transitivity ($D$) by starting from a sparse Erd\"os-R\'enyi graph with $n = 50$ nodes and $p = 0.1$. We find that, across or models, temperatures and variations of the prompts, the resulting network exhibits higher marginal transitivity ($D$) compared to the random null model, which makes connections at random starting from the same Erd\"os-R\'enyi graph, and the results are statistically significant (cf. \cref{fig:principle_2}; $P < 0.001$; t-test comparing the marginal transitivity of the resulting LLM-generated networks and the random null model).}

In summary, these findings show that most LLMs exhibit a consistent tendency for triadic closure across various configurations, temperatures, and environments. This behavior mirrors human network dynamics, highlighting the models' ability to simulate realistic social and structural networks and reinforcing their alignment with social principles observed in real-world communities.

\subsubsection*{Principle 3: Homophily} \label{sec:principle_3}

\begin{figure*}[ht!]
    \centering
    \includegraphics[width=0.9\textwidth]{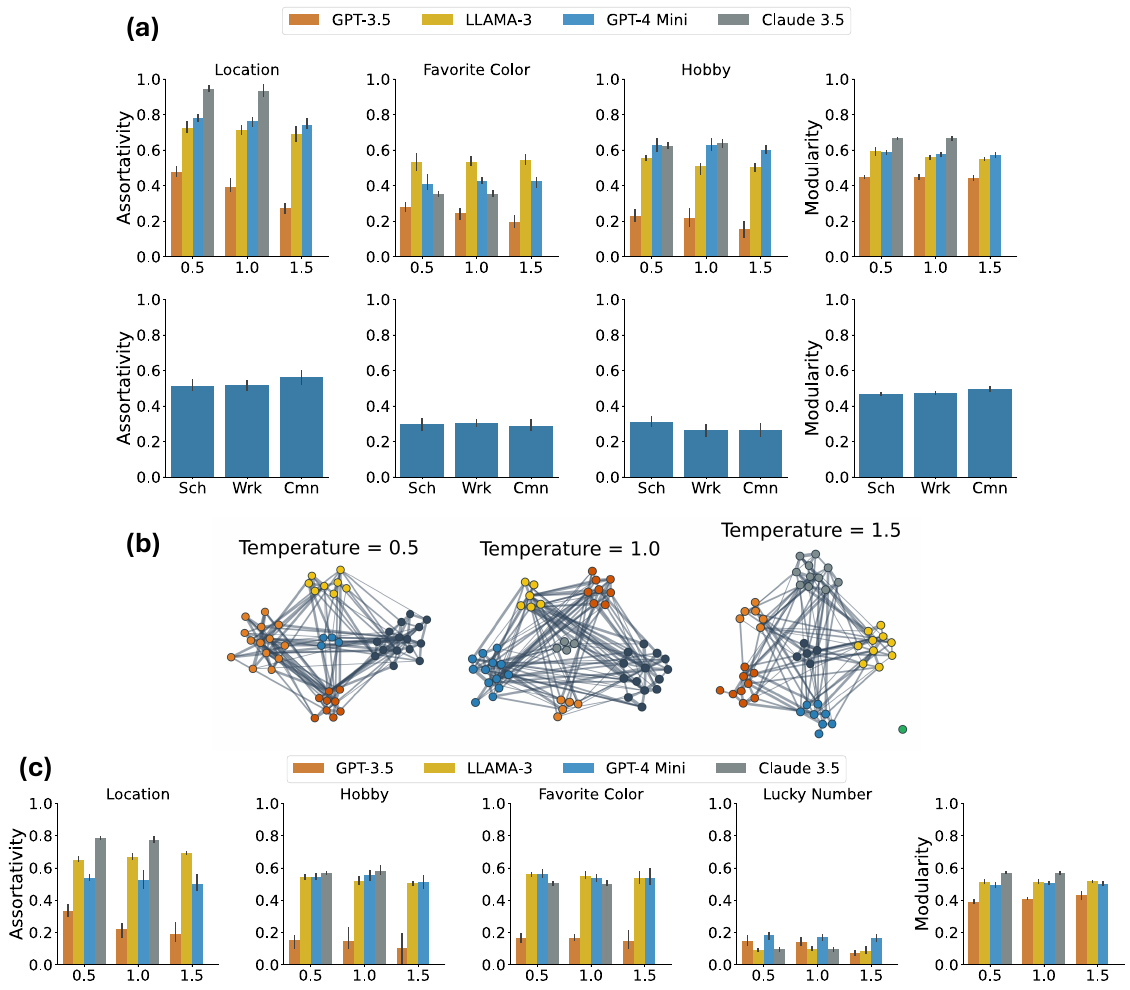}
    \caption{\textbf{Results for Principle 3 (Homophily) and Principle 4 (Community structure due to homophily)}. \textbf{(a):} Assortativity and Louvain modularity by Principle 3 ($n = 50$, 5 runs per row) across school, work, and community settings. All comparisons to the random null model ($R = 0$) are statistically significant ($P < 0.0003$, t-tests with \redit{Bonferroni correction for three tests}). Modularity is also significantly greater than 0 ($P < 0.001$).  
    \textbf{(b):} Network examples and communities for GPT-3.5 agents. \redit{Compared to a null model where agents connect randomly ($R = 0$).}  
    \textbf{(c):} Influence of distractor features (favorite color, lucky number) on homophily. \redit{Compared to a random null model with $R = 0$.} \redit{All results are statistically significant ($P < 0.00025$), with Bonferroni correction over 3 tests (location, favorite color, hobby).}}
    \label{fig:principle_3}
\end{figure*}

Homophily reflects the tendency for nodes with similar characteristics or attributes to form connections and associate with each other. This phenomenon is based on the principle that individuals in a network are more likely to connect with others who share similar traits, interests, or demographics \cite{mcpherson2001birds}.

To test whether LLM agents exhibit homophily, we perform the following experiment: We generate nodes with randomly generated attributes regarding a hobby (randomly chosen among three hobbies), a favorite color (randomly chosen among three colors), and a location within the US (randomly chosen among three US locations) and provide the attributes of the other nodes and the node's own attributes, and each node is tasked to form up to $\delta = 5$ links with others. For each node $i$, we provide it with the features $x_{j}$ of all non-neighbors $j$ of $i$. The seed network is taken to be the empty graph. We run ten simulations
for networks with $n = 50$ nodes and $\delta = 5$. 

% We compare with the \texttt{Random} null model, which chooses $\delta = 5$ links at random. 

To evaluate homophily, we calculate the attribute assortativity coefficient for each of the features. For each property $P$ which takes $K$ distinct values $P_1, \dots, P_{K}$ (indexed by $k$ or $l$), its \textit{assortativity coefficient $R$} is defined as 
\begin{align*}
    R = \frac {\sum_{k = 1}^K M_{kk} - \sum_{k = 1}^K a_k b_k} {1 - \sum_{k = 1}^K a_k b_k}.
\end{align*}

Here, $M$ represents the mixing matrix. Its elements $M_{kl}$ reflect the proportion of edges connecting two nodes with values $P_k$ and $P_l$, respectively. We define $a_k = \sum_{l = 1}^K M_{kl}$ and $b_k = \sum_{l = 1}^K M_{lk}$. Assortativity ranges from $-1$ to $+1$. A positive assortativity indicates nodes preferentially connect to similar ones, forming a homophilous network. Conversely, a negative assortativity suggests connections primarily occur between dissimilar nodes, indicating heterophily.

From \cref{fig:principle_3}, we observe that different attributes exhibit varying levels of assortativity. First, homophily is present across all LLMs, regardless of the specific model or configuration (e.g., temperature settings), all show positive assortativity for all four attributes, \redit{where they obtain statistically significant results $P < 0.0003$ (t-test with 0, Bonferroni correction across three tests across the three features)}. This aligns with human societies, where homophily is a primary driver of network formation \cite{mcpherson2001birds}\footnote{As an additional robustness check, we also tested mutual agreement connections. In that setting, after a node $j$ is chosen by node $i$, $j$ has to confirm the creation of the link from itself to $i$ ($j \to i$). We ran several experiments with different models and temperatures and we found the results not to be affected, namely, the proposed connections were always bilateral.}.

Moreover, to test the effect of the features on homophily as indicated by the assortativity coefficient for each attribute, we introduce a \textit{distractor feature}, which corresponds to a lucky number that is randomly chosen between 0 and 9. We repeat the simulations for all models and measure the effect of each feature on  \cref{fig:principle_3}. \redit{We still observe strong homophily effects with high statistical significance after applying Bonferroni correction for four tests ($P < 0.00025$, t-test with the random null model where connections are done at random which has assortativity $R = 0$).} 

However, we show that lucky numbers consistently show lower assortativity coefficients, indicating they are less considered when forming homophilous connections. This is consistent with our prior expectation that humans typically do not prioritize shared lucky numbers when establishing relationships.

Surprisingly, even though the lucky number does not seem to impact homophily much, the favorite color exhibits a similar level of homophily as hobbies. One might expect that hobbies, being substantive interests, would have a stronger influence on social connections than favorite colors, which are more arbitrary preferences. However, this finding aligns with the social identity theory and the minimal group paradigm \cite{tajfel1971social}. According to this paradigm, even minimal and arbitrary group distinctions -- such as a preference for certain colors -- can lead to in-group favoritism and influence social connections. This suggests that LLM agents, akin to humans, may form connections based on even trivial shared attributes, reflecting inherent tendencies toward group formation based on minimal commonalities. 

All in all, LLMs can capture and reproduce subtle human social behaviors, not just linguistic patterns. This underscores their potential as powerful tools for social simulation. However, these findings may also raise important considerations regarding bias, fairness, and the ethical design of AI systems (cf. Discussion Section). 

\subsection*{Macro-Level Principles} \label{sec:macro_level}

\subsubsection*{Principle 4: Community Structure}

The community structure of networks refers to the organization of nodes or individuals within a network into distinct and densely interconnected groups or clusters~\cite{newman2004finding,newman2006modularity,blondel2008fast,clauset2004finding}.
Identifying community structures is crucial for understanding the overall dynamics of a network, as it reveals patterns of relationships and interactions that might not be apparent at the global level.\footnote{As an example, we present only the results from GPT-3.5 for Principle 4 (Community Structure) and Principle 5 (Small-World).}

Both triadic closure and homophily contribute to the formation of community structures. By examining how these two factors contribute to network formation, we aim to gain insights into the underlying mechanisms driving community dynamics in LLM-generated networks.
We employ the simulation results presented in the synthetic networks to determine whether community structure in networks generated by LLMs emerges from triadic closure or homophily. 

First, we consider the networks generated in \cref{fig:principle_2}. We examine how LLM agents' choices strengthen the network's community structure. 
Specifically, we leverage the fact that the SBM graph has a preexisting community structure and measure how the newly formed links reinforce such a structure. Visual inspection shows that the newly added links, represented by the bold edges, happen mostly within each cluster, reinforcing the community structure. 

\redit{To measure the emergence of communities in the triadic closure case, we initially examine the probability of forming an edge within the same community ($\hat{p}$). The quantity $\hat p$ is calculated by the ratio of edges in $G_T \setminus G_1$ (newly formed edges) connecting nodes within the same cluster:
\begin{align*}
    \hat p = \frac {\left | \left \{ \{ i, j \} \in E (G_T) \setminus E(G_1) : y_i = y_j \right \}\right |} {\left | E(G_T) \setminus E(G_1) \right |},
\end{align*}
\noindent
where $y_i, y_j \in { A, B }$ denote the community memberships of nodes $i$ and $j$, respectively. A value of $\hat{p}$ exceeding $0.5$ suggests a triadic closure tendency and community structure.

\cref{fig:principle_2} shows that $\hat p$ is significantly higher than 0.5 ($P < 0.001$, t-test compared to 0.5), and is significantly bigger compared to a random null model where connections are made at random. All in all, this indicates that most edges are within the same community, strengthening the community structure.} 

Next, we investigate the community structure resulting from homophily using modularity maximization \cite{blondel2008fast}. Modularity quantifies the discrepancy between the actual number of edges within communities and the expected number in a random network with identical node count and degree distribution, following the Chung-Lu model \cite{chung2004average}. This model presumes that nodes maintain their weighted degree, with edges randomly distributed. The weighted modularity $Q$ \cite{clauset2004finding} for a graph with edge weights $w_{ij}$ and $C$ communities is defined as

\begin{align*}
    Q = \sum_{c = 1}^C \left [ \frac {L_c} {W} - r \left ( \frac {k_c} {2W} \right )^2 \right ].
\end{align*}

Here $W$ represents the total edge weights, $L_c$ the intra-community link weights for community $c$, $k_c$ the total weighted degree within community $c$, and $r$ the resolution parameter, set to 1 for our analysis. High modularity values (e.g., greater than 0.5) indicate significant community structuring, diverging from the random model.\footnote{Given the NP-Hard nature of maximizing $Q$, we employ the Louvain algorithm \cite{blondel2008fast} to approximate the highest possible modularity.}

\redit{Firstly, we note that when the experimental setting for Principle 2 (cf. \cref{fig:principle_2}) is initialized with an SBM or an  Erd\"os-Renyi network, we obtain positive modularity $Q > 0$ ($P < 0.001$; t-test comparing with 0).}

Secondly, regarding the homophily experiment (cf. \cref{fig:principle_3}), for the network's weights, we use the number of common attributes shared between each pair of nodes: $w_{ij} = \left| \left\{ k: x_{i}^{(k)} = x_{j}^{(k)} \right\} \right|$ for each link $(i, j)$ in the final network. Here, $x_{i}^{(k)}$ and $x_j^{(k)}$ correspond to the $k$-th features of $x_i$ and $x_j$, respectively.

In \cref{fig:principle_3}, various colors represent the communities identified by the Louvain algorithm at different temperatures for GPT-3.5. Notably, communities appear more distinct at lower temperatures, likely due to reduced randomness in decision-making at these temperatures.

\cref{fig:principle_3} presents the distribution of Louvain modularity values across simulations accross different LLM models and different environments, indicating consistent community structure with positive modularity at all temperatures, confirmed by a t-test against a modularity of $Q = 0$ for a random graph ($P < 0.001$).

Our results demonstrate that community structures manifest in networks generated by LLMs, driven by both triadic closure and homophily.
% In the real-world networks, we further illustrate that community structures are reinforced in real-world networks, where LLMs consider a combination of factors, including homophily and triadic closure, in their decisions to form connections.

\subsubsection*{Principle 5: Small-World} 

% \begin{figure*}
%     \centering
%     \subfigure[$\beta = 0.25$\label{subfig:principle_4_instance_0.25}]{\includegraphics[width=0.9\textwidth]{figures/principle_5_final_graphs_0.25.png}}
%     \subfigure[$\beta = 0.5$\label{subfig:principle_4_instance_0.5}]{\includegraphics[width=0.9\textwidth]{figures/principle_5_final_graphs_0.5.png}}
%     \subfigure[$\beta = 0.75$\label{subfig:principle_4_instance_0.75}]{\includegraphics[width=0.9\textwidth]{figures/principle_5_final_graphs_0.75.png}}
%     \caption{\textbf{Simulation results for Principle 5 (small world).} Network instances for the networks created according to Principle 5 using the altered Watts-Strogatz Model for node count $n = 50$, average degree $k = 5$, rewriting probability $\beta \in \{ 0.25, 0.5, 0.75 \}$, together with plots of the \textbf{average clustering coefficient} $C$ and the \textbf{average shortest path length} $L$. The comparison is made with respect to a Watts-Strogatz graph with $n = 50, k = 5, \beta \in \{ 0.25, 0.5, 0.75 \}$. The error bars correspond to 95\% confidence intervals. \redit{The results are compared against the Watts-Strogatz model with the same parameters $k$ and $\beta$ as a null model. The t-test comparing $L$ and $C$ for the LLM-generated networks and Watts-Strogatz networks yields $P > 0.05$ (Bonferroni correction for two tests).}}
% \end{figure*}

\begin{figure*}[!ht]
    \centering
    \includegraphics[width=\textwidth]{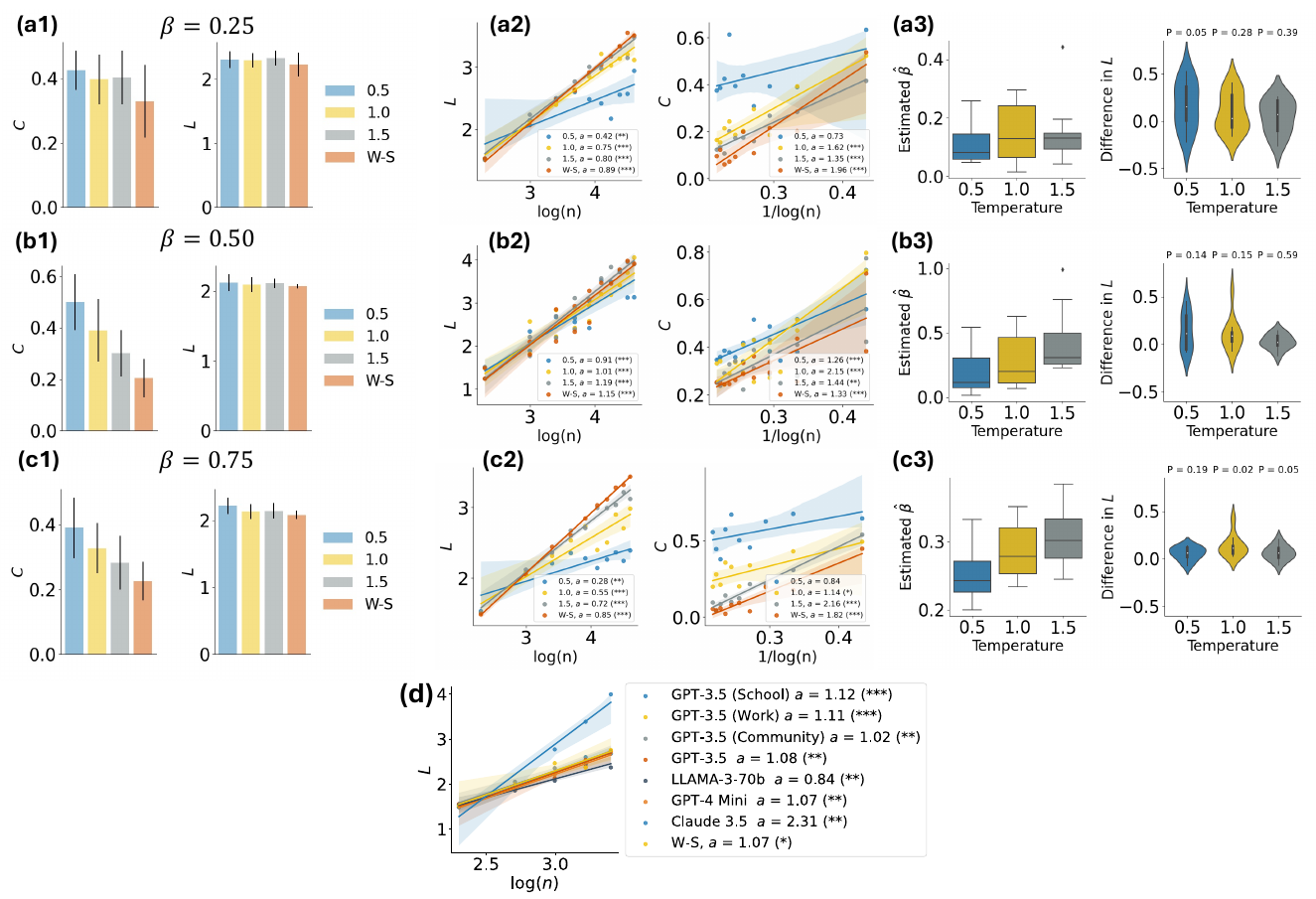}
    \caption{\textbf{Fitted results for Principle 5 (small world) for $\beta = 0.25, k = 5$ (a1-a3), $\beta = 0.5, k =5$ (b1-b3), and $\beta = 0.75, k=5$ (c1-c3).} \textbf{(a1-c1):} \emph{Average clustering coefficient} $C$ the \emph{average shortest path length} $L$. The comparison is made with respect to a Watts-Strogatz graph with $n = 50, k = 5, \beta \in \{ 0.25, 0.5, 0.75 \}$. The error bars correspond to 95\% confidence intervals. \redit{The results are compared against the Watts-Strogatz model with the same parameters $k$ and $\beta$ as a null model. The t-test comparing $L$ and $C$ for the LLM-generated networks and Watts-Strogatz networks yields $P > 0.05$ (Bonferroni correction for two tests).} \textbf{(a2-c2):} Regression plots relating \emph{average shortest path length} ($L$) and \emph{average clustering coefficient} ($C$) with $n$. The value $a$ in legends represents the effect size (slope of the regression lines). \textbf{(a3-c3):} Estimated values $\hat \beta$ of $\beta \in \{ 0.25, 0.5, 0.75 \}$ for LLM-generated networks based on matching the average clustering coefficient and difference in the average shortest path between LLM-generated networks and Watts-Strogatz with the \textbf{estimated rewiring probability} $\hat \beta$ for GPT-3.5 agents. We report the $P$-values of the t-test comparing the average shortest path length of the LLM-generated networks and the average shortest path length of the Watts-Strogatz graphs with rewiring probability $\hat \beta$. \textbf{(d):} Regression plot for the relation $L \sim \log (n)$ for different LLM models and environments (school, work, community) for $\beta = 0.25$ and $k = 5$. The legend shows the effect size ($a$) and the $P$-value. \redit{The results are compared against the Watts-Strogatz model with the same parameters $k$ and $\beta$ as a null model.} (*: $P < 0.025$; **: $P < 0.005$, and ***: $P < 0.0005$, \redit{Bonferroni correction for two tests; $L$ and $C$}.).}
    \label{fig:principle_5}
\end{figure*}

The small-world phenomenon is characterized by networks where nodes are interconnected in tight clusters, yet the average distance between any two nodes remains relatively short, typically scaling logarithmically with the network size \cite{watts1998collective, kleinberg2000small}. This balance between high clustering and short path lengths characterizes small-world networks.

A small-world network is defined by its \textit{average shortest path length $L$}, which grows logarithmically with the size of the network $n$,\footnote{As per the definition in \cite{jacobs2021large}.} expressed as
\begin{align*}
    L \sim \log(n).
\end{align*}

Our analysis utilizes the Watts-Strogatz model~\cite{watts1998collective} as a benchmark to investigate whether LLMs can generate networks exhibiting small-world characteristics.
This model has a delicate balance between local clustering and short average path lengths: Nodes tend to form clusters or groups (triadic closure), exhibiting a high level of interconnectedness within these local neighborhoods, whereas at the same time, the existence of a few long-range connections ensures that the entire network is reachable with relatively few steps~\cite{park2018strength,lyu2022investigating,jahani2023long}. 

We employ a modified version of the model, where edge rewiring is informed by LLM queries, based on the current network structure. The generation process is parametrized by the number of nodes ($n$), average degree ($k$), and the rewiring probability ($\beta$). See details in \cref{app:small_world_construction}

We generated networks of various sizes, ranging from $n = 10$ to $n = 100$, to explore the relationship between the network size ($n$) and two key metrics: the average shortest path length ($L$) and the average clustering coefficient ($C$). For this analysis, we considered values of $\beta$ set at 0.25, 0.5, and 0.75, with a fixed $k = 5$ to serve as a consistent parameter. 

However, when directly compared with the Watts-Strogatz model, the networks generated by the LLMs do not precisely replicate the characteristics of Watts-Strogatz networks for the corresponding rewiring probabilities ($\beta$). As illustrated in panels (a1-c1) of \cref{fig:principle_5}, \redit{we fail to reject the null hypothesis at level $0.05$ that the LLM-generated networks have the same average shortest path length as the Watts-Strogatz model for the rewiring probabilities ($\beta$) of 0.25, 0.5, and 0.75 (t-test comparing the average shortest path lengths, Bonferroni correction for two tests). Additionally,  LLM-generated networks also fail to reject the null hypothesis at level $0.05$ that the LLM-generated networks have the same average clustering coefficient as the Watts-Strogatz model for the same rewiring probabilities $\beta$ (Bonferroni correction for two tests). These results may suggest similarities in the network structure and connectivity patterns between the LLM-generated networks and the classical Watts-Strogatz model.}

We also provide regression analysis by examining the correlation between the average shortest path length and average clustering coefficient versus $\log(n)$ (refer to \cref{fig:principle_5}). We found that across all tested temperatures, the relationships were statistically significant, with most regressions yielding \redit{statistically significant results after applying Bonferroni correction for two tests ($P < 0.0005$)}. This indicates that the average shortest path length increases proportionally with $\log(n)$. Similarly, for the average clustering coefficient, we demonstrated that it inversely scales with $1 / \log(n)$, with the majority of regression analyses also \redit{showing high statistical significance after Bonferroni correction for the two tests ($P < 0.0005$)}. These findings align with the small-world properties of organizational networks as documented in the study by~\cite{jacobs2021large}, suggesting that these characteristics are not only prevalent but also predictable across different network sizes.

To quantify how LLM-generated networks resemble Watts-Strogatz networks, we fit the estimated $\hat{\beta}$ values for each LLM-generated network.\footnote{We conducted a binary search to identify the $\hat{\beta}$ values for which the Watts-Strogatz networks' average clustering coefficients match those of the LLM-generated networks.}
In \cref{fig:principle_5}, we plot the estimated values for $\hat \beta$ for each value of $\beta$ and each temperature. Here, $P$-values result from a t-test comparing with the average shortest path length of Watts-Strogatz with rewiring probability $\hat \beta$.
 These results show that while the average shortest path lengths are not identical, they are sufficiently close, with the differences not being statistically significant at the $0.1$ level for most temperature settings. Finally, as \cref{fig:principle_5} shows, the relation $L \sim \log (n)$ holds for different LLM models and environments. 
 
In conclusion, our analysis demonstrates that LLM-generated networks exhibit key small-world properties, with logarithmic scaling of average shortest path lengths and inverse logarithmic scaling of average clustering coefficients. While these networks do not perfectly align with the Watts-Strogatz model, they exhibit similar structural characteristics.

\subsection*{Decisions on Real-World Networks with Heterogeneous Agents} \label{sec:real_world_networks}

We investigate the behavior of LLMs in real-world network formation contexts with four datasets in two differing real-world domains. 
Despite the significant advancements in social network analysis over recent years, the availability of fully complete and comprehensive network datasets remains exceptionally rare~\cite{yuan2018interpretable}.
We employ three datasets from the \textit{Facebook100} collection \cite{traud2012social} and the telecommunication (\textit{Andorra}) and the employment (\textit{MobileD}) datasets from \cite{yuan2018interpretable}. The Facebook100 data correspond to ``friendship'' networks from one hundred American colleges and universities, captured at a specific moment from Facebook's online social network. The Andorra dataset contains nationwide call records in Andorra from July 2015 to June 2016, where calls correspond to mutual calls between Andorran residents containing information about the caller's and the callee's location, phone type, and usage. Finally, the MobileD dataset corresponds to a company network where relations correspond to call or text communication, and each employee is either a manager or a subordinate.

For all network datasets, the agents have \textit{heterogeneous} profiles (i.e., profiles with different features) whose statistics (degree distribution, clustering coefficient distribution, assortativity) we report in \cref{fig:statistics}.

To infer the models' tendencies, we employ a discrete choice modeling framework \cite{overgoor2019choosing,mcfadden1972conditional}. Specifically, we model the network formation process as a discrete choice process, wherein nodes are sequentially prompted to form connections from a set of available alternatives (see \cref{app:dcm}).

\subsubsection*{Candidate Set Construction for Network Decisions}

At each decision step $t$, a query node $i_t$ selects a link from a set of candidate nodes $A_t$ with size $|A_t| = A$. Given the limited context window of LLMs, we consider two alternative strategies for constructing the candidate set $A_t$:

\paragraph{Uniform Sampling.} We uniformly sample $A$ non-neighbor nodes from the graph. This approach serves as a neutral baseline, ensuring that the alternatives presented to the model are selected without structural or feature-based bias. Uniform sampling reflects a scenario where the agent has no \emph{a priori} ranking or filtering of candidates and evaluates all choices purely based on the features provided in the prompt.

\paragraph{Recommendation-Based Sampling.} \redit{We also consider a more realistic and structured candidate selection method that mimics the behavior of recommender systems. In this approach, we use a supervised link prediction model based on logistic regression to compute the likelihood of a link between each candidate pair $(u, v)$ (cf. \cite{liben2003link}). The model takes as input common structural features known to be predictive of link formation, such as similarity, the number of common neighbors between $u$ and $v$, the preferential attachment score between $u$ and $v$, the Jaccard similarity between the neighborhoods of $u$ and $v$, and the Adamic-Adar index (see \cref{app:sampling_strategies} for a description of the recommendation system and \cref{app:link_prediction} for the effect sizes and AUC of the recommendation system). We then select the top-$A$ highest-scoring nodes as the candidate set $A_t$ for each query node $i_t$. This method mirrors how real-world systems (e.g., social media friend suggestions, hiring portals, content feeds) narrow down decision spaces through algorithmic filtering based on network and user features.}

\paragraph{Hyperparameters.} For the three datasets from Facebook100 are Caltech36 ($n = 769$) Swarthmore42 ($n = 1,659$), and UChicago30 ($n = 6,591$), we set the number of alternatives to be $A = 15$ and randomly sampled from the existing network. For the UChicago30 dataset, we consider a randomly sampled subset of $N = 2,000$ nodes because of the limited context window of the LLM models. For Andorra ($n = 32,812$) and MobileD ($n = 1,982$), we set the number of alternatives to $A = 5$ and consider a randomly sampled subset of $N = 1,000$ nodes.

\input{table_1}

\begin{figure*}[!ht]
    \centering
    \includegraphics[width=0.9\linewidth]{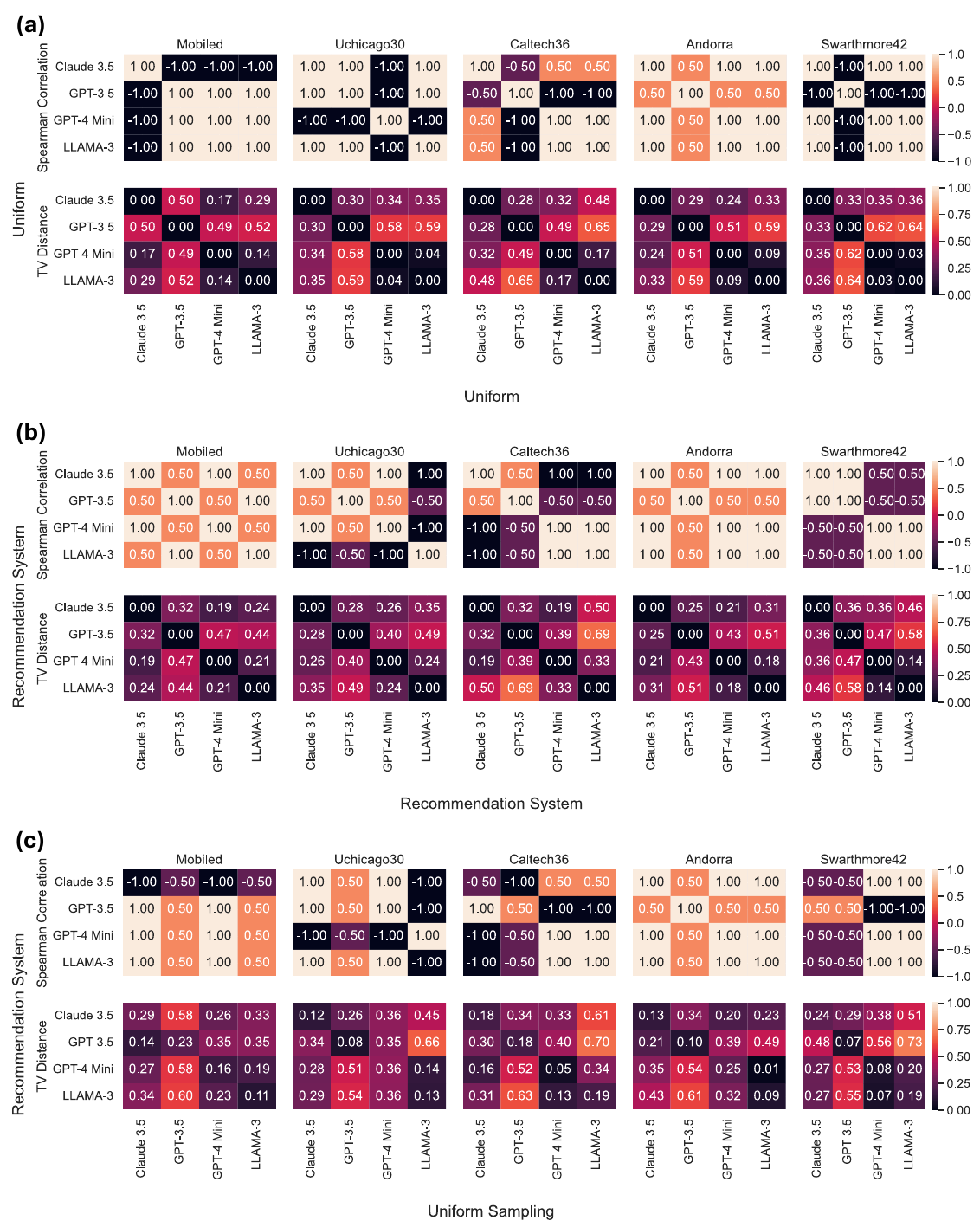}
    \caption{\redit{\textbf{Comparison between the network formation decisions among different models for the uniform and the recommendation system sampling strategies}. We report the Spearman correlation between the effects corresponding to the fits as well as the total variation (TV) distance between the corresponding fitted models (cf. \cref{app:alignment} for a detailed description of the metrics).}}
    \label{fig:models_comparison}
\end{figure*}

\subsubsection*{Regression Coefficients and Model Alignment} 

\paragraph{Uniform Sampling.} We regress network formation decisions on standardized scores reflecting the three micro-level principles. We present the regression results in \cref{tab:real_world_data_environments}.\footnote{More detailed results can be found in \cref{app:regression}.} First, we observe a dominant effect of homophily across all datasets and models. The coefficients for homophily ($\hat \theta_{\mathrm{H}}$) are consistently the largest and highly significant ($P < 0.05$) in almost all cases. For instance, in the Caltech36 dataset, the homophily coefficients for GPT-3.5, GPT-4, and Llama 3 70b Instruct are 0.65, 1.95, and 2.43, respectively ($P<0.001$). The emphasis on homophily suggests that LLMs, much like humans, prioritize forming connections based on shared characteristics.

Second, we find that {preferential attachment plays a secondary role} in the network formation decisions of LLMs. While the coefficients for preferential attachment ($\hat \theta_{\mathrm{PA}}$) are generally positive and statistically significant across most models and datasets, they are notably smaller than those for homophily. For example, in the Swarthmore42 dataset, GPT-3.5 and Llama 3 70b Instruct have preferential attachment coefficients of 0.19 and 0.39, respectively ($P<0.001$). This suggests that while LLMs do consider the degree of potential connection nodes--favoring connections to well-connected nodes--the influence of this factor is less evident compared to homophily.

The influence of triadic closure appears to vary across different datasets and models. In most cases, the coefficients for triadic closure ($\hat \theta_{\mathrm{TC}}$) are positive and significant, indicating that LLMs consider the number of mutual connections when forming new links. However, in some instances, such as with GPT-3.5 on the Andorra dataset, the triadic closure coefficient is negative ($-0.24$) and significant, suggesting a structure-dependent role of this principle as shown by the low clustering coefficient of the network, which is dominated by preferential attachment (cf. \cref{app:statistics}).
This variability implies that while triadic closure is a factor in LLMs' decision-making, its impact may be influenced by the specific characteristics of the dataset or the model used.

\redit{Additionally, in \cref{app:context_window} we report the results for larger context windows ($A \in \{ 50, 100 \}$), and find that our results are robust to larger context windows.}

Thus, our analysis demonstrates that while homophily, triadic closure, and preferential attachment are integral to the network formation behaviors of LLMs, homophily is the dominant factor. 

\paragraph{Recommendation-Based Sampling and Model Alignment.} \redit{In \cref{fig:models_comparison}, we present both the Spearman correlations of these effects and the total variation (TV) distances between the distributions of agent decisions under the fitted discrete choice models (see \cref{app:sampling_strategies} for details on the comparison metrics). 

Our analysis reveals that the key behavioral patterns of LLM agents, namely, the relative strength of their preferences for homophily, triadic closure, and preferential attachment, are largely consistent across different candidate selection strategies. Specifically, we observe high Spearman correlations and low TV distances in the majority of comparisons, both when using the same sampling strategy and when comparing the uniform sampling method to the recommendation-based approach. Homophily still remains the dominant factor in the agent's decisions and is context-dependent even when the agents interact with the recommendation system. 

These results suggest that the observed LLM behaviors are, in the majority, robust to variations in how the candidate set $A_t$ is constructed.}

\paragraph{Average Marginal Effects.} \redit{Due to space constraints, we report the average marginal effects (AMEs) in \cref{app:ame}. The AMEs provide a complementary view of the regression results by quantifying the expected change in choice probability associated with a one-unit change in each standardized feature. Consistent with the coefficient estimates, homophily exhibits the largest AMEs in most datasets, often exceeding 1.0 and reaching above 2.5 under the recommendation-based strategy (e.g., GPT-4 Mini on UChicago30). Preferential attachment also shows consistently positive AMEs, though typically smaller in magnitude than those for homophily, indicating a weaker, yet still significant, propensity to connect to high-degree nodes. Triadic closure effects are more variable: in several Facebook100 networks they are positive, reinforcing local clustering, while in datasets such as Andorra and MobileD they are frequently negative, indicating a tendency to form cross-community links rather than closing triangles. Across datasets, the recommendation-based strategy tends to amplify the dominant mechanism, i.e., most often homophily in Facebook100 networks and preferential attachment in MobileD, highlighting that sampling strategy can strengthen the prevailing behavioral bias while leaving the relative ranking of mechanisms broadly consistent.}

\paragraph{Change of Graph Statistics.} \redit{Finally, the networks resulting from the newly added links preserve the graph statistics (cf. \cite{leskovec2006sampling}), such as the degree distribution, the adjacency matrix spectrum, and the distribution of the sizes of the connected components (see \cref{app:graph_stats_change_sampling} and \cref{app:graph_stats_change}). KS test results show that adding these new edges ($\le 5\%$ new edges) minimally affects global metrics -- degree distribution, spectrum, and component sizes remain largely unchanged --  while significant shifts occur mainly in local clustering, especially under the Uniform strategy. The Recommendation System produces even fewer local changes, indicating that LLM-driven edge additions at this scale largely preserve overall network structure, with strategy choice shaping only localized patterns. The results are also robust to temperature changes (cf. \cref{app:graph_stats_change})}

\begin{figure*}[ht!]
    \centering
    \includegraphics[width=0.9\linewidth]{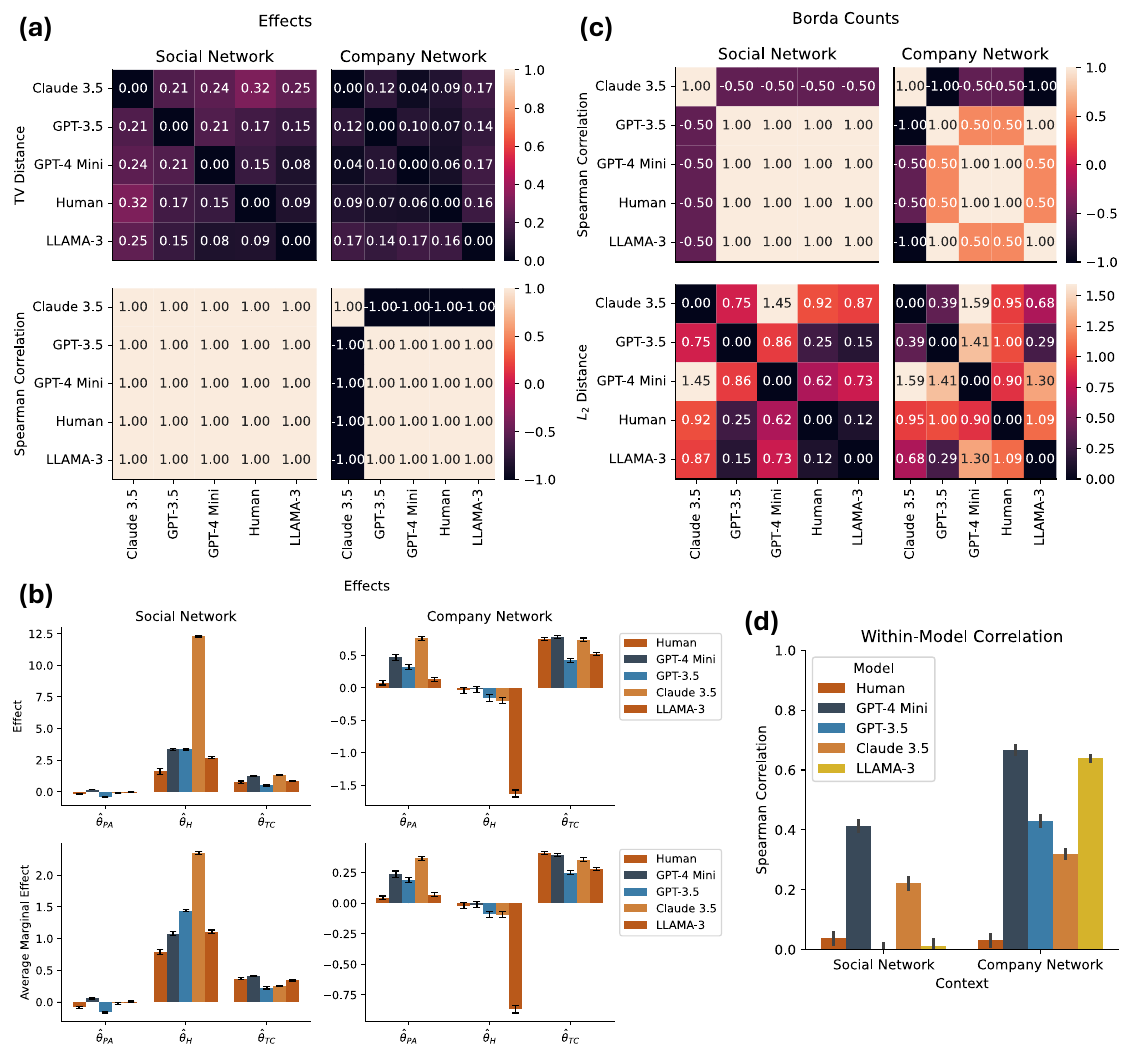}
    \caption{\redit{\textbf{Measurement of LLM-Human alignment for two contexts: Social network ($n = 100)$, and company network ($n = 103$).} \textbf{(a)} TV distance between fitted models and Spearman correlation between estimated effects. \textbf{(b)} Effects and average marginal effects with standard errors to measure alignment between how different models and humans rank preferential attachment, homophily, and triadic closure. \textbf{(c)} Spearman correlation and $L_2$ distance between the average Borda count vectors to measure alignment between how different models and humans rank preferential attachment, homophily, and triadic closure. \textbf{(d)} Within-model Spearman correlations between the decisions within each model to measure alignment within the decisions of one model. \cref{app:human_baseline} contains detailed information on the construction of the baseline.}}
    \label{fig:human_baseline}
\end{figure*}

\subsection*{Human Baseline} 

\paragraph{Survey.} \redit{To assess how closely the network formation preferences of LLM agents align with human decision-making, we conducted a controlled survey-based experiment involving both human participants and LLMs. The experiment was designed to elicit link formation choices in two distinct social contexts: \textit{(i)} a social network where participants assumed the role of a student forming friendships within a college social network, and \textit{(ii)} a company network where participants assumed the role of an employee making professional connections within a company network.

For each scenario, the participant was presented with a focal node (representing themselves) and $A = 3$ candidate profiles containing a value indicating similarity as well as relevant network statistics (degree, common neighbors) with the focal profile. The participant was then asked to select exactly one candidate with whom to form a connection and rate the criteria -- among similarity, degree, and common neighbors -- that they considered when choosing the specific profile. \cref{app:human_baseline} contains more information about the experimental setup.

We recruited human participants via the Prolific platform, ensuring diversity in demographic backgrounds. We obtain $n = 100$ and $n = 103$ responses for the social and company network contexts, respectively. To compare with LLM models, we administered the same dynamically-generated survey inputs that we gave to each participant to an LLM -- preserving the wording, structure, and candidate attributes -- for the five LLM models.  
 
This experimental design allows for a direct, context-controlled comparison between human and LLM decision patterns. By presenting an identical set of alternatives, we can measure alignment in two complementary ways: \textit{(i)} at the \emph{principle level}, by estimating discrete choice models for both humans and LLMs and comparing the inferred effect sizes for homophily, triadic closure, and preferential attachment, and \textit{(ii)} at the \emph{choice level}, by comparing the Borda count vectors between how the participants ranked each criterion in their choices. Furthermore, the two scenarios are chosen to reflect socially and professionally relevant settings as in the case of the real-world datasets.}

\paragraph{LLM-Human Alignment in Network Formation Contexts.} \redit{\Cref{fig:human_baseline} summarizes the results of our alignment study. Across all scenarios, the rankings of network formation principles based on effect sizes are perfectly correlated between humans and LLMs, and the Borda vector averages are likewise almost perfectly correlated (with the exception of Claude~3.5). The total variation (TV) distances between the discrete choice models inferred from human and LLM choices are consistently small -- below $0.32$ in all cases and typically below $0.10$ -- indicating high distributional similarity (cf. panels~(a)-(c)). In the vast majority of cases, the signs of the estimated effects also match between humans and LLMs. Consistent with our earlier observations in \cref{tab:real_world_data_environments}, we find strong homophily in the social network context ($\hat \theta_{\mathrm{H}} > 0; P < 0.001$) and strong heterophily in the company network context ($\hat \theta_{\mathrm {H}} < 0; P < 0.001$). Additionally, the ranking of the effects is consistent with our findings in \cref{tab:real_world_data_environments}. 

While aggregate-level alignment between human and LLM decisions is high across models (cf. \cref{fig:human_baseline}a-c), notable differences emerge in \emph{within-model} agreement. Specifically, LLMs exhibit substantially higher internal consistency in their decision rankings than humans, who display greater variability in their within-model rankings (cf. \cref{fig:human_baseline}d).}

\section*{Discussion} \label{sec:discussion}

\subsubsection*{Summary of Findings and Broader Impact} In this study, we conducted a comprehensive evaluation of LLMs' network formation preferences, examining both micro-level network principles--such as preferential attachment, triadic closure, and homophily--and macro-level network properties like community structure and the small-world phenomenon. Our findings indicate that networks generated by multiple LLMs exhibit these properties, particularly when the models are primed with network statistics like the number of mutual friends or the degrees of potential connections.
Furthermore, using discrete choice modeling, we explored the emergence of these properties in simulations based on real-world networks. Our results reveal that the LLM agents' selections are predominantly driven by homophily, followed by triadic closure and preferential attachment. 

On the one hand, our study enhances our understanding of how multiple LLMs behave in networked settings. Specifically, our findings reveal varying strengths in network formation properties among LLMs, suggesting that when these models are employed to coordinate social networks in social or work environments, they may exhibit human-like behaviors. This has important implications for applications like agent-based modeling, where a realistic simulation of human behavior is crucial.
Traditionally, agent-based models rely on ``formalized'' rules or heuristics to represent individual behaviors, which may not adequately capture the complexity of human decision-making. Given the abilities of LLMs to solve complex tasks that require extensive reasoning, which may be beyond the scope of a traditional agent-based modeling framework, LLM agents provide the potential to simulate more nuanced and context-aware interactions that can potentially resemble human social behavior closely without the need to specify decision rules or heuristics rigidly.

\redit{In addition, our work expands and provides several novel results to the existing literature of findings on LLM-based network simulations  \cite{ashery2025emergent,ferraro2024agent,orlando2025can}, and is the first, to the best of our knowledge, to include a human baseline to verify the findings. Specifically, in our work, we show the emergence of \emph{several topological collective biases}, which are complementary to the linguistic social conventions found in \cite{ashery2025emergent} and \cite{ferraro2024agent}, and the emergence of the friendship paradox on LLM-generated networks, as shown in \cite{orlando2025can}.}  

\subsubsection*{Limitations and Boundary Conditions} On the other hand, our results suggest that we should exercise caution when leveraging LLMs in networking scenarios. The model family, configuration, and prompts can subtly affect the models' behavior, resulting in qualitatively similar but quantitatively different outcomes. 

For example, newer models such as GPT-4 and Claude 3.5 exhibit stronger biases compared to prior models such as GPT-3.5. For instance, in the preferential attachment principle, newer models such as GPT-4 and Claude 3.5 have stronger biases -- i.e., connect to \textit{highest}-degree-nodes yielding star-like networks -- compared to GPT-3.5 and LLama 3, which had weaker biases -- i.e., connect to \textit{high}-degree-nodes. Similar results can be found in homophily, as larger biases towards homophily and triadic closure. Thus, even though LLMs exhibit these principles, we should be cautious of such biases when designing simulations. 
    
Similarly, in our experiments with real-world data, we find an interesting phase transition from homophily to heterophily: In the Facebook100 data, the LLMs generally exhibited positive biases toward homophily ($\hat \theta_{\mathrm H} > 0$). However, in employment networks, agents were either managers or subordinates, and we discovered that LLM agents were heterophilous in such a case ($\hat \theta_{\mathrm H} < 0$), which aligns with career advancement dynamics (i.e., employees want to form links with managers because of better career prospects). 

\redit{Additionally, even though on aggregate LLM and human responses are aligned, the fine-grained picture on the individual level is different: When compared to human responses, the responses given by LLMs are significantly more correlated than the responses obtained by humans (cf. \cref{fig:human_baseline}).} 

The above underscores the need for researchers to provide oversight and ensure that LLM behaviors align with human expectations when employing them in scientific research methods, such as agent-based modeling and even prototypical human subject research with LLMs.

Thus, although we find that LLMs resemble human network formation behaviors, we should consider whether these models should exhibit such behaviors when serving as assistants to humans in work and social lives. Biases like homophily, triadic closure, or preferential attachment may lead to network structures that overemphasize certain individuals or fragment information flow. 

% Humans form networks with communities and central nodes partly due to limited social capacities; however, LLMs do not share these limitations. 

Therefore, when used as social assistants, LLMs may not necessarily need to mirror human networking behaviors and could be personalized to promote more equitable and efficient information dissemination. Our study highlights the need for more deliberate efforts to align LLM behavior in this domain.

\subsubsection*{Future Research Directions} Looking ahead, we identify three promising directions for future research: First, investigating LLM behavior in more complex interactions, such as simulated dialogues, could provide deeper insights. By examining the specific dialogues of LLMs during interactions, we can better understand their network formation preferences and how they adapt to different social dynamics.
Second, we could explore how LLMs can be integrated into real-world settings, such as social media platforms. LLM-assisted bots might be employed to facilitate interactions, break echo chambers, and moderate democratic discussions (see, for example, \cite{papachristou2023leveraging,anthis2025llm}).

\redit{An important direction for future work is to more explicitly integrate the synthetic and empirical analyses. While our current study shows that the same micro-level principles (homophily, triadic closure, preferential attachment) emerge across both settings, a systematic structural comparison would further validate external robustness. For instance, one can quantify the similarity of LLM-generated synthetic graphs and empirical benchmarks using metrics such as degree distribution, clustering coefficient, assortativity, modularity, or spectral properties. Moreover, causal inference techniques (e.g., counterfactual link removal or intervention-based analyses) could help identify whether the same underlying mechanisms explain network growth in both idealized and real-world scenarios. Such approaches would not only improve interpretability but also establish stronger guarantees that LLM-driven behaviors generalize beyond controlled synthetic environments.}

Finally, we can use our methods to create realistic synthetic networks. These synthetic datasets can serve as benchmarks for evaluating graph learning methods, thereby addressing the scarcity of existing graph benchmarks (cf. \cite{palowitch2022graphworld}). By adjusting parameters like temperature and environmental settings or selecting different models, we can generate diverse networks to test graph neural network performance under various conditions. Importantly, our approach allows for the generation of artificial data that resembles real-world data while adhering to privacy regulations.

% Furthermore, our study may pave the way for innovative applications of LLMs in network science and computational social science research. Given that LLMs adhere to fundamental network principles and exhibit behaviors akin to human network formation, they hold significant potential for use in simulations of social systems and agent-based modeling. As LLM agents gain increasing ability to exhibit more complex and nuanced behaviors, such as negotiation, collaboration, and competition, this presents opportunities to replicate networked environments like marketplaces or organizational structures. In these settings, LLMs could effectively model decision-making processes and interaction dynamics.

% This research could shed light on the nuances of LLM behavior in realistic settings and inform their integration into diverse social environments.
% Second, we suggest exploring the application of our findings in specific settings, particularly in organizational contexts. For example, considering the potential use of LLMs to assist HR professionals, our study's insights could guide the alignment of LLMs with organizational goals. By leveraging network information, LLMs could help HR departments identify their firms' best fit or talent, optimizing recruitment and talent management processes.

% This has the potential to expedite research in graph learning and trustworthy machine learning by providing realistic network datasets generated by LLMs.

\subsection*{Acknowledgements}

The authors would like to thank Nikhil Garg, Jon Kleinberg, Yanbang Wang, Yuanqi Du, the attendees of the Learning on Graphs NYC meetup, and the participants of the Cornell AI, Policy, and Practice working group for their valuable feedback on the current version of the paper. 

\bibliographystyle{abbrv}
\bibliography{references}

\clearpage
\clearpage

\onecolumn

\setcounter{page}{1}

\begin{center}
    {\huge \textbf{Supplementary Material for \\ ``Network Formation and Dynamics among Multi-LLMs}} \\ \medskip
    {\large Marios Papachristou\footnote{Correspondence to: mpapachr@asu.edu}, Yuan Yuan}
\end{center}

% \appendix

\setcounter{table}{0}
\setcounter{figure}{0}

\renewcommand\thefigure{SI.\arabic{figure}}    
\renewcommand\thetable{SI.\arabic{table}}
\renewcommand\thealgorithm{SI.\arabic{algorithm}}
\renewcommand{\thesection}{\Alph{section}}

\section{Methods and Materials} \label{app:methods}

\subsection{Experimental Procedure} \label{sec:link_formation}

In our study, we performed experiments to assess whether key network principles at both the micro-level (such as preferential attachment, triadic closure, and homophily) and the macro-level (including community structure and weak ties) align with classical network models. Subsequently, we utilized real-world networks to determine the factors that are most heavily weighted by LLMs.

\subsubsection{Network Formation Process} Our experiments span a time series of $T$ steps, with a sequence of network structures denoted as $G_1, G_2, \dots, G_{T}$ with vertex sets $V_1, \dots, V_T$. The initial network, $G_1$, is referred to as the \emph{seed network}. At each step $t$, we select a \emph{query node} $i_t$ (which may either be a new arrival or an existing node in the graph) and assign it the task of forming new links. This is accomplished by selecting nodes from a set of alternatives $A_t$ (meaning potential candidates for link formation) and initiating a query call $\mathcal Q(A_t, i_t, \delta)$ to the LLM (as outlined in \cref{alg:prompt}) to create up to $\delta$ new links. The edge set selection process involves presenting the LLMs with personal or network features of the alternatives, denoted as $F(A_t) = \left\{f_a : a \in A_t \right\}$, which may include information such as the neighbors of the nodes, node degrees, common connections with $i_t$, and community memberships, formatted in JSON. We adopt a zero-shot learning approach, avoiding the provision of examples to the model to prevent bias, in line with relevant studies such as \cite{brookins2023playing}. This approach allows for the exploration of the innate preferences of LLMs.

We employ multiple temperatures to account for the variability in response generation by LLM systems, which is also observed in classical statistical models of network formation \cite{jackson2008social}. Our study conducts experiments using three temperatures for all models except Claude 3.5: 0.5, 1.0, and 1.5. For Claude 3.5 the temperature range is between 0 and 1, and we run experiments with two temperatures: 0.5, and 1.0.

Moreover, the model is tasked with outputting a JSON object indicating the node chosen for link formation and the rationale behind the choice. This approach is adopted because LLMs have demonstrated proficiency in processing code-like structures, such as HTML and JSON.

\subsubsection{Prompts} 

The general prompt we use is given in \cref{alg:prompt}. An example of this prompt is given at \cref{alg:prompt_example}.

\begin{algorithm}[htpb!]
    \footnotesize
    \caption{Example prompt regarding social network data.}
    \label{alg:prompt_example}
    \begin{lstlisting}[mathescape=true]
    # Task
    You are located in a school. Your task is to select a set of people to be friends with.

    # Profile
    Your profile is given below after chevrons:
    <PROFILE>
        {
            "name" : "Person 0",
            "favorite subject" : "Chemistry", 
            "neighbors" : ["Person 3", "Person 432", "Person 4", "Person 3", "Person 32"]
        }
    </PROFILE>

    # Candidate Profiles
    The candidate profiles to be friends with are given below after chevrons:

    <PROFILES>
    [
        {
            "name" : "Person 1",
            "favorite subject" : "Mathematics",
            "neighbors" : ["Person 3", "Person 4", "Person 23", "Person 65"]
        },
        {
            "name" : "Person 33",
            "favorite subject" : "History",
            "neighbors" : ["Person 342", "Person 2", "Person 12"]
        }, ...
    ]
    
    </PROFILES>

    # Output
    The output should be given a list of JSON objects with the following structure

    [
        {{
            "name" : name of the person you selected,
            "reason" : reason for selecting the person
        }}, ...
    ]

    # Notes
    - The output must be a list of JSON objects ranked in the order of preference.
    - You can make at most 1 selection.
    
    \end{lstlisting}

\end{algorithm}

\begin{algorithm}[htpb!]
    \footnotesize
    \caption{General Prompt used to implement $\mathcal Q(A_t, i_t, \delta)$.}
    \label{alg:prompt}
    \begin{lstlisting}[mathescape=true]
    # Task
    Your task is to select a set of people to be friends with.

    # Profile
    Your profile is given below after chevrons:
    <PROFILE>$F(\{ i_t \})$</PROFILE>

    # Candidate Profiles
    The candidate profiles to be friends with are given below after chevrons:

    <PROFILES>$F(A_t)$</PROFILES>

    # Output
    The output should be given a list of JSON objects with the following structure

    [
        {{
            "name" : name of the person you selected,
            "reason" : reason for selecting the person
        }}, ...
    ]

    # Notes
    - The output must be a list of JSON objects ranked in the order of preference.
    - You can make at most $\delta$ selections.
    
    \end{lstlisting}

\end{algorithm}

\newpage

\subsubsection{Feature Representations for Prompts} \label{app:prompt_features}

Below, we give examples of the features used in the prompt presented in \cref{alg:prompt}. The features are formatted as a list of JSON objects which are provided to the prompt. 

\noindent \underline{\textit{Principle 1: Preferential Attachment.}} We have the following features:

\begin{lstlisting}
    [
        {
            "name" : 0,
            "neighbors" : [5, 7, 1, 6]
        }, 
        ...
    ]
\end{lstlisting}

% And for the degree-based information (Figure 1(b)) we have the following features: 

% \begin{lstlisting}
%     [
%         {
%             "name" : 0,
%             "degree" : 4
%         }, 
%         ...
%     ]
% \end{lstlisting}

\noindent \underline{\textit{Principle 2: Triadic Closure.}} We have the following features:

\begin{lstlisting}
    [
        {
            "name" : 0,
            "common_neighbors" : [5, 7, 1, 6]
        }, 
        ...
    ]
\end{lstlisting}

% And for the degree-based information (Figure 2(b)) we have the following features: 

% \begin{lstlisting}
%     [
%         {
%             "name" : 0,
%             "num_common_neighbors" : 4
%         }, 
%         ...
%     ]
% \end{lstlisting}

\noindent \underline{\textit{Principle 3: Homophily.}} We have the following features:

\begin{lstlisting}
    [
        {
            "name" : 0,
            "favorite_color" : "red",
            "hobby" : "hiking", 
            "location" : "Boston"
        }, 
        ...
    ]
\end{lstlisting}

\noindent \underline{\textit{Principle 5: Small-World.}} We have the following features:

\begin{lstlisting}
    [
        {
            "name" : 0,
            "neighbors" : [5, 7, 1, 6]
        }, 
        ...
    ]
\end{lstlisting}

\noindent \underline{\textit{Real-World Data.}} We have the following features:

\begin{lstlisting}
    [
        {
            "name" : 0,
            "status" : "student",
            "major" : 10,
            "second major" : 93,
            "accommodation" : "house",
            "high_school" : 5,
            "graduation_year" : 2008
        }, 
        ...
    ]
\end{lstlisting}

We note that the initial Facebook100 dataset included gender information as a feature. We chose not to include gender as one of the features, as it has been shown that language models exhibit gender bias \cite{vig2020investigating,kotek2023gender,bordia2019identifying}. An example of the prompt using real-world social network data is given at \cref{alg:prompt_example}.

\subsubsection{Robustness Checks and Prompt Sensitivity} \label{sec:robustness_checks_ablations} 

We tried the following LLM models: 

\begin{itemize}
    \item GPT-3.5 (\texttt{gpt-3.5-turbo})
    \item GPT-4o Mini (\texttt{gpt-4o-mini})
    \item Llama 3 (\texttt{llama-3-70b-instruct})
    \item Claude 3.5 Sonnet (\texttt{claude-3-5-sonnet-20240620}). 
\end{itemize}

For each of the models except Claude 3.5 we used three temperatures: 0.5, 1.0, 1.5. For the Claude 3.5 model we used temperatures 0.5 and 1.0 (since the model does not allow temperatures above 1.0). \redit{Finally, we experimented with different contexts (e.g., friendship, collaboration, community) to test prompt sensitivity.}

\subsection{Details for Small-World Experiments} \label{app:small_world_construction}

The algorithm for the altered Watts-Strogatz model is described as follows:

\begin{compactenum}
    \item Similarly to Watts-Strogatz, we first create a ring network with $n$ nodes. After that, for each node $[n]$, we create $k$ edges where $k / 2$ edges connect to its rightmost neighbors and $k / 2$ edges connect to its leftmost neighbors.
    \item To create $G_t$, for each node $[n]$, we take its $k / 2$ rightmost neighbors and rewire them with probability $\beta$. For each of the $k / 2$ rightmost neighbors that are to be rewired, we make one query to the LLM, which indicates how the edge will be rewired. The choice is made by providing the LLM with all the network nodes and each node's neighbors (i.e., the network structure). 
\end{compactenum}

The model closely resembles the Watts-Strogatz model, with the primary distinction being the method of edge rewiring. Instead of randomly selecting edges for rewiring, as in the Watts-Strogatz model, we determine the rewiring of an edge by inquiring about the LLM and providing it with the current network structure.

\subsection{Real-World Network Experiments} \label{app:real_world_experiments_fit}

\subsubsection{The Discrete Choice Model in Real-World Network Experiments} \label{app:dcm} 

For each node $i_t$ that we consider at time $t$, we randomly remove one of its current friends from the real-world network. After we remove a neighbor for each of $i_1, \dots, i_T$, we end up with the network $G_1$, which we use as a seed network for the LLM agents. 

Subsequently, during the link formation process, we present each node $i_t$ with a set of candidate nodes (denoted by $A_t$), comprising one of the previously removed friends and other nodes that are not their friends. We then instruct the LLM to form a link with one of the candidates, providing the attributes of the candidates and the social network structure to aid its decision-making. These choices are made sequentially. 

We use the \textit{utility} of the model for each node for each sequential decision of network formation:
\begin{align*}
    U_{ij, t} = \theta_{\mathrm{PA}} \log d_{j, t} + \theta_{\mathrm{H}} \log w_{ij} + \theta_{\mathrm{TC}} \log c_{ij, t} + \epsilon_{ij, t}.
\end{align*}

\noindent In this equation, $\theta_{\mathrm{PA}}$ measures the strength of preferential attachment based on the degree $d_{j, t}$ of $j$ at step $t$,  $\theta_{\mathrm{H}}$ measures the strength of homophily based on the similarity $w_{ij}$ (i.e. number of common attributes) between $i$ and $j$, and $\theta_{\mathrm{TC}}$ measures the strength of triadic closure, based on the number of common neighbors $c_{ij, t}$ between $i$ and $j$ at step $t$. The error term $\epsilon_{ij, t}$ is distributed as i.i.d. standard Gumbel.\footnote{The standard Gumbel distribution has CDF $e^{e^{-x}}$.}
All variables are first normalized based on their range, and then the log transformation is taken.

The multinomial logit model (MNL) indicates that the probability that $i$ links to $j$ at step $t$ is given by

{\small
\begin{align} \label{eq:mnl_prob}
    p_{ij, t} = \Pr \left [ \mathrm{argmax}_{r \in A_t} U_{ir, t} = j \right ] = \frac {d_{j, t}^{\theta_{\mathrm{PA}}} w_{ij}^{\theta_{\mathrm{H}}} c_{ij, t}^{\theta_{\mathrm{TC}}}}{\sum_{r \in A_t} d_{r, t}^{\theta_{\mathrm{PA}}} w_{ir}^{\theta_{\mathrm{H}}} c_{ir, t}^{\theta_{\mathrm{TC}}}}.
\end{align}}

Given a sequence of nodes $i_1, \dots, i_T \in V$ and choices (denoted by subscripted $j$) $j_1 \in A_1, \dots, j_T \in A_T$, the parameters can be found by maximizing the log-likelihood function. To get the standard errors of the coefficients and the corresponding $P$-values, we follow the process outlined in \cite{overgoor2019choosing}.

\subsubsection{Estimating the Parameters of the Discrete Choice Model}

To estimate the parameters of the discrete choice model, we optimize the following log-likelihood function: 

{\small
\begin{align} \label{eq:mnl_likelihood}
    (\hat \theta_{\mathrm{PA}}, \hat \theta_{\mathrm{TC}}, \hat \theta_{\mathrm{H}}) =  \argmax_{(\theta_{\mathrm{PA}}, \theta_{\mathrm{TC}}, \theta_{\mathrm{H}}) \in \mathbf R^3} \sum_{t = 1}^T \bigg (  \theta_{\mathrm{PA}} \log d_{j_t, t} + \theta_{\mathrm{H}} \log w_{i_t j_t} + \theta_{\mathrm{TC}} \log c_{i_t j_t, t} -  \log \big (  \sum_{r \in A_t} d_{r, t}^{\theta_{\mathrm{PA}}} w_{i_t r}^{\theta_{\mathrm{H}}} c_{i_t r, t}^{\theta_{\mathrm{TC}}} \big )  \bigg ),
\end{align}
}

\noindent
where $i_1, \dots, i_T$ are the chooser nodes (i.e., the LLM agents who want to form a link), and $j_1, \dots, j_T$ are the nodes which are chosen from the alternative sets $A_1, \dots, A_T$. The likelihood function is convex, and we optimize it with the L-BFGS-B method~\cite{liu1989limited}. The standard errors of the coefficients are approximated as $\sqrt {-H^{-1} / N}$ where $H$ is the Hessian matrix of the log-likelihood at $(\hat \theta_{\mathrm{PA}}, \hat \theta_{\mathrm{TC}}, \hat \theta_{\mathrm{H}})$ and $N$ is the number of data points (cf. \cite{overgoor2019choosing,train2009discrete}).

\subsubsection{Candidate Set Construction for Network Decisions} \label{app:sampling_strategies}

At each decision step $t$, a query node $i_t$ selects a link from a set of candidate nodes $A_t$. Given the limited context window of LLMs, we consider two alternative strategies for constructing the candidate set $A_t$:

\begin{itemize}
    \item \textbf{Uniform Sampling.} We uniformly sample $A$ non-neighbor nodes from the graph. This approach serves as a neutral baseline, ensuring that the alternatives presented to the model are selected without structural or feature-based bias. Uniform sampling reflects a scenario where the agent has no \emph{a priori} ranking or filtering of candidates and evaluates all choices purely based on the features provided in the prompt.
    
    \item \textbf{Recommendation-Based Sampling.} We use a supervised link prediction model based on logistic regression to compute the likelihood of a link between each candidate pair $\{ i, j \}$ following the work of \cite{liben2003link}. The probability of $\{ i, j \}$ being an edge equals $\sigma \left ( \psi^T z_{ij} \right )$ where $\sigma (\cdot)$ is the sigmoid function, $\psi$ is a vector of trainable parameters, and $z_{ij, t} = \left (w_{ij}, c_{ij, t}, d_{i, t} \cdot d_{j, t}, \sum_{k \in N_t(i) \cap N_t(j)} \frac {1} {\log |d_{k, t}|}, \frac {|N_t(i) \cap N_t(j)|} {|N_t(i) \cup N_t(j)|} \right )$ is the feature vector for the pair $\{i, j\}$ with the following entries: 
    
    \begin{itemize}
        \item \textbf{Similarity} between node attributes $w_{ij}$.
        \item \textbf{Number of common neighbors} $c_{ij, t}$.
        \item \textbf{Preferential attachment score} $d_{i, t} \cdot d_{j, t}$.
        \item \textbf{Adamic-Adar index} $\sum_{k \in N_t(i) \cap N_t(j)} \frac {1} {\log |d_{k, t}|}$.
        \item \textbf{Jaccard Similaririty} between node neighborhoods  $\frac {|N_t(i) \cap N_t(j)|} {|N_t(i) \cup N_t(j)|}$.
    \end{itemize}

    To avoid structural bias due to the LLM agents, the logistic regression model is trained at $t = 1$, i.e., prior to any link formation by the LLM agents, as:

    \begin{align*}
        \hat \psi = \argmax_{\psi \in \mathbf R^5} \left \{ \sum_{\{ i, j \} \in E_+} \log \sigma \left ( \psi^T z_{ij, 1} \right )  + \sum_{\{i, j\} \in E_-} \log \left ( 1 - \sigma(\psi^T z_{ij, 1}) \right )  \right \},
    \end{align*}

    where $E_+$ and $E_-$ are equally sized sets of positive and negative edges sampled from $G_1$. For each $t$, the set $A_t$, which is presented to the LLM agent, is constructed as the top-$A$ highest-scoring nodes concerning node $i_t$, where the scores are computed as $\hat y_{i_t, j} = \sigma(\hat \psi^T z_{i_t j, t})$. To train the logistic regression model, we use the \texttt{statsmodels} Python package. Information about the parameters of the link prediction algorithms can be found in \cref{app:link_prediction}. 
\end{itemize}

\subsubsection{Measuring Alignment Between Models} \label{app:alignment}

The plots of \cref{fig:models_comparison} are constructed by measuring similarity between models. Specifically, for two LLM models $\mathcal M$ and $\mathcal M'$, we fit two discrete choice models $\hat \theta^{\mathcal M}$ and $\hat \theta^{\mathcal M'}$ respectively according to \cref{eq:mnl_prob,eq:mnl_likelihood} and measure the following: 

\begin{itemize}
    \item \textbf{Spearman correlation Between Effects.} We measure $\mathrm{Spearman} \left ( \hat \theta^{\mathcal M}, \hat \theta^{\mathcal M'} \right )$, which measures the difference in how the different LLM models rank preferential attachment, homophily, and triadic closure. 
    \item \textbf{TV Distance Between Fitted Models.} We measure the total variation distance between the probabilities that each LLM model assigns to nodes parametrized by the MNL model. Specifically, the TV distance is calculated by sampling a time index $t$ uniformly in $\{1, \dots, T \}$ and a pair of a node $u_t$ and an alternative set $A_t$ from $G_t$ as follows:

    {\small
    \begin{align} \label{eq:tvd}
        d_{TV} \left ( {\mathcal M}, {\mathcal M'} \right ) = \frac 1 2 \mathbb E_{t \sim \mathcal U \left (\{1, \dots, T\} \right )} \mathbb E_{(u_t, A_t) \sim G_t} \left [ \sum_{v \in A_t} \left | \frac {d_{v, t}^{\hat \theta^{\mathcal M}_{\mathrm{PA}}} w_{uv}^{\hat \theta^{\mathcal M}_{\mathrm{H}}} c_{uv, t}^{\hat \theta^{\mathcal M}_{\mathrm{TC}}}}{\sum_{r \in A} d_{r, t}^{\hat \theta^{\mathcal M}_{\mathrm{PA}}} w_{ur}^{\hat \theta^{\mathcal M}_{\mathrm{H}}} c_{ur, t}^{\hat \theta^{\mathcal M}_{\mathrm{TC}}}} - \frac {d_{u, t}^{\hat \theta^j_{\mathrm{PA}}} w_{uv}^{\hat \theta^{\mathcal M'}_{\mathrm{H}}} c_{uv, t}^{\hat \theta^{\mathcal M'}_{\mathrm{TC}}}}{\sum_{r \in A} d_{r, t}^{\hat \theta^{\mathcal M'}_{\mathrm{PA}}} w_{uv}^{\hat \theta^{\mathcal M'}_{\mathrm{H}}} c_{uv, t}^{\hat \theta^{\mathcal M'}_{\mathrm{TC}}}} \right | \right ].
    \end{align}
    }

    In our implementation, we report the Monte-Carlo estimate of \cref{eq:tvd}. 
    
\end{itemize}

\subsection{Human Baseline} \label{app:human_baseline}

\subsubsection{Survey} We construct the survey by presenting three alternatives ($A = 3$) to each participant in two different contexts (social network, company network). For each participant $i$, and each context $\omega \in \{ \text{Social Network},\\ \text{Company} \}$, we generate a dynamic survey with alternative set $A_{i, \omega}$. For each alternative $j \in A_{i, \omega}$ we generate: 

\begin{enumerate}
    \item \textbf{Number of friends (degree) $d_{j, \omega}$ .} Uniformly sampled integer between 0 and 1000.
    \item \textbf{Number of common friends (common neighbors) $c_{ij, \omega}$.} Uniformly sampled integer between 0 and the number of friends (degree).
    \item \textbf{Similarity $w_{ij, \omega}$.} Uniformly sampled integer between 0 and 10 for the social network context, indicating the number of common interests. Uniformly sampled binary variable determining the role of the person (0 for co-worker, 1 for manager).
\end{enumerate}

For the social network context, we ask the participant (focus profile) to pretend to be a college student. For the company network context, we ask the participant to pretend to be an employee (non-manager). The participants are asked how they ranked -- from 1 to 3 (1 = worst, 3 = best) -- the three attributes when forming their decision and optionally give the reasoning behind their decisions.

\paragraph{IRB Approval.} The study protocol was approved by the Institutional Review Board of Cornell University (Protocol \#IRB0150009), and all participants provided informed consent prior to participation.

\subsubsection{Sample Size Construction on Prolific} 

We distribute the survey on the Prolific platform on participants located in the United States, who are at least 18 years old and speak English as their primary language. We obtain $n = 100$ responses for the social network context, and $n = 103$ responses for the company context. We use the standard sampling settings for prolific. Participants are allowed to submit the survey only once. 

\subsubsection{Measuring Alignment between models and Humans}

For each participant $i$, context $\omega \in \{ \text{Social Network}, \text{Company} \}$ and alternative set $A_{i, \omega}$ we construct $z_{ij, \omega} = \left (  d_{j, \omega}, c_{ij, \omega}, w_{ij, \omega} \right )$ for each $j \in A_{i, \omega}$. We let $y_{i, \omega}^{\mathcal M} \in A_{i, \omega}$ to be the decision of model/human $\mathcal M$ abd $h_{i, \omega}^{\mathcal M}$ to be the vector of Borda counts for the choice $y_{i, \omega}^{\mathcal M}$. For instance, if a human participant ranked homophily over triadic closure over preferential attachment, their Borda score vector is $h_{i, \omega}^{\mathrm{Human}} = (1, 2, 3)$ assuming that the first dimension corresponds to preferential attachment (number of friends), the second dimension corresponds to triadic closure (number of common friends), and the third dimension corresponds to homophily (similarity).  

\paragraph{Between-Model Agreement.} For each model $\mathcal M$ we fit a discrete choice model $\hat \theta^{\mathcal M}$ similarly to \cref{app:dcm}. We also calculate the average Borda Score vector as $\bar h_{\omega}^{\mathcal M} = \frac 1 n \sum_{i = 1}^n h_{i, \omega}^{\mathcal M}$

To measure between-model agreement we compare the pairs of models according to the following criteria: 

\begin{enumerate}
    \item \textbf{Principle-Level.} For each pair of models we measure $\mathrm {Spearman} \left ( \hat \theta^{\mathcal M}, \hat \theta^{\mathcal M'} \right )$ and $d_{TV}(\mathcal M, \mathcal M')$ similarly to \cref{eq:tvd}. 
    \item \textbf{Choice-Level.} For each pair of models we measure $\mathrm {Spearman} \left ( \bar h_{\omega}^{\mathcal M}, \bar h_{\omega}^{\mathcal M'} \right )$ and the distance $\left \| \bar h_{\omega}^{\mathcal M} - \bar h_{\omega}^{\mathcal M'} \right \|_2$
\end{enumerate}

\paragraph{Within-Model Agreement.} To measure the within-model agreement for model $\mathcal M$ we measure the average Spearman correlation of the Borda count vectors between pairs of participants, i.e. 

\begin{align*}
    \frac {1} {\binom n 2} \sum_{i, j : i < j} \mathrm{Spearman} \left ( h_{i, \omega}^\mathcal M, h_{j, \omega}^\mathcal M \right ).
\end{align*}

\paragraph{Survey Prompt.} We use the prompt described in \cref{alg:prompt_survey} for the survey data.

\begin{algorithm}[htpb!]
    \footnotesize
    \caption{Prompt used to implement the survey. When the context $\omega$ is a social network the \texttt{profile\_text} variable is set to be \texttt{'You are an undergraduate student at a university. You are looking for friends to connect with on a social network.'}. When the context is the company network the \texttt{profile\_text} variable is set to \texttt{'You are an employee at a company. You are looking for colleagues to connect with on a company network.'}}
    \label{alg:prompt_survey}
    \begin{lstlisting}[mathescape=true]
    # Task
    Your task is to select a set of people to be friends with.

    # Profile
    {profile_text}

    # Candidate Profiles
    The candidate profiles to be friends with are given below after chevrons:

    <PROFILES>$F(A_{i, \omega})$</PROFILES>

    # Output
    The output should be as a JSON object with the following structure

    {{
        "name" : name of the person you selected (integer format),
        "reason" : reason for selecting the person,
        "ranking_degree" : ranking of how much you based your decision on the degree of the person (1 = most important, 2 = average important, 3 = least important),
        "ranking_similarity" : ranking of how much you based your decision on the similarity of the person (1 = most important, 2 = average important, 3 = least important),
        "ranking_common_friends" : ranking of how much you based your decision on the number of common friends with the person (1 = most important, 2 = average important, 3 = least important)
    }}

    # Notes
    * The output must be a single JSON object ranked in the order of preference.
    * You can make at most 1 selection.
    * Your output must be contained within the json markdown cue.
    * Rankings must be mutually exclusive, i.e. you cannot have the same ranking for two different attributes.

    \end{lstlisting}

\end{algorithm}

\subsubsection{Robustness Checks and Prompt Sensitivity} \label{sec:robustness_checks_ablations_real_world} 

\paragraph{Models.} We tried the following LLM models: 

\begin{itemize}
    \item GPT-3.5 (\texttt{gpt-3.5-turbo})
    \item GPT-4o Mini (\texttt{gpt-4o-mini})
    \item Llama 3 (\texttt{llama-3-70b-instruct})
    \item Claude 3.5 Sonnet (\texttt{claude-3-5-sonnet-20240620}). 
\end{itemize}

Additionally, to measure alignment with humans we use data collected from humans through the Prolific platform

\paragraph{Temperatures.} For each of the models except Claude 3.5 we used three temperatures: 0.5, 1.0, 1.5. For the Claude 3.5 model we used temperatures 0.5 and 1.0 (since the model does not allow temperatures above 1.0). In the main paper we report results with temperature equal to 0.5. The results are robust to temperature change (see \cref{tab:regression} for results with other temperatures).

\paragraph{Context Window Length.} We perform robustness checks regarding the size of the context window. We use the large context window model gpt-4.1-mini and context windows $A \in \{ 50, 100 \}$ (see \cref{app:context_window}). 

\paragraph{Sampling Strategies.} We perform robustness checks with two policies for generating the alternative sets $A_t$: (i) uniform sampling and (ii) sampling based on a recommendation system based on logistic regression.

\paragraph{Prompt Sensitivity.} We perform prompt sensitivity experiments based on the different contexts and the type of experiment: (i) decisions in real-world networks with real-world data, where we use \cref{alg:prompt}, (ii) survey data, where we use \cref{alg:prompt_survey}. Our findings are robust to different prompts.

\subsection{Data and Code Availability} \label{app:data_code}

Data and code are openly available on GitHub at the following link: 

\begin{center}
\url{https://github.com/papachristoumarios/llm-network-formation}
\end{center}

\noindent The real-world social network data have been taken from the sources of \cite{traud2012social} and \cite{yuan2018interpretable}. 

\newpage

\newpage

\section{Real-World Datasets} 

\subsection{Statistics of Real-World Datasets} \label{app:statistics}

\begin{figure*}[!h]
    \centering
    \includegraphics[width=0.49\linewidth]{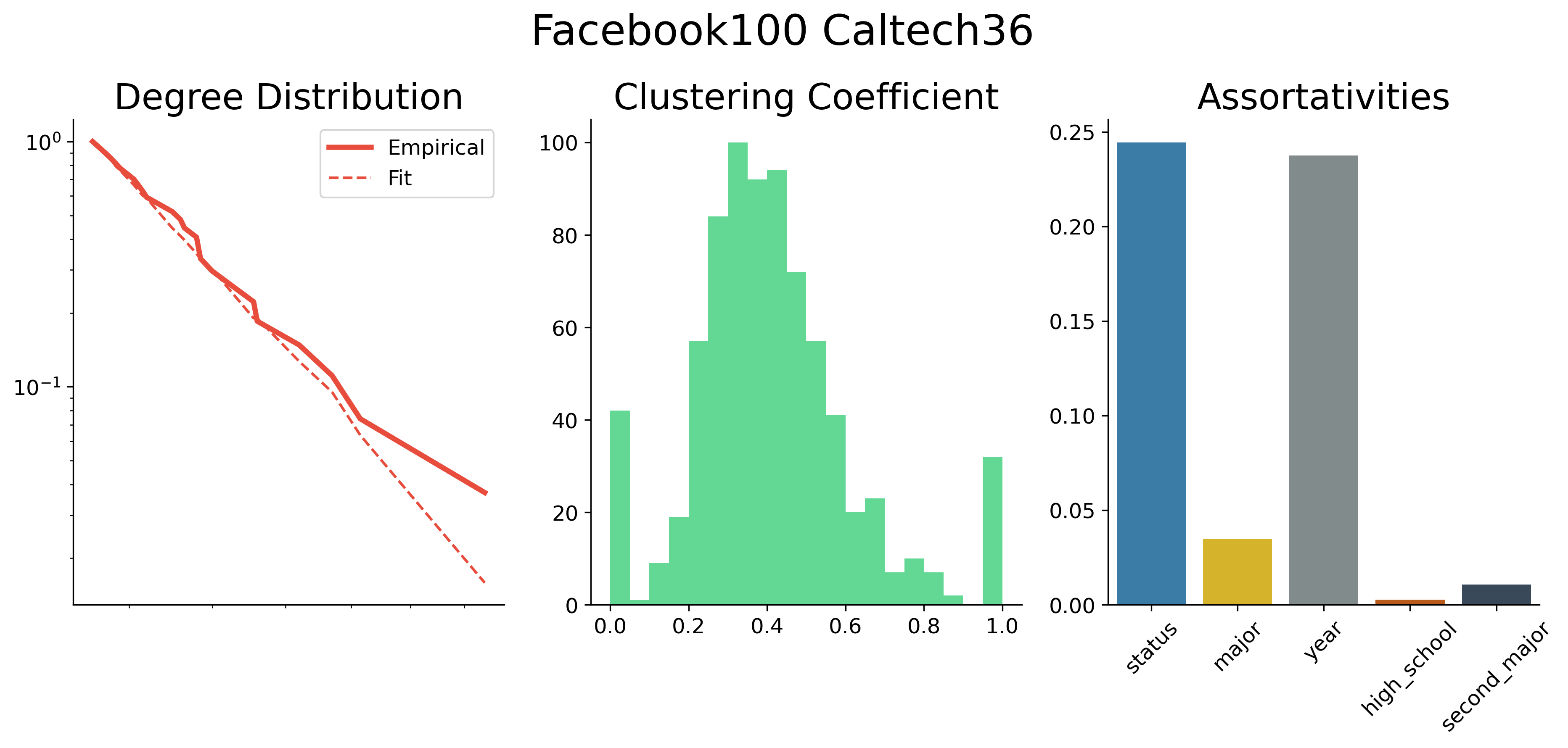}
    \includegraphics[width=0.49\linewidth]{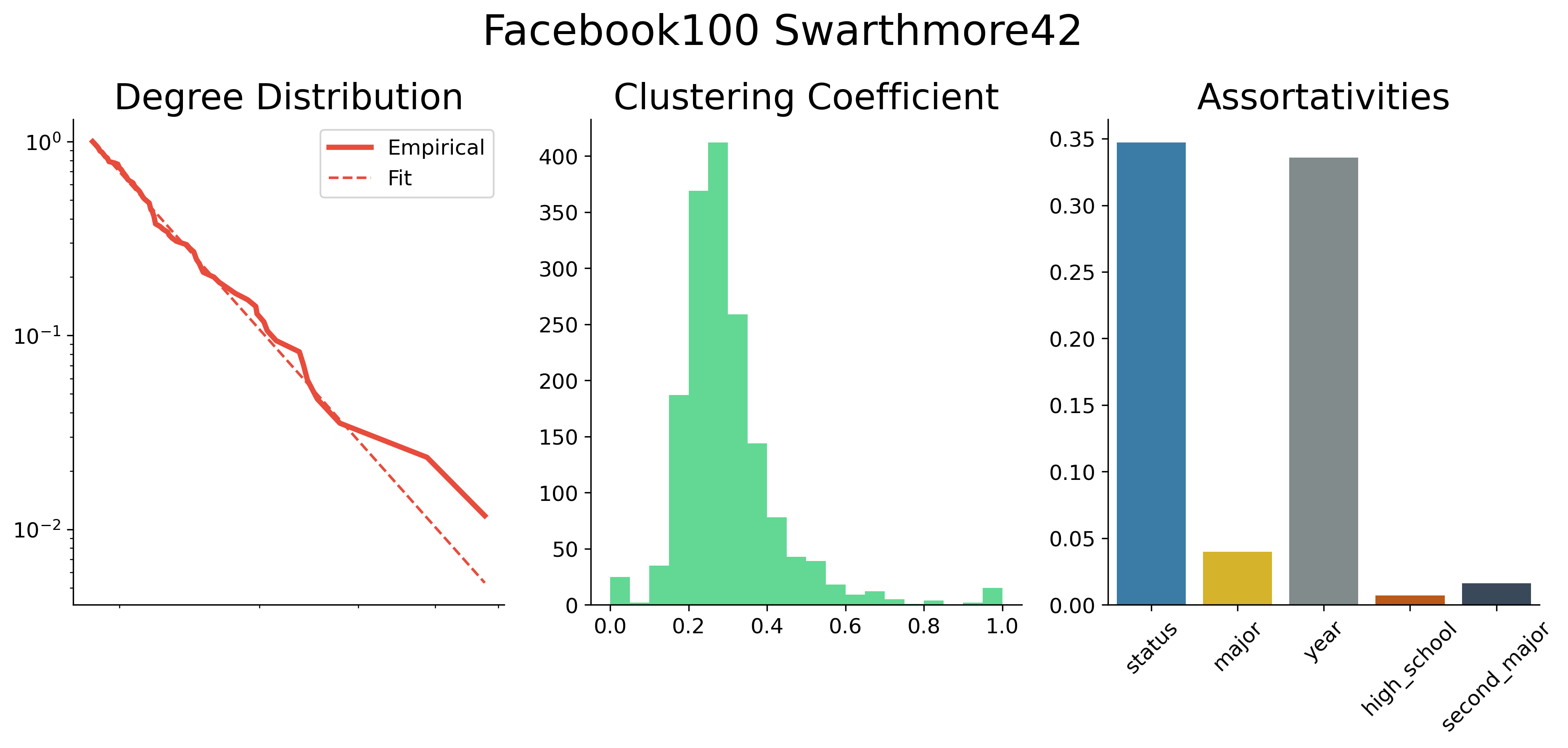}
    \includegraphics[width=0.49\linewidth]{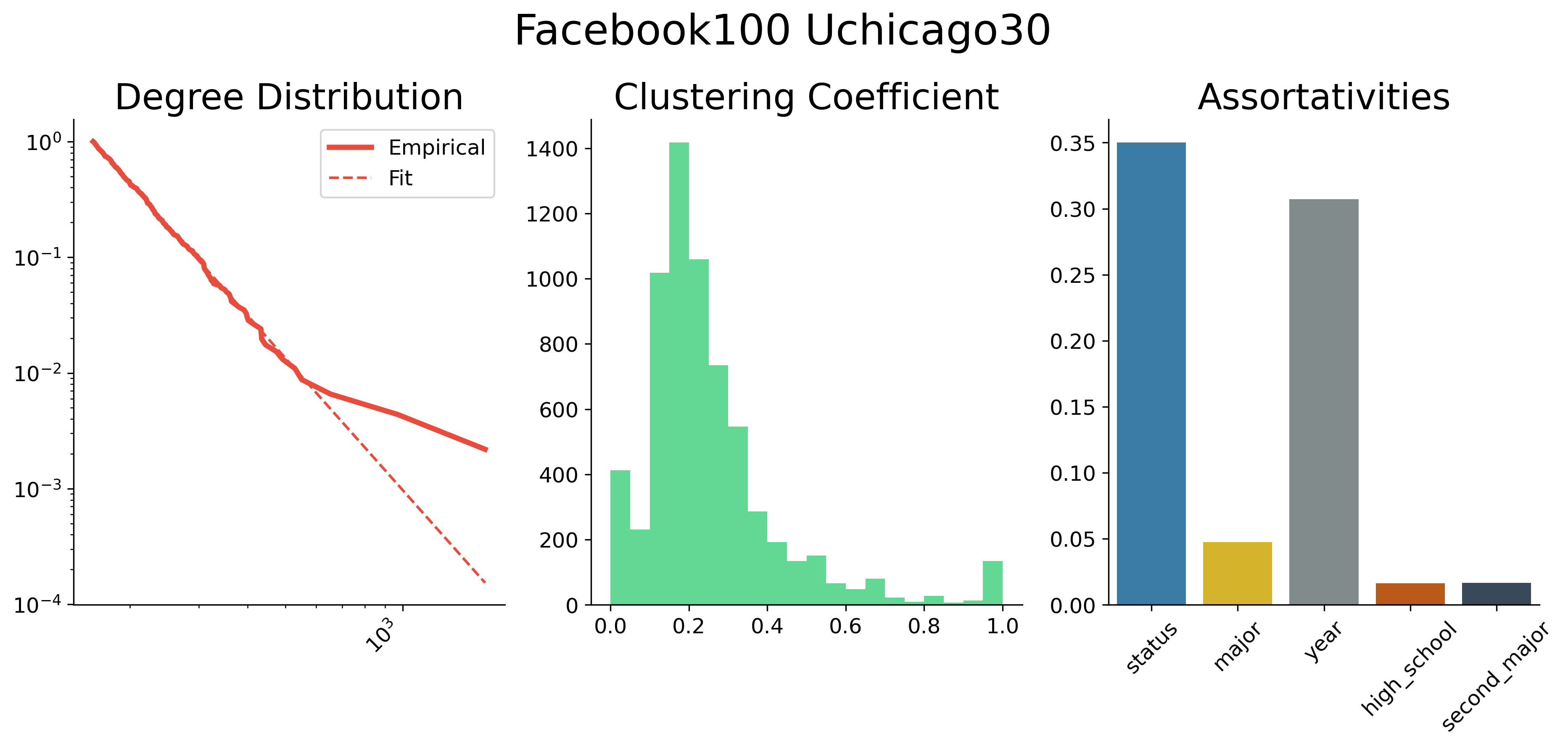}
    \includegraphics[width=0.49\linewidth]{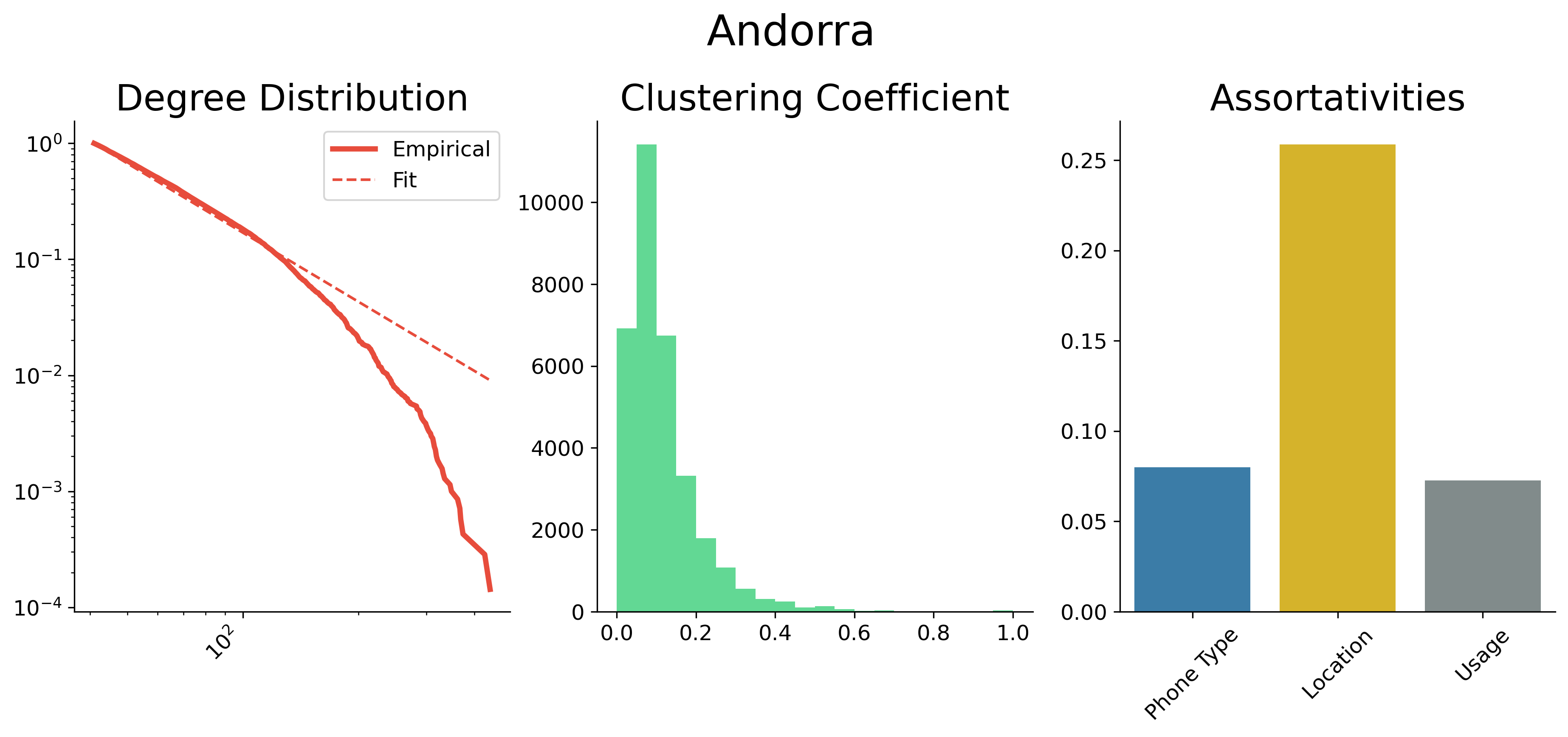}
    \caption{Distributions of real-world datasets analyzed in our study, including degree, clustering coefficients, and the assortativities of the attributes included in the datasets.}
    \label{fig:statistics}
\end{figure*}

\newpage

\subsection{Robustness of Results to Sampling Strategies} 

\subsubsection{Recommendation System Parameters} \label{app:link_prediction}

\input{table_2.tex}

\Cref{tab:link_prediction} shows the effects and the AUC score for the recommender system based on logistic regression (cf. \cref{app:sampling_strategies}).

\newpage

\subsubsection{Change in Graph Statistics due to Different Sampling Strategies} \label{app:graph_stats_change_sampling}

\redit{Across all datasets and models, the Kolmogorov–Smirnov statistics indicate that adding a small fraction of new edges ($\le 5\%$) produces only minor shifts in the degree distribution, spectrum, and sizes of connected components, with most p-values far above the 0.05 threshold. Significant changes arise primarily in the local clustering coefficient—particularly under the Uniform strategy—suggesting localized structural effects without major disruption to global network properties. In contrast, the Recommendation System strategy yields even fewer significant differences, with most metrics remaining statistically indistinguishable from the original graphs. These results indicate that, at this perturbation scale, LLM-driven edge additions preserve the overall network structure, with strategy choice influencing the extent of local structural change.}

\input{table_3}

\newpage

\subsubsection{Average Marginal Effects} \label{app:ame}

\redit{In \cref{tab:ame} we report the average marginal effects (AMEs) per feature for the experiments of \cref{tab:real_world_data_environments}. Our analysis reveals that LLM-driven edge formation is consistently shaped by preferential attachment and homophily, with homophily often exhibiting the largest marginal effects -- frequently exceeding 1.0 and reaching above 2.5 under the Recommendation System strategy. Preferential attachment is positive across all datasets and models, indicating a systematic tendency to link to high-degree nodes. Triadic closure effects are more variable, sometimes reinforcing local clustering and sometimes favoring cross-community connections, particularly under the Recommendation System. Compared to Uniform edge additions, the Recommendation System generally amplifies both preferential attachment and homophily, suggesting that recommendation-driven link formation intensifies these social-network-like biases.}

\input{table_4}

\newpage

\subsection{Robustness of Results to Temperature} 

\subsubsection{Regression Coefficients} \label{app:regression}

In \cref{tab:regression}, we report the regression coefficient for the regression in the real-world network data for all temperatures and GPT-4 (gpt-4-1106-preview). The first column corresponds to the temperature, the next three columns correspond to the fitted coefficients from the regression model of Section 1.C (also shown in Figure 5) accompanied by the standard errors (in parentheses) and the $P$-values indicated by stars (the null hypothesis corresponds to the parameters being set to 0). Next, LL corresponds to the log-likelihood of the fitted model, and AIC corresponds to the Akaike Information Criterion. Finally, we report the percent change in the accuracy compared to random guessing, the percent change in the average path length (as a measure of the small-world phenomenon), and the clustering coefficient (as a measure of the small-world phenomenon and the triadic closure), as well as the $t$-statistic for the change in modularity ($Q$) between the ground truth network dataset (before the edge deletions) and the network after the network formation process.

We observe that $\hat \theta_{\mathrm{H}} > \hat \theta_{\mathrm{TC}} > \hat \theta_{\mathrm{PA}} > 0$ accross all settings. LLM agents do better than random guessing, reinforce the small-world phenomenon, and weaken the triadic closure, though the changes are very small, 0-1\% change for the average path length and up to 10\% change for the clustering coefficient. Finally, the community structure is strengthened after new links are formed.

\input{table_5}

\newpage

\subsubsection{Change in Graph Statistics due to Different Temperatures} \label{app:graph_stats_change}

\redit{Further, to measure the changes in the network statistics, we use the metrics presented in \cite{leskovec2006sampling} to quantify the changes. Specifically, we measure the Kolmogorov-Smirnov statistic and the corresponding $P$-value for the degree distribution, the distribution of the sizes of the connected components, the distribution of the singular values of the adjacency matrix, and the distribution of the local clustering coefficient, (CC) for the real-world Facebook100 networks and the gpt-4-1106-preview model (the results are similar for the other networks and models we examined). Except the changes on the distribution of the sizes of the connected components for UChicago30, we find that most KS statistics are negligible and the corresponding $P$-values are large (e.g. $P \gg 0.5$), indicating that most network statistics are not affected. \cref{tab:graph_stats} summarizes the results:}

\input{table_6}

\newpage

\subsection{Robustness of Results to Large Context Windows} \label{app:context_window}

\redit{We perform experiments with the large-context model gpt-4.1-mini. We set the temperature to 0.5. \cref{tab:context_window} shows the effect sizes. We observe that for social networks (Caltech36, Swarthmore42, UChicago30) homophily still remains the dominant force for large context windows and that in most cases $\hat \theta_H > \hat \theta_{TC} > \hat \theta_{PA} > 0$, agreeing with the results of \cref{tab:real_world_data_environments}. Additionally, for the MobileD network we observe heterophily ($\hat \theta_H < 0$). On the other hand, we observe that for the Andorra dataset, for larger contexts, the triadic closure has a positive effect ($\hat \theta_{TC} > 0$; $P < 0.001$) compared to negative weight in \cref{tab:real_world_data_environments} ($P < 0.001$).}

\input{table_7}

\newpage
\section{Network Evolution and Omitted Simulations} \label{app:omitted}

Here we depict the evolution of the networks generated by the LLM agents, as well as omitted simulations.

\subsection{Principle 1: Preferential Attachment}

\subsubsection*{Network Evolution} We plot the evolution of the LLM-based preferential attachment networks at three timesteps, together with the degree distribution alongside the degree distribution of a BA graph with the same number of nodes. We observe that for the temperature being 0.5 we have a core-periphery-like formation which diverges from the BA model, whereas for the temperature being 1.5 the network has the same degree distribution as the BA model. 

\begin{figure*}[!h]
    \centering
    \includegraphics[width=0.8\textwidth]{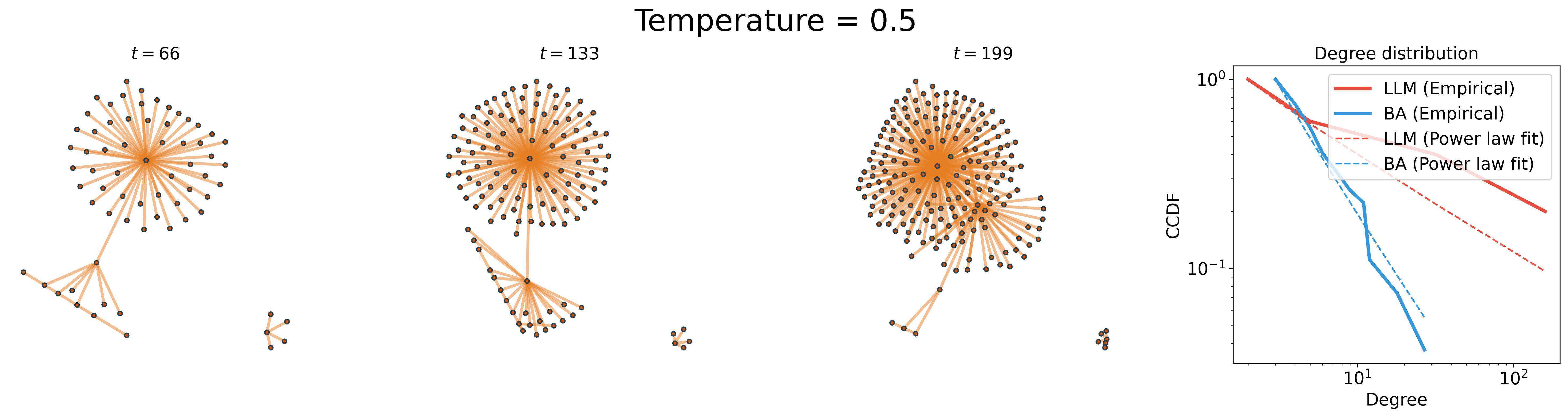}
    \includegraphics[width=0.8\textwidth]{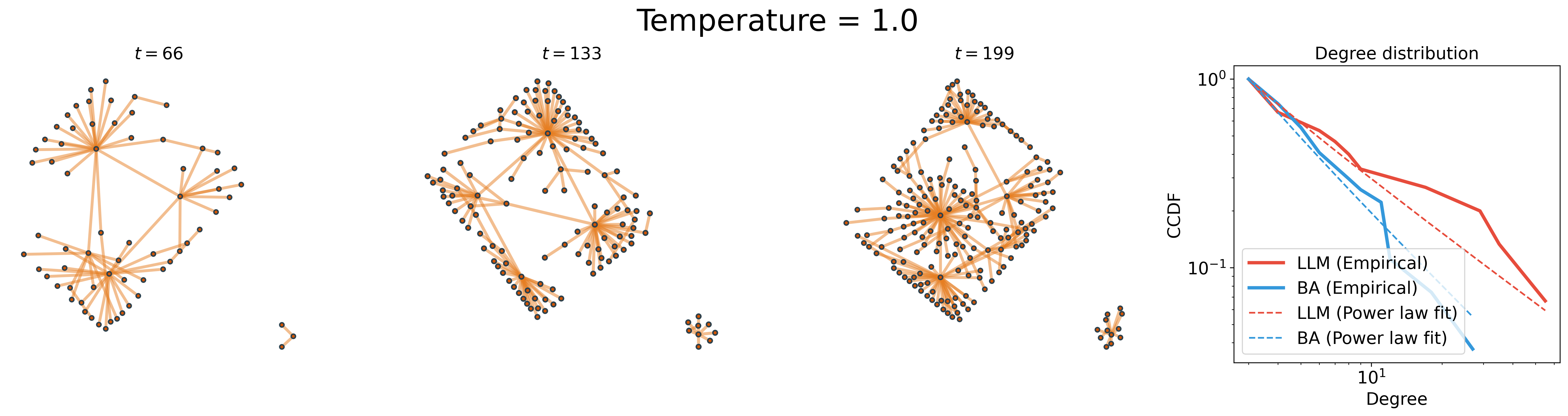}
    \includegraphics[width=0.8\textwidth]{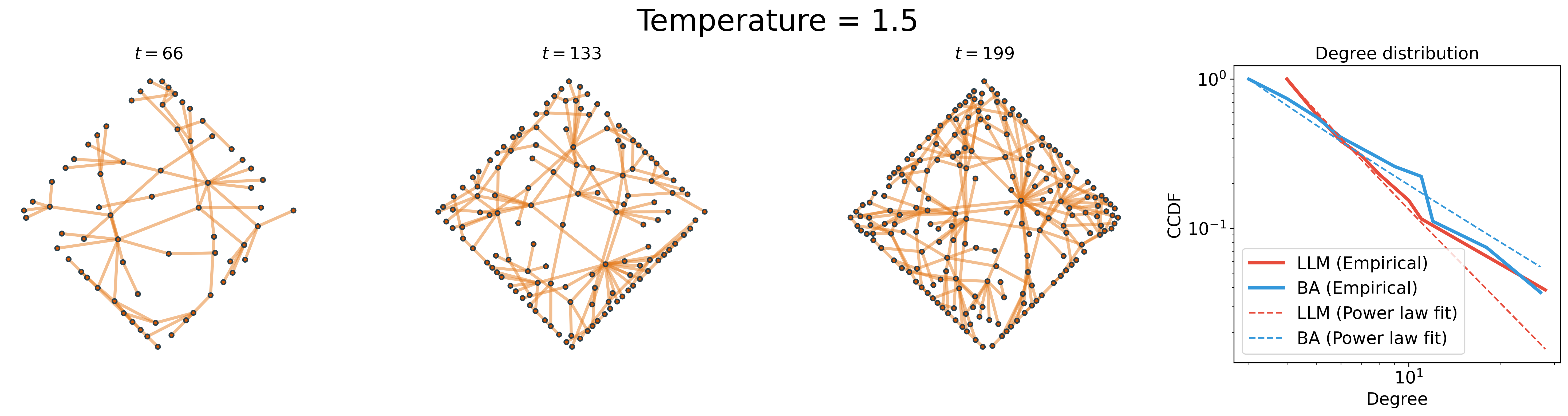}
    \caption{Dynamic evolution of networks created based on Principle 1.}
    \label{fig:princple_1_evolution}
\end{figure*}

\subsubsection*{Simulations with Degree Information}  
\begin{figure*}[t]

    \centering
    {\includegraphics[width=0.9\textwidth]{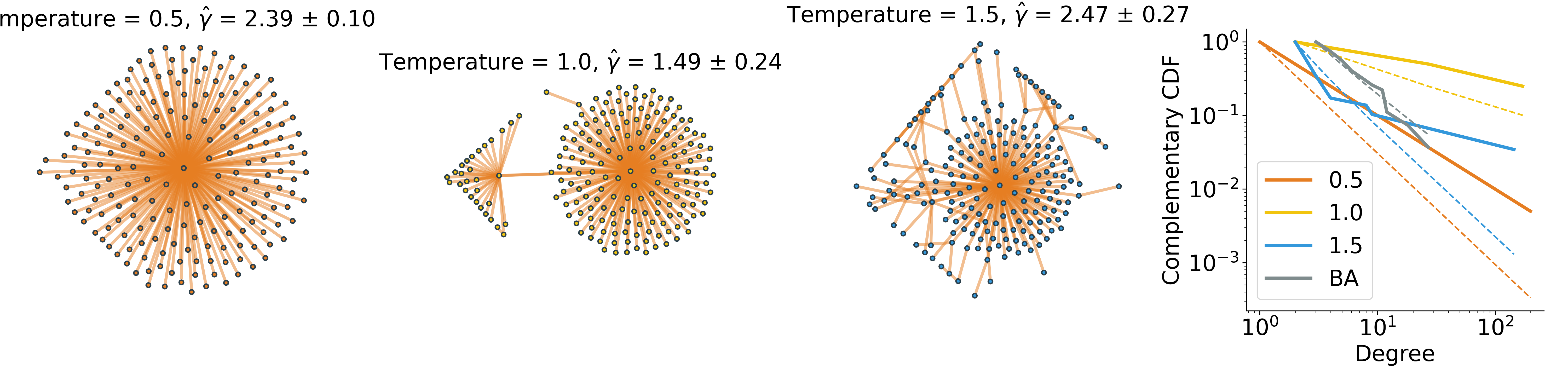}}
    \caption{\textbf{Results for Principle 1 (preferential attachment):} We display simulated networks comprising 200 nodes across different temperatures. For the degree-based simulations, node degree data $\{ d_{j, t} : j \in V_t \}$ was provided ($V_t$ corresponds to the vertex set of the network $G_t$ at round $t$). With degree information only, the networks form more unrealistic star-like structures, diverging from scale-free configurations and more closely mirroring a core-periphery network structure.
    }
    \label{fig:app_principle_1_final_graphs}
\end{figure*}

In \cref{fig:app_principle_1_final_graphs} we provide the results with degree-information only. We observe that the agents form connections around high-degree nodes only (see \cref{fig:app_principle_1_final_graphs}). The same result (star-like networks) holds for the other LLM models and temperatures.

\subsection{Principle 2: Triadic Closure}

\subsubsection*{Network Evolution} We plot the evolution of the LLM-generated networks based on the triadic closure principle, together with the transitivity measure and the algebraic connectivity (which corresponds to the second-smallest eigenvalue of the graph Laplacian). We observe that the algebraic connectivity gradually increases as new edges between the clusters are created. Specifically, the algebraic connectivity reaches a higher value for higher temperatures, indicating the more frequent creation of new intra-cluster edges. Moreover, we observe that the transitivity initially increases and then decreases until it reaches its final value. 

\begin{figure*}[!h]
    \centering
    \includegraphics[width=0.8\textwidth]{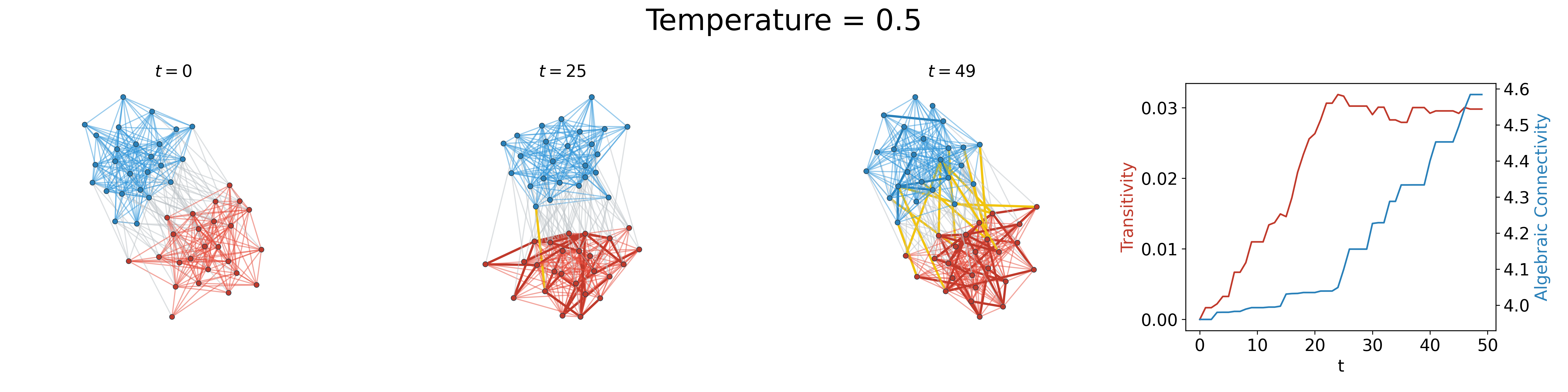}
    \includegraphics[width=0.8\textwidth]{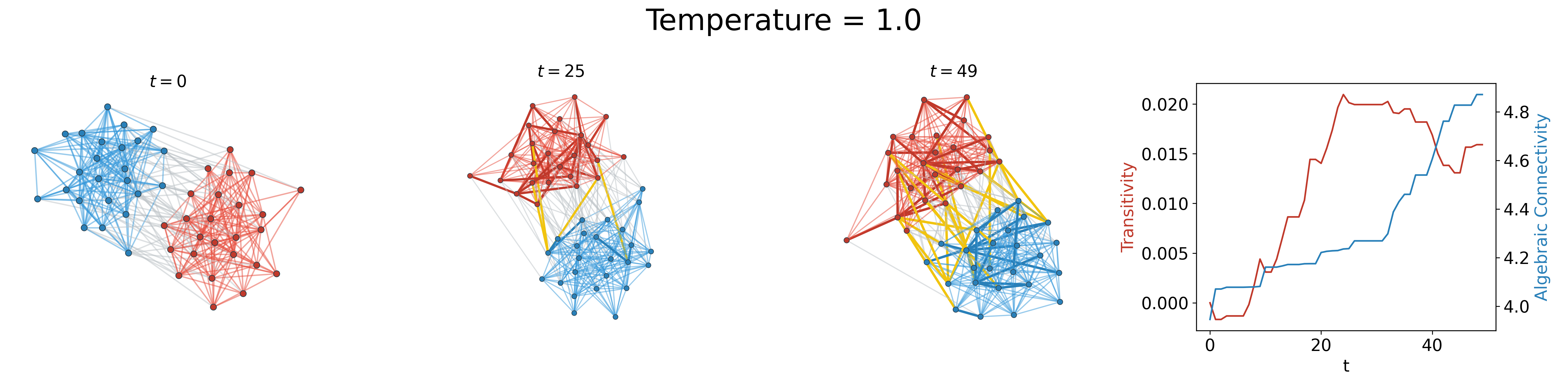}
    \includegraphics[width=0.8\textwidth]{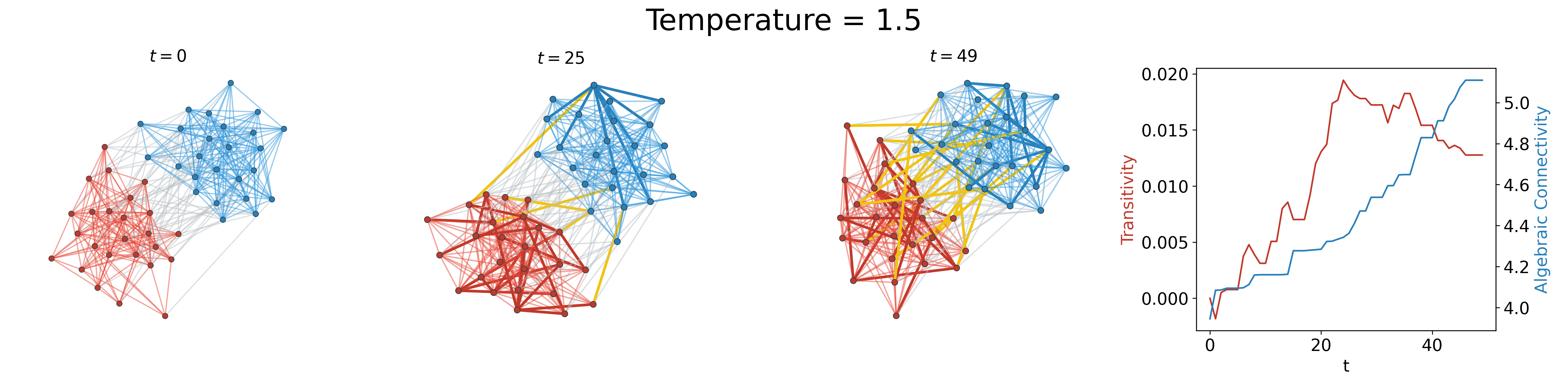}
    \caption{Dynamic evolution of networks created based on Principle 2.}
    \label{fig:princple_2_evolution}
\end{figure*}

\subsubsection*{Simulations with the Number of Common Neighbors} 
\begin{figure*}[!h]
    \centering
    {\includegraphics[width=0.9\textwidth]{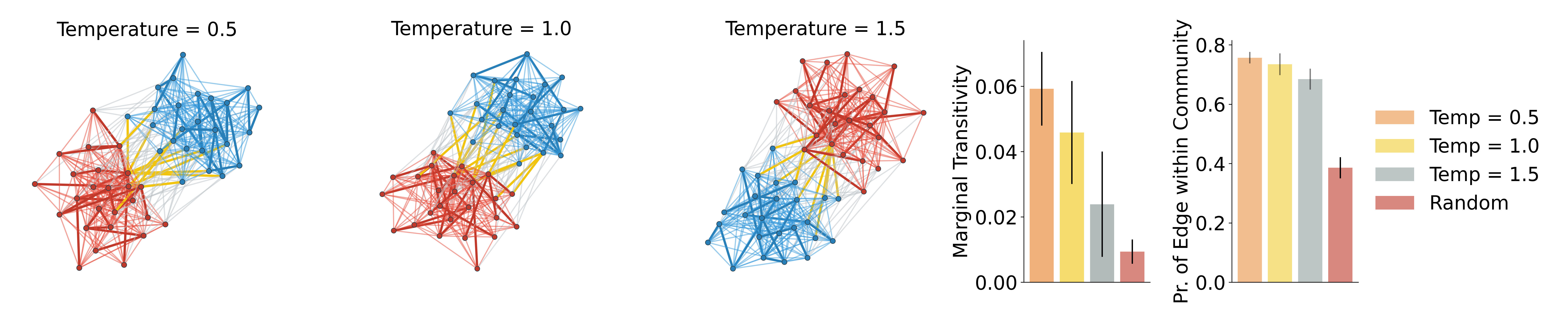}}

    \caption{\textbf{Results for Principle 2 (triadic closure).}  The figure shows the same networks as in \cref{fig:principle_2} with the only change that instead of the intersection of neighborhoods between the query node and each alternative, we provide the number of common neighbors (i.e., the size of the intersection) between the query node and each alternative. Similarly, we observe that the probability of forming an edge within the same community and the marginal transitivity, which indicate triadic closure, is significantly larger than randomly creating links ($P < 0.001$, t-test). The error bars correspond to 95\% confidence intervals.}
    \label{fig:app_principle_2_final}
\end{figure*}

Instead of giving the neighborhood information, the simulations presented in \cref{fig:app_principle_2_final} use the number of common neighbors. We observe behavior similar to \cref{fig:principle_2}. 

\newpage

\subsection{Principle 5: Small-World Phenomenon}

\begin{figure*}[!h]
    \centering
    \subfigure[$\beta = 0.25$\label{subfig:principle_4_instance_0.25}]{\includegraphics[width=0.9\textwidth]{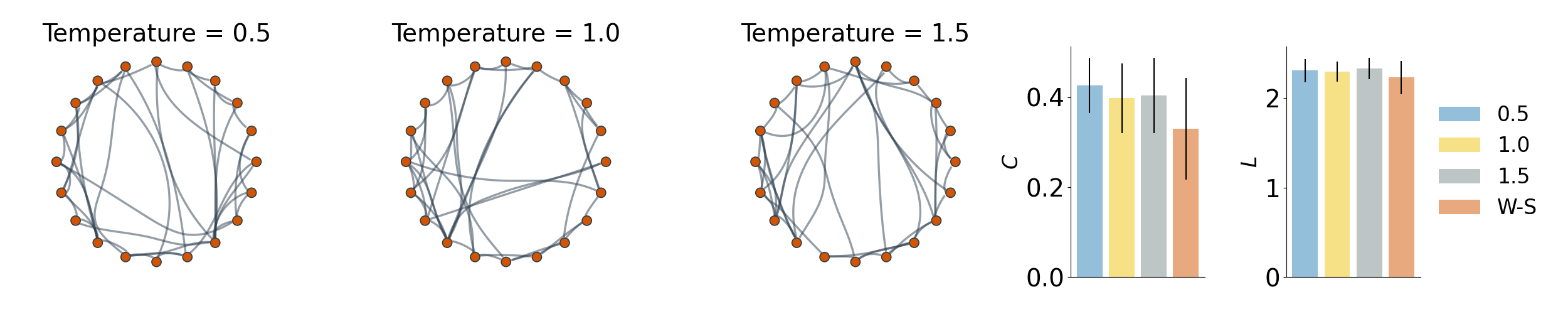}}
    \subfigure[$\beta = 0.5$\label{subfig:principle_4_instance_0.5}]{\includegraphics[width=0.9\textwidth]{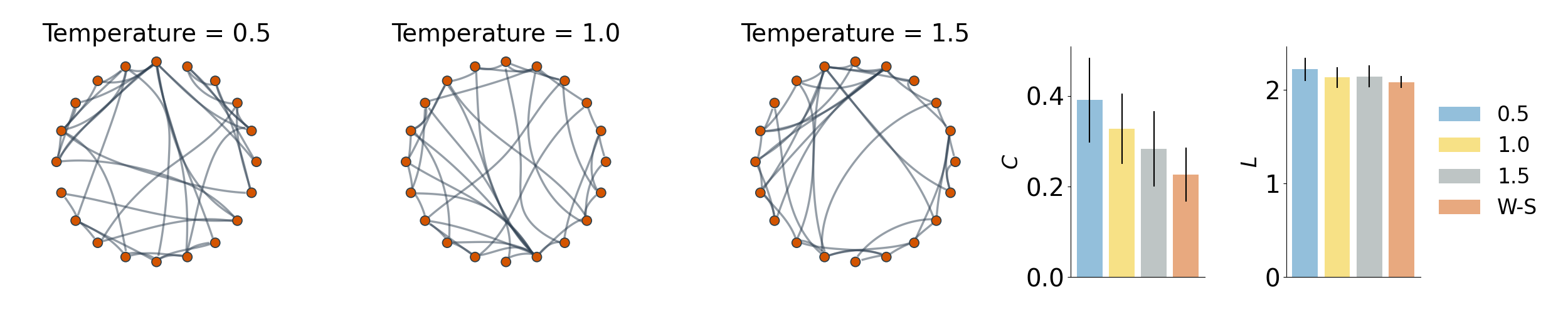}}
    \subfigure[$\beta = 0.75$\label{subfig:principle_4_instance_0.75}]{\includegraphics[width=0.9\textwidth]{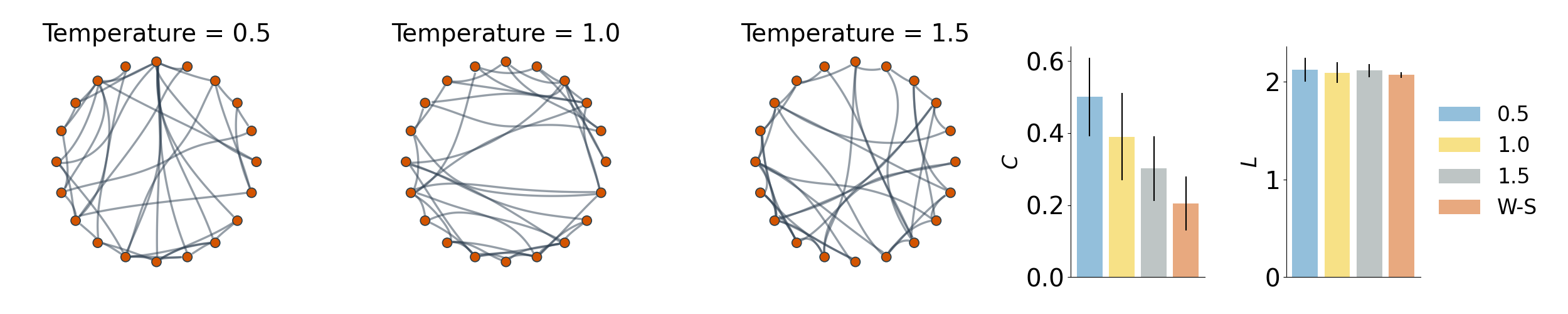}}
    \caption{\textbf{Simulation results for Principle 5 (small world).} Network instances for the networks created according to Principle 5 using the altered Watts-Strogatz Model for node count $n = 50$, average degree $k = 5$, rewriting probability $\beta \in \{ 0.25, 0.5, 0.75 \}$, together with plots of the \textbf{average clustering coefficient} $C$ and the \textbf{average shortest path length} $L$. The comparison is made with respect to a Watts-Strogatz graph with $n = 50, k = 5, \beta \in \{ 0.25, 0.5, 0.75 \}$. The error bars correspond to 95\% confidence intervals. \redit{The results are compared against the Watts-Strogatz model with the same parameters $k$ and $\beta$ as a null model. The t-test comparing $L$ and $C$ for the LLM-generated networks and Watts-Strogatz networks yields $P > 0.05$ (Bonferroni correction for two tests).}}
    \label{fig:small_world_networks}
\end{figure*}

\cref{fig:small_world_networks} shows LLM-generated small-world networks for $\beta \in \{0.25, 0.5, 0.75 \}$ and compares them with the Watts-Strogatz networks with the same parameters. 

\newpage

\section{Chain-of-Thought Experiments} \label{app:cot}

\begin{algorithm}[t]
    \footnotesize
    \caption{Example prompt regarding social network data with Chain-of-Thought reasoning. Note that compared to \cref{alg:prompt_example} the order of the fields \texttt{name} and \texttt{reason} in the output format is reversed.}
    \label{alg:prompt_example_cot}
    \begin{lstlisting}[mathescape=true]
    # Task
    You are located in a school. Your task is to select a set of people to be friends with.

    # Profile
    Your profile is given below after chevrons:
    <PROFILE>
        {
            "name" : "Person 0",
            "favorite subject" : "Chemistry", 
            "neighbors" : ["Person 3", "Person 432", "Person 4", "Person 3", "Person 32"]
        }
    </PROFILE>

    # Candidate Profiles
    The candidate profiles to be friends with are given below after chevrons:

    <PROFILES>
    [
        {
            "name" : "Person 1",
            "favorite subject" : "Mathematics",
            "neighbors" : ["Person 3", "Person 4", "Person 23", "Person 65"]
        },
        {
            "name" : "Person 33",
            "favorite subject" : "History",
            "neighbors" : ["Person 342", "Person 2", "Person 12"]
        }, ...
    ]
    
    </PROFILES>

    # Output
    The output should be given a list of JSON objects with the following structure

    [
        {{
            "reason" : reason for selecting the person,
            "name" : name of the person you selected
        }}, ...
    ]

    # Notes
    - The output must be a list of JSON objects ranked in the order of preference.
    - You can make at most 1 selection.
    
    \end{lstlisting}

\end{algorithm}

We experiment with Chain-of-Thought (CoT) reasoning \cite{wei2022chain}. To induce CoT reasoning we ask the LLM agents to output the reason and then their choice (i.e. by reversing the order of \texttt{reason} and \texttt{name} in \cref{alg:prompt}). The resulting prompt can be found at \cref{alg:prompt_example_cot}. In the following figures, we show the results from the same experiments as the ones we of the main text with the different that CoT is used.

\begin{figure*}[!h]
    \centering
    \subfigure[Probability of connecting to top-$k$ nodes for different models, temperatures, and environments]{\includegraphics[width=0.6\linewidth]{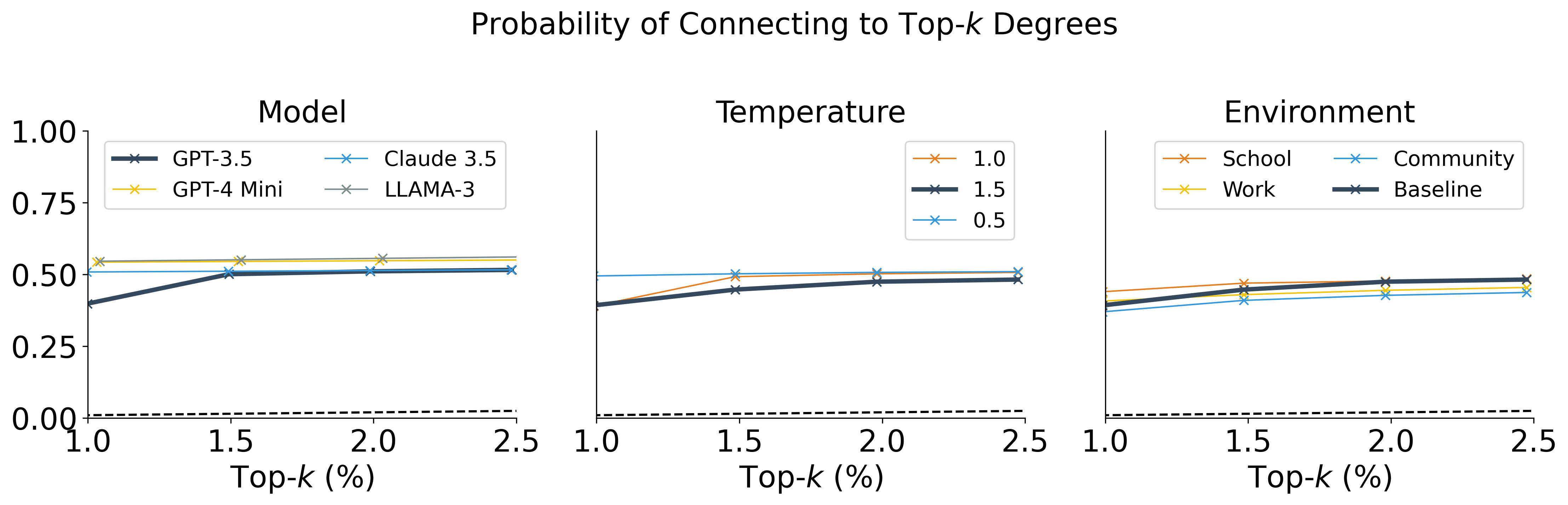}\label{subfig:principle_1_models_environments_topk_cot}}
    \subfigure[Power law fits ($\hat \gamma$) and standard errors for different models, temperatures, and environments]{\includegraphics[width=0.6\linewidth]{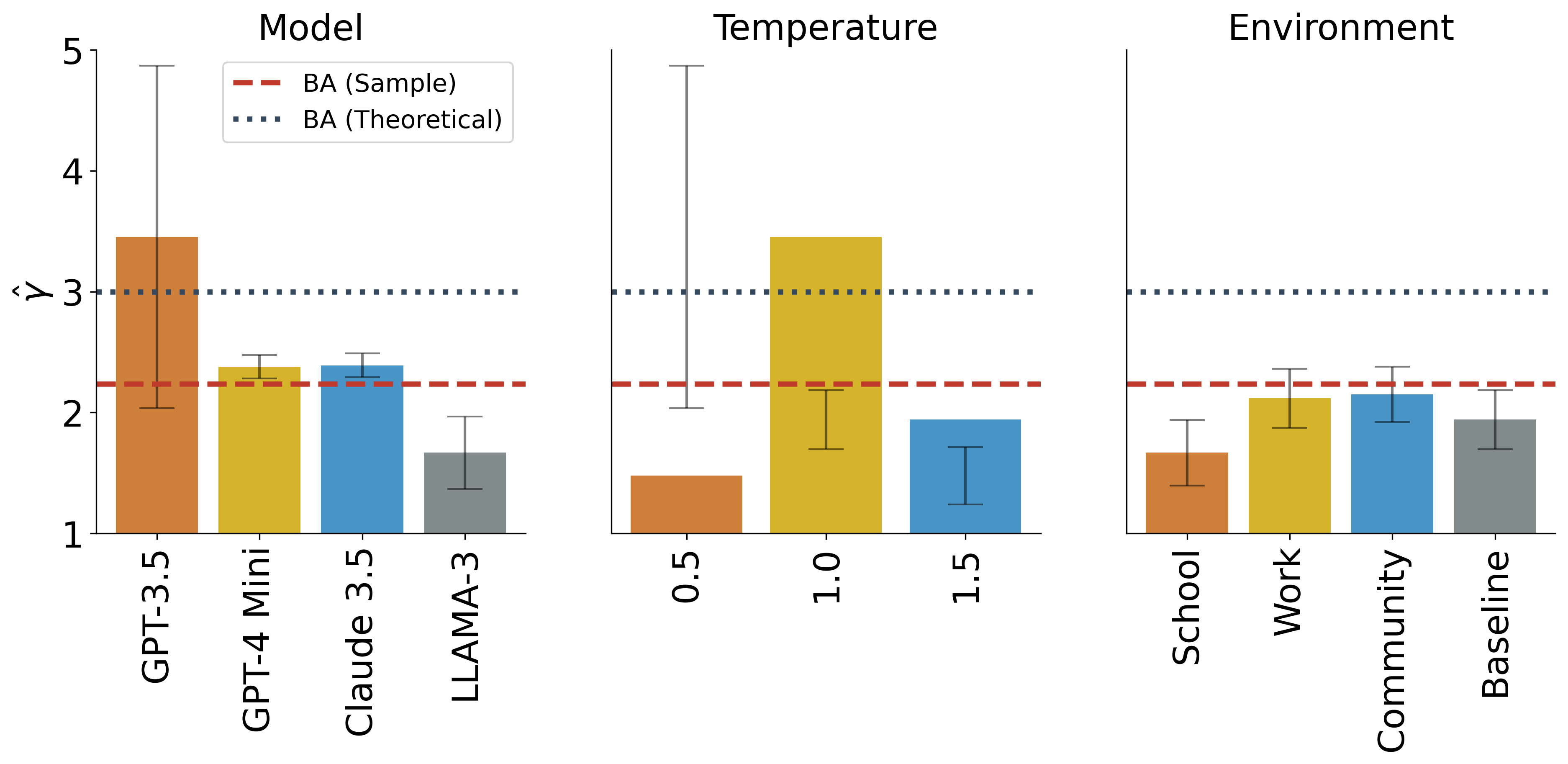}\label{subfig:principle_1_models_environments_cot}}
    \caption{\textbf{Results for Principle 1 with CoT reasoning (preferential attachment)} The multi-LLM setup was given neighborhood information $\{ N_{j, t} : j \in V_t \}$. \textbf{Top:} Probability of connecting to top-$k$-degree nodes for varying model (temperature is fixed to 1.0 and environment to baseline), temperature (model fixed to GPT-3.5 and environment to baseline) and environment (model fixed to GPT-3.5 and environment temperature to 1.5) for networks generated according to Principle 1 with $n = 200$ nodes. \textbf{Bottom:} Power Law exponents and standard errors for varying model, temperature, and environment.}
    \label{fig:principle_1_cot}
\end{figure*}

 \begin{figure*}[!h]
    \centering
    \subfigure[Probability of connecting to top-$k$ for different models, temperatures, and environments]{\includegraphics[width=0.675\linewidth]{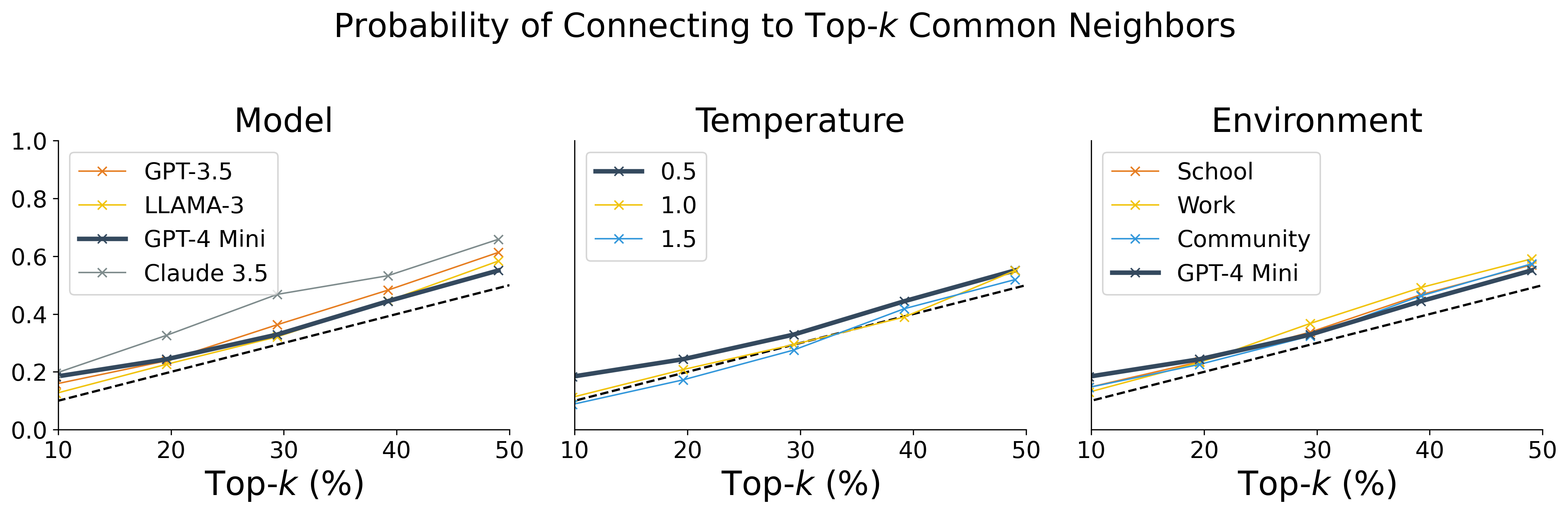} \label{subfig:principle_2_models_environments_topk_cot}}    
    \subfigure[Marginal transitivity ($D$) and probability of an edge within a community ($\hat p$) for different models, temperatures, and environments]{\includegraphics[width=0.675\linewidth]{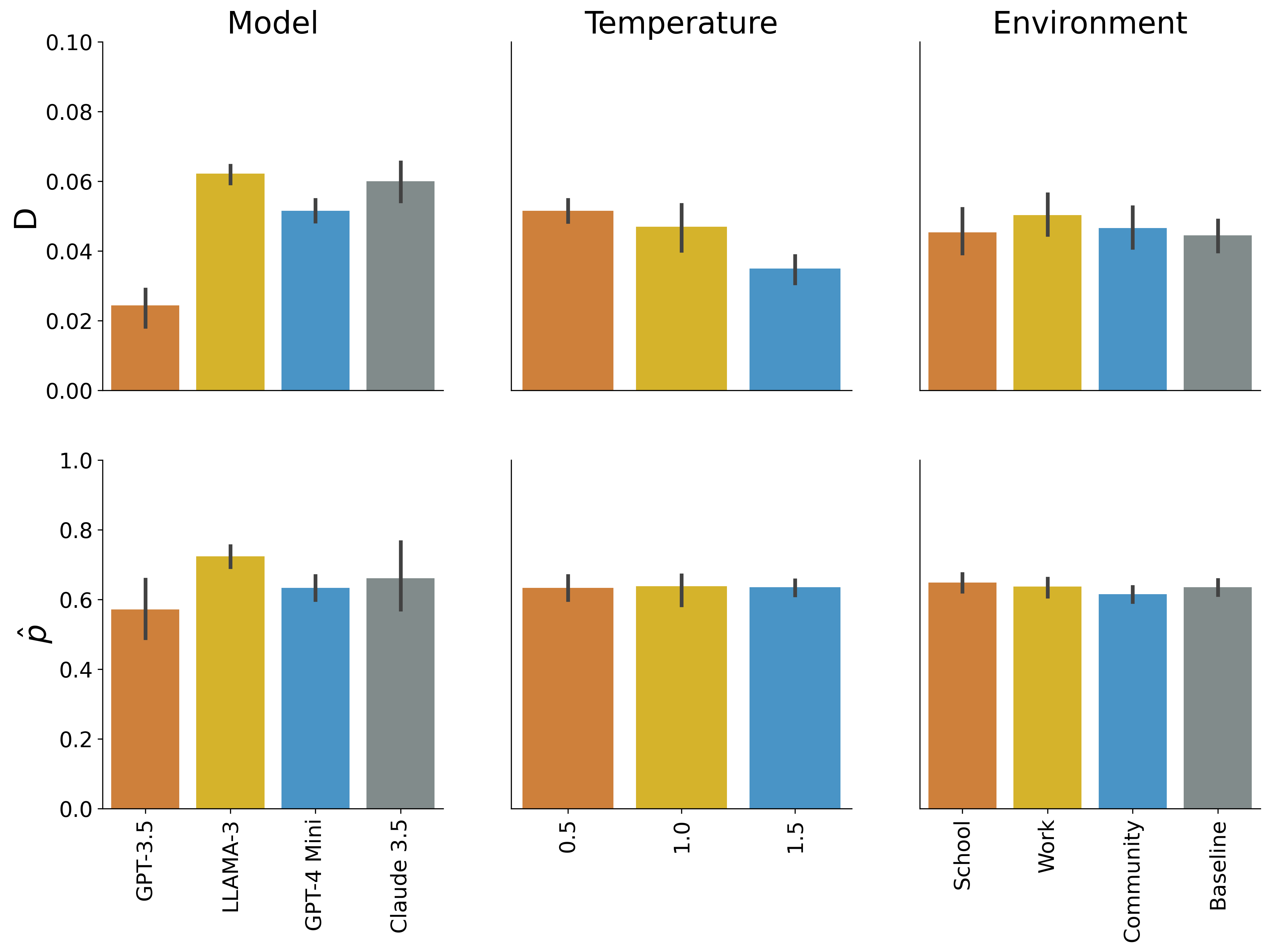} \label{subfig:principle_2_models_environments_cot}}
    \caption{\textbf{Results for Principle 2 with CoT reasoning (triadic closure).}  \textbf{Top:} Probability of connecting to top-$k$ nodes (in terms of common neighbors) for varying model (temperature is fixed to 1.0 and environment to baseline), temperature (model fixed to GPT-4 Mini and environment to baseline) and environment (model fixed to GPT-4 Mini and environment temperature to 0.5) for networks generated according to Principle 2 ($n = 50$, 10 simulations for each model, environment and temperature). \textbf{Bottom:} Marginal transitivity ($D$) and probability of an edge within a community ($\hat p$) for networks generated according to Principle 2 in different models, temperatures, and environments.}
    \label{fig:principle_2_cot}
\end{figure*}

\begin{figure*}[!h]
    \centering
    \subfigure[Assortativity and Louvain Modularity with different LLM models and environments]{\includegraphics[width=0.9\linewidth]{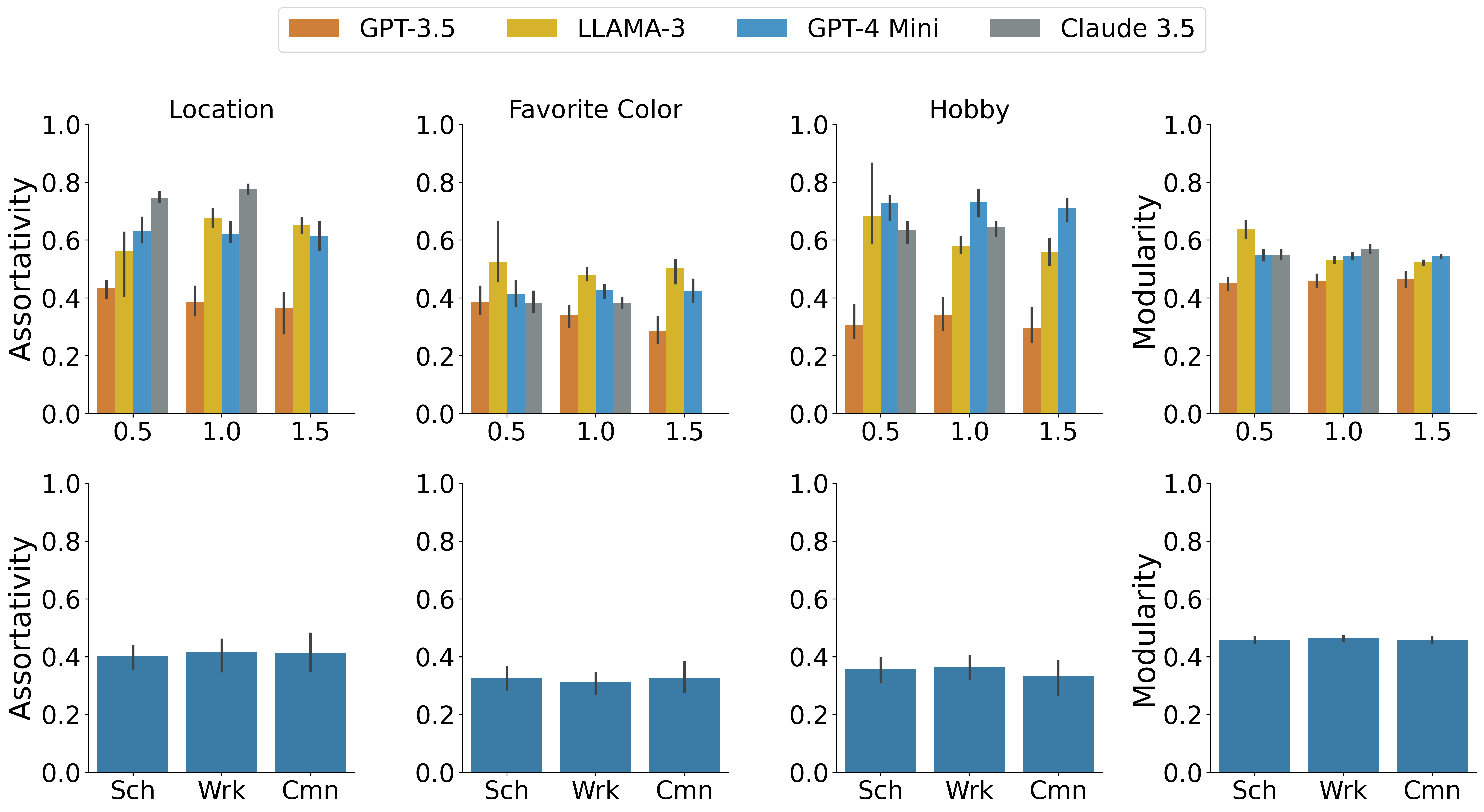}\label{subfig:principle_3_models_environments_cot}}
    % \subfigure[Effect of features on homophily]{\includegraphics[width=\linewidth]{assortativities_lucky_number.png}\label{subfig:principle_3_models_environments_distractor}}
    % \subfigure[Network instances generated by GPT-3.5 agents \label{subfig:principle_3_instances}]{\includegraphics[width=0.75\linewidth]{principle_3_profiles_final_graphs.png}}
    \caption{\textbf{Results for Principle 3 (Homophily) and Principle 4 (Community structure due to homophily) with CoT reasoning.} 
    \textbf{Top:} Assortativities and Louvain modularity according to Principle 3 ($n = 50$, 5 simulations for each row) in different environments (school, work, community) using different models. The statistical significance is $P < 0.0003$ for all t-tests (comparing with 0, \redit{Bonferroni correction for three tests)}.}
    \label{fig:principle_3_cot}
\end{figure*}

\begin{figure*}[!h]
    \captionsetup[subfigure]{font=scriptsize,labelfont=scriptsize}
    \centering
    \subfigure[Regression plot for different models and environments for $\beta = 0.25$ and $k = 5$.]{\includegraphics[width=0.6\linewidth]{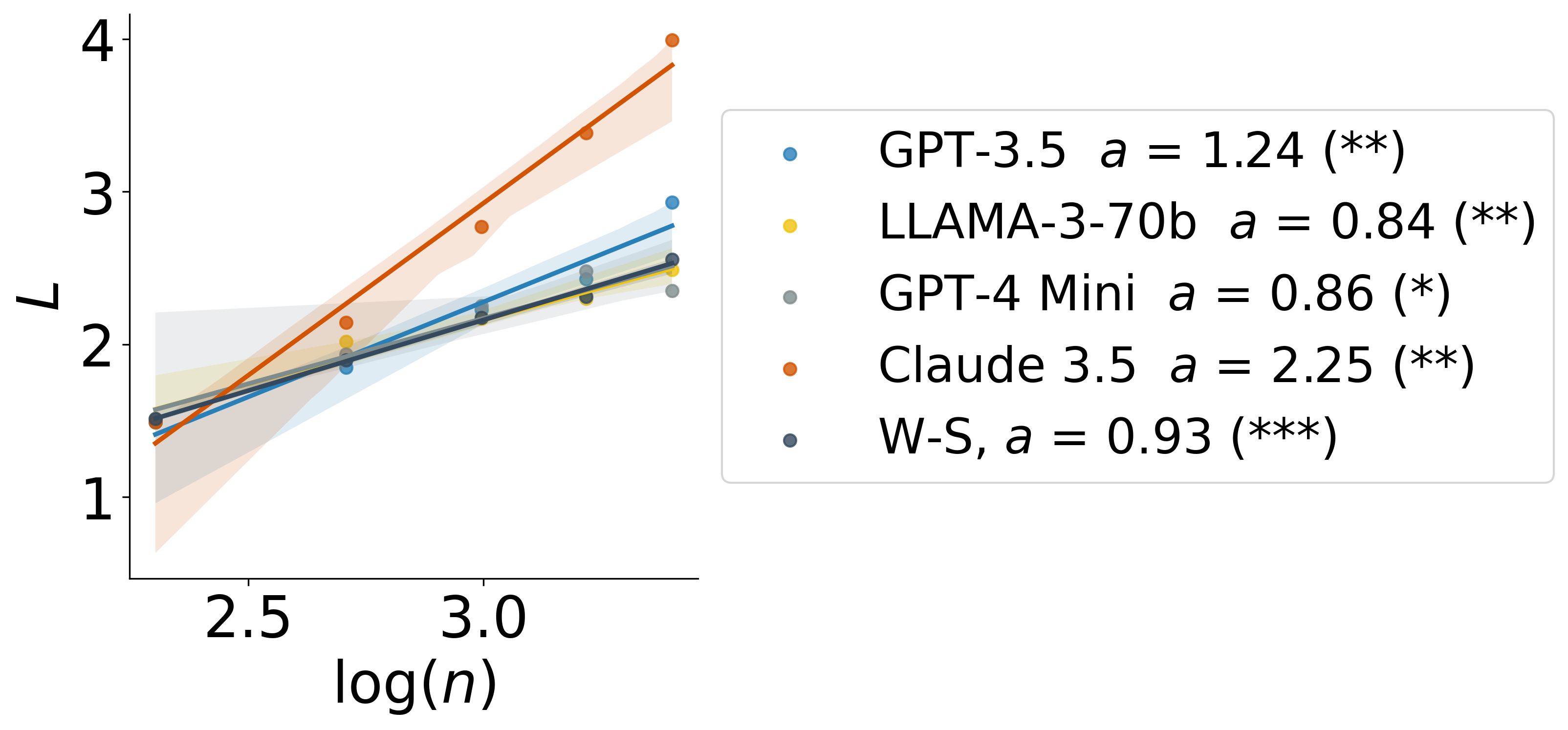}\label{subfig:principle_5_models_environments_cot}}
    \caption{\textbf{Fitted results for Principle 5 with CoT reasoning (small world).} Regression plot for the relation $L \sim \log (n)$ for different LLM models for $\beta = 0.25$ and $k = 5$. The legend shows the effect size ($a$) and the $P$-value. (*: $P < 0.0025$; **: $P < 0.005$, and ***: $P < 0.0005$, \redit{Bonferroni correction for two tests}).}
    \label{fig:principle_5_cot}
\end{figure*}

\end{document}

%% file: table_1.tex
\begin{table*}[htpb!]
    \centering
    \tiny
    \begin{tabular}{lllllll}
    \toprule
    Model & Preferential Attachment ($\hat \theta_{\mathrm{PA}}$) & Homophily ($\hat \theta_{\mathrm{H}}$) & Triadic Closure ($\hat \theta_{\mathrm{TC}}$) & Log Likelihood & AIC \\
    \midrule
    \multicolumn{6}{c}{Caltech36 ($n = 769$ nodes, $m = 33,312$ edges, $N = 769$ samples, $A = 15$ alternatives each)} \\
    \midrule 
     GPT-3.5 + Uniform & 0.20*** (0.002) & 0.65*** (0.005) & -0.06 (0.006) & -2,088.21 & 4,184.41  \\
     GPT-4o Mini + Uniform & 0.34*** (0.006) & 2.13*** (0.03) & 0.44*** (0.02) & -1,201.27 & 2,410.55 \\
     Claude 3.5 + Uniform & 0.46*** (0.005) & 0.55*** (0.01) & 0.55*** (0.007) & -1,748.19 & 3,504.38 \\ 
     Llama 3 70b + Uniform & 0.28*** (0.006) & 2.43*** (0.02) & 0.84*** (0.01) & -809.57 & 1,627.15 \\ \midrule
     GPT-3.5 + RecSys & 0.15** (0.002) & 0.08 (0.007) & -0.60*** (0.02) & -2,114.41 & 4,236.82 \\
     GPT-4o Mini + RecSys & 0.21*** (0.004) & 2.32*** (0.005) & 0.33** (0.005) & -1,611.38 & 3,230.77 \\
     Claude 3.5 + RecSys & 0.65*** (0.002) & 1.86*** (0.01) & 0.20 (0.01) & -1,852.96 & 3,713.91 \\
     Llama 3 70b + RecSys & 0.23*** (0.003) & 4.13*** (0.01) & 0.68*** (0.01) & -919.15 & 1,846.30 \\
     \midrule
     \multicolumn{6}{c}{Swarthmore42 ($n = 1,659$ nodes, $m = 122,100$ edges, $N = 1,659$ samples, $A = 15$ alternatives each)} \\ \midrule
    GPT-3.5 + Uniform & 0.19*** (0.008) & 0.47*** (0.01) & 0.00 (0.009) & -4,484.45 & 8,976.90 \\ 
    GPT-4o Mini + Uniform & 0.27*** (0.21) & 2.22*** (0.78) & 0.57*** (0.43) & -1,899.09 & 3,806.19 \\
    Claude 3.5 + Uniform & 0.36*** (0.002) & 0.75*** (0.006) & 0.55*** (0.004) & -3,563.02 & 7,134.03 \\ 
    Llama 3 70b + Uniform & 0.39*** (0.003) & 2.31*** (0.005) & 0.62*** (0.004) & -1,820.26 & 3,648.52  \\
    \midrule
    GPT-3.5 + RecSys & 0.14** (0.001) & 0.11 (0.002) & -0.08 (0.002) & -4,564.89 & 9,137.78 \\
    GPT-4o Mini + RecSys & 0.33*** (0.007) & 2.94*** (0.01) & 0.45*** (0.006) & -2,723.78 & 5,455.57 \\
    Claude 3.5 + RecSys & 1.26*** (0.004) & 1.22*** (0.007) & 0.95*** (0.006) & -2,281.61 & 4,571.22 \\
    LLama 3 70b + RecSys & 0.09 (0.007) & 2.58*** (0.02) & 1.18*** (0.009) & -610.00 & 1,228.00 \\
    \midrule
    \multicolumn{6}{c}{UChicago30 ($n = 6,951$ nodes, $m = 416,206$ edges, $N = 2,000$ samples, $A = 15$ alternatives each)} \\ \midrule
    GPT-3.5 + Uniform & 0.22*** (0.001) & 0.48*** (0.004) & -0.02 (0.0005) & -8,157.38 & 16,322.77 \\
    GPT-4o Mini + Uniform  & 0.27*** (0.005) & 2.22*** (0.019) & 0.57*** (0.011) & -1,899.09 & 3,806.19 \\
    Claude 3.5 + Uniform &  0.43*** (0.003) & 0.78*** (0.005) & 0.39*** (0.002) & -6,604.77 & 13,217.54 \\
    Llama 3 70b + Uniform & 0.43*** (0.007) & 2.57*** (0.014) & 0.32*** (0.005) & -3,689.00 & 7,386.00 \\ 
    \midrule
    GPT-3.5 + RecSys &0.14*** (0.002) & -0.08 (0.006) & 0.18** (0.007) & -4,459.64 & 8,927.28 \\
    GPT-4o Mini + RecSys & 0.32*** (0.001) & 3.44*** (0.01) & -0.74*** (0.005) & -3,154.27 & 6,316.53 \\
    Claude 3.5 + RecSys & 0.75*** (0.002) & 1.68*** (0.005) & 0.17* (0.003) & -2,386.04 & 4,780.09 \\
    LLama 3 70b + RecSys & 0.27*** (0.004) & 2.81*** (0.02) & 0.53** (0.01) & -661.75 & 1,331.50 \\
    \midrule
    \multicolumn{6}{c}{Andorra ($n = 32,812$ nodes,  $m = 513,931$ edges, $N = 1,000$ samples, $A = 5$ alternatives each)} \\ \midrule
    GPT-3.5 + Uniform & 0.53*** (0.001) & 0.21* (0.01) & -0.24*** (0.002) & -1,712.91 & 3,433.83 \\
    GPT-4o Mini + Uniform & 0.54*** (0.004) & 3.47*** (0.06) & -0.09* (0.01) & -1,002.11 & 2,012.22 \\ 
    Claude 3.5 + Uniform & 0.54*** (0.003) & 1.94*** (0.009) & -0.15*** (0.003) & -1,541.77 & 3,091.55 \\ 
    Llama 3 70b + Uniform & 0.38*** (0.003) & 3.92*** (0.02) & -0.04 (0.01) & -985.95 & 1,979.91 \\ 
    \midrule
    GPT-3.5 + RecSys & 0.31*** (0.03) & -0.07 (0.009) & -0.43*** (0.007) & -1,722.42 & 3,452.84 \\
    GPT-4o Mini + RecSys & 0.11 (0.003) & 3.68*** (0.008) & -0.64*** (0.006) & -938.74 & 1,885.47 \\
    Claude 3.5 + RecSys & 0.38*** (0.003) & 1.78*** (0.01) & -0.41*** (0.005) & -1,651.67 & 3,311.33 \\
    Llama 3 70b + RecSys & 0.53*** (0.002) & 3.63*** (0.01) & -0.12* (0.003) & -1,238.22 & 2,484.45 \\
    \midrule 
    \multicolumn{6}{c}{MobileD ($n = 1,982$ nodes, $m = 25,470$ edges, $N = 1,000$ samples, $A = 5$ alternatives each)} \\ \midrule
    GPT-3.5 + Uniform & 1.06*** (0.003) & -0.94*** (0.009) & -0.02 (0.001) & -1,663.42 & 3,334.84 \\
    GPT-4o Mini + Uniform & 1.38*** (0.02) & -0.85*** (0.02) & 0.87*** (0.01) & -880.39 & 1,768.78 \\
    Claude 3.5 + Uniform & 0.71*** (0.009) & -2.44*** (0.02) & 1.13*** (0.005) & -1,197.92 & 2,403.83 \\
    Llama 3 70b + Uniform & 1.04*** (0.005) & -0.36** (0.01) & 0.71*** (0.002) & -1,269.42 & 2,546.83 \\
    \midrule
    GPT 3.5 + RecSys & 1.68*** (0.006) & -0.35** (0.008) & -0.91*** (0.006) & -1,613.85 & 3,235.69 \\
    GPT-4o Mini + RecSys & 3.16*** (0.01) & -0.49* (0.02) & 0.67*** (0.01) & -681.39 & 1,370.78 \\
    Claude 3.5 + RecSys & 1.85*** (0.01) & -0.87*** (0.006) & 0.16** (0.007) & -1,542.90 & 3,093.80 \\
    Llama 3 70b + RecSys & 1.43*** (0.01) & 1.05*** (0.007) & 0.36*** (0.005) & -1,468.61 & 2,945.22 \\
    \midrule
    \multicolumn{6}{l}{\emph{Note: *: $P < 0.05$, **: $P < 0.01$, ***: $P < 0.001$}} \\    
    \bottomrule
    \end{tabular}
    \caption{Effect sizes for real-world networks from Facebook100 \cite{traud2012social}, the Andorra dataset \cite{yuan2018interpretable}, and the MobileD dataset \cite{yuan2018interpretable} for several LLMs for temperature set to 0.5. \redit{We test two sampling strategies: a \textit{uniform} strategy where $A_t$ is sampled uniformly from the set of nodes, and a \textit{recommender system} (RecSys) based on logistic regression and trained on pairwise node similarities and network characteristics (number of common neighbors, Jaccard similarity, preferential attachment score, and the Adamic-Adar index). See \cref{app:sampling_strategies} for more information on the sampling strategies. Average marginal effects (cf. \cref{app:ame}) show that homophily is the strongest driver of link formation, with recommendation-based sampling amplifying the dominant mechanism in each dataset while preserving the overall ranking of behavioral factors.}}
    \label{tab:real_world_data_environments}
\end{table*}

%% file: table_2.tex
\begin{table*}[!h]
\centering
\scriptsize
\begin{tabular}{lccccccc}
\toprule
 & Constant & Similarity & Common Neighbors & Jaccard & Adamic-Adar & PA Score & AUC Score \\
\midrule
Caltech36 & -2.9861$^{***}$ & 0.5103$^{***}$ & -0.9384$^{***}$ & 9.9513$^{***}$ & 4.9751$^{***}$ & -4.28$\times 10^{-5}$$^{***}$ & 95.2 \\
Swarthmore42 & -3.0713$^{***}$ & 0.3323$^{***}$ & -0.8603$^{***}$ & 29.0475$^{***}$ & 4.5827$^{***}$ & -4.596$\times 10^{-6}$ & 94.8 \\
UChicago30 & -3.1467$^{***}$ & 0.5493$^{***}$ & -1.5431$^{***}$ & 59.5934$^{***}$ & 8.3355$^{***}$ & 3.575$\times 10^{-5}$$^{***}$ & 97.05 \\
Andorra & -2.1337$^{***}$ & 0.2247$^{***}$ & -2.0713$^{***}$ & 96.1618$^{***}$ & 13.2652$^{***}$ & 3.00$\times 10^{-4}$$^{***}$ & 92.4 \\
MobileD & -4.6000$^{***}$ & 0.4075$^{*}$ & -1.9072$^{***}$ & 24.7593$^{**}$ & 14.6024$^{***}$ & -0.0058$^{***}$ & 98.9 \\
\midrule
\emph{Note} & \multicolumn{7}{r}{$^*: P < 0.05, \; ^{**}: P < 0.01, \; ^{***}: P < 0.001$} \\
\bottomrule
\end{tabular}
\caption{Recommendation System Parameters.}
\label{tab:link_prediction}
\end{table*}

%% file: table_3.tex
\begin{table*}[!h]
\tiny
\begin{tabular}{lcHccccccccc}
\toprule
Name & Model & Temp & Degrees (KS) & P-value & Sizes of CCs (KS) & P-value & Spectrum (KS) & P-value & LCC (KS) & P-value & \% New Edges \\
\midrule
\multicolumn{12}{c}{Uniform} \\
\midrule
Caltech36 & GPT-4 Mini & 0.5 & 0.05 & 0.3 & 0.1 & 1 & 0.06 & 0.2 & 0.04 & 0.6 & 5 \\
Swarthmore42 & GPT-4 Mini & 0.5 & 0.02 & 0.8 & 0.2 & 1 & 0.02 & 1 & 0.02 & 1 & 3 \\
UChicago30 & GPT-4 Mini & 0.5 & 0.02 & 0.3 & 0.06 & 1 & 0.02 & 0.2 & 0.006 & 1 & 1 \\
Caltech36 & GPT-3.5 & 0.5 & 0.06 & 0.1 & 0.5 & 0.6 & 0.07 & 0.04 & 0.1 & 0.0005 & 5 \\
Swarthmore42 & GPT-3.5 & 0.5 & 0.02 & 0.8 & 0.4 & 0.5 & 0.02 & 0.9 & 0.07 & 0.0008 & 3 \\
UChicago30 & GPT-3.5 & 0.5 & 0.02 & 0.2 & 0.1 & 1 & 0.02 & 0.1 & 0.03 & 0.003 & 1 \\
Caltech36 & LLAMA-3 & 0.5 & 0.06 & 0.1 & 0.08 & 1 & 0.07 & 0.04 & 0.02 & 1 & 5 \\
Swarthmore42 & LLAMA-3 & 0.5 & 0.02 & 0.9 & 0.09 & 1 & 0.02 & 0.9 & 0.02 & 1 & 3 \\
UChicago30 & LLAMA-3 & 0.5 & 0.02 & 0.2 & 0.02 & 1 & 0.02 & 0.07 & 0.007 & 1 & 1 \\
Caltech36 & Claude 3.5 & 0.5 & 0.05 & 0.3 & 0.1 & 1 & 0.05 & 0.2 & 0.03 & 0.9 & 5 \\
Swarthmore42 & Claude 3.5 & 0.5 & 0.02 & 0.9 & 0.3 & 0.9 & 0.01 & 1 & 0.02 & 0.9 & 3 \\
UChicago30 & Claude 3.5 & 0.5 & 0.02 & 0.2 & 0.2 & 0.3 & 0.02 & 0.2 & 0.01 & 0.9 & 1 \\
Andorra & GPT-4 Mini & 0.5 & 0.002 & 1 & 0 & 1 & 0.1 & 1 & 0.001 & 1 & 0.2 \\
MobileD & GPT-4 Mini & 0.5 & 0.03 & 0.3 & 0 & 1 & 0.08 & 3e-06 & 0.06 & 0.002 & 3 \\
Andorra & GPT-3.5 & 0.5 & 0.002 & 1 & 0 & 1 & 0.1 & 1 & 0.002 & 1 & 0.2 \\
MobileD & GPT-3.5 & 0.5 & 0.04 & 0.06 & 0 & 1 & 0.05 & 0.02 & 0.2 & 6e-29 & 4 \\
Andorra & LLAMA-3 & 0.5 & 0.002 & 1 & 0 & 1 & 0.1 & 1 & 0.0009 & 1 & 0.2 \\
MobileD & LLAMA-3 & 0.5 & 0.04 & 0.08 & 0 & 1 & 0.09 & 8e-07 & 0.04 & 0.06 & 4 \\
Andorra & Claude 3.5 & 0.5 & 0.002 & 1 & 0 & 1 & 0.1 & 1 & 0.001 & 1 & 0.2 \\
MobileD & Claude 3.5 & 0.5 & 0.04 & 0.04 & 0 & 1 & 0.2 & 5e-39 & 0.1 & 9e-19 & 4 \\
\midrule
\multicolumn{12}{c}{Recommendation System} \\
\midrule
Caltech36 & GPT-4 Mini & 0.5 & 0.05 & 0.2 & 0.1 & 1 & 0.06 & 0.2 & 0.05 & 0.2 & 5 \\
Swarthmore42 & GPT-4 Mini & 0.5 & 0.02 & 0.9 & 0.1 & 1 & 0.02 & 0.9 & 0.04 & 0.1 & 2 \\
UChicago30 & GPT-4 Mini & 0.5 & 0.01 & 0.6 & 0.3 & 0.03 & 0.01 & 0.7 & 0.02 & 0.4 & 0.7 \\
Caltech36 & GPT-3.5 & 0.5 & 0.05 & 0.2 & 0.5 & 0.6 & 0.06 & 0.1 & 0.07 & 0.03 & 5 \\
Swarthmore42 & GPT-3.5 & 0.5 & 0.02 & 0.8 & 0.3 & 0.9 & 0.02 & 1 & 0.07 & 0.0005 & 3 \\
UChicago30 & GPT-3.5 & 0.5 & 0.01 & 0.5 & 0.5 & 0.002 & 0.01 & 0.6 & 0.02 & 0.08 & 0.9 \\
Caltech36 & LLAMA-3 & 0.5 & 0.06 & 0.2 & 0.1 & 1 & 0.07 & 0.06 & 0.03 & 0.9 & 5 \\
Swarthmore42 & LLAMA-3 & 0.5 & 0.01 & 1 & 0.2 & 1 & 0.01 & 1 & 0.02 & 1 & 1 \\
UChicago30 & LLAMA-3 & 0.5 & 0.006 & 1 & 0.2 & 0.4 & 0.007 & 1 & 0.005 & 1 & 0.3 \\
Caltech36 & Claude 3.5 & 0.5 & 0.05 & 0.2 & 0.2 & 1 & 0.04 & 0.4 & 0.08 & 0.02 & 5 \\
Swarthmore42 & Claude 3.5 & 0.5 & 0.01 & 1 & 0.2 & 1 & 0.008 & 1 & 0.05 & 0.02 & 2 \\
UChicago30 & Claude 3.5 & 0.5 & 0.006 & 1 & 0.4 & 0.02 & 0.006 & 1 & 0.01 & 0.6 & 0.5 \\
Andorra & GPT-4 Mini & 0.5 & 0.002 & 1 & 0 & 1 & 0.1 & 1 & 0.003 & 1 & 0.1 \\
MobileD & GPT-4 Mini & 0.5 & 0.03 & 0.6 & 0 & 1 & 0.07 & 0.0003 & 0.08 & 2e-05 & 2 \\
Andorra & GPT-3.5 & 0.5 & 0.002 & 1 & 0 & 1 & 0.1 & 1 & 0.004 & 0.9 & 0.2 \\
MobileD & GPT-3.5 & 0.5 & 0.05 & 0.01 & 0 & 1 & 0.04 & 0.08 & 0.07 & 0.0004 & 4 \\
Andorra & LLAMA-3 & 0.5 & 0.002 & 1 & 0 & 1 & 0.1 & 1 & 0.003 & 1 & 0.2 \\
MobileD & LLAMA-3 & 0.5 & 0.04 & 0.05 & 0 & 1 & 0.07 & 6e-05 & 0.08 & 9e-06 & 4 \\
Andorra & Claude 3.5 & 0.5 & 0.002 & 1 & 0 & 1 & 0.1 & 1 & 0.005 & 0.7 & 0.2 \\
MobileD & Claude 3.5 & 0.5 & 0.04 & 0.06 & 0 & 1 & 0.1 & 2e-17 & 0.1 & 6e-14 & 4 \\
\bottomrule
\end{tabular}
\caption{\redit{\textbf{Change in Graph Statistics for the experiments of \cref{tab:real_world_data_environments}.} We report the KS statistic and the P-values for the following quantities (see \cite{leskovec2006sampling} for more information on the statistics): (i) degree distribution, (ii) distribution of the sizes of strongly connected components, (iii) adjacency matrix spectrum, (iv) local clustering coefficient. The last column reports the percentage of new edges added. Adding $\le 5\%$ of edges based on LLM decisions leaves global graph properties largely unchanged, with only occasional local clustering increases—more frequent under the Uniform strategy than the Recommendation System.}}
\label{tab:graph_statistics_change_sampling}
\end{table*}

%% file: table_4.tex
\begin{table*}[!h]
    \centering
    \tiny
    \begin{tabular}{llllllll}
    \toprule
     & & \multicolumn{2}{c}{$\hat \theta_{\mathrm{PA}}$} & \multicolumn{2}{c}{$\hat \theta_{\mathrm{H}}$} & \multicolumn{2}{c}{$\hat \theta_{\mathrm{TC}}$} \\
     & Algorithm & Uniform & RecSys & Uniform & RecSys & Uniform & RecSys \\
    Name & Model &  &  &  &  &  &  \\
    \midrule
    {Caltech36} & Claude 3.5 & 0.39*** (0.00) & 0.57*** (0.00) & 0.47*** (0.01) & 1.62*** (0.01) & 0.47*** (0.01) & 0.18*** (0.01) \\
     & GPT-3.5 & 0.18*** (0.00) & 0.14*** (0.00) & 0.61*** (0.00) & 0.07*** (0.01) & -0.05*** (0.01) & -0.56*** (0.01) \\
     & GPT-4 Mini & 0.21*** (0.00) & 0.17*** (0.00) & 1.30*** (0.02) & 1.85*** (0.00) & 0.27*** (0.01) & 0.27*** (0.00) \\
     & LLAMA-3 & 0.11*** (0.00) & 0.10*** (0.00) & 0.96*** (0.01) & 1.82*** (0.01) & 0.33*** (0.00) & 0.30*** (0.01) \\
    {Swarthmore42} & Claude 3.5 & 0.29*** (0.00) & 1.07*** (0.00) & 0.60*** (0.01) & 1.03*** (0.01) & 0.44*** (0.00) & 0.80*** (0.01) \\
     & GPT-3.5 & 0.18*** (0.01) & 0.13*** (0.00) & 0.44*** (0.01) & 0.10*** (0.00) & 0.00 (0.01) & -0.08*** (0.00) \\
     & GPT-4 Mini & 0.12*** (0.00) & 0.23*** (0.01) & 1.01*** (0.01) & 2.03*** (0.01) & 0.26*** (0.00) & 0.31*** (0.00) \\
     & LLAMA-3 & 0.17*** (0.00) & 0.05*** (0.00) & 0.99*** (0.00) & 1.52*** (0.01) & 0.26*** (0.00) & 0.69*** (0.01) \\
    {Uchicago30} & Claude 3.5 & 0.35*** (0.00) & 0.66*** (0.00) & 0.64*** (0.00) & 1.47*** (0.00) & 0.32*** (0.00) & 0.15*** (0.00) \\
     & GPT-3.5 & 0.21*** (0.00) & 0.13*** (0.00) & 0.45*** (0.00) & -0.08*** (0.01) & -0.02*** (0.00) & 0.17*** (0.01) \\
     & GPT-4 Mini & 0.13*** (0.00) & 0.27*** (0.00) & 1.03*** (0.01) & 2.82*** (0.01) & 0.24*** (0.00) & -0.60*** (0.00) \\
     & LLAMA-3 & 0.21*** (0.00) & 0.19*** (0.00) & 1.27*** (0.01) & 1.97*** (0.01) & 0.16*** (0.00) & 0.37*** (0.01) \\
    {MobileD} & Claude 3.5 & 0.43*** (0.01) & 1.39*** (0.01) & -1.47*** (0.01) & -0.66*** (0.00) & 0.68*** (0.00) & 0.12*** (0.01) \\
     & GPT-3.5 & 0.83*** (0.00) & 1.30*** (0.00) & -0.74*** (0.01) & -0.27*** (0.01) & -0.02*** (0.00) & -0.70*** (0.01) \\
     & GPT-4 Mini & 0.89*** (0.01) & 2.06*** (0.01) & -0.55*** (0.01) & -0.32*** (0.01) & 0.56*** (0.01) & 0.44*** (0.01) \\
     & LLAMA-3 & 0.66*** (0.00) & 1.02*** (0.01) & 0.23*** (0.01) & 0.75*** (0.01) & 0.45*** (0.00) & 0.26*** (0.00) \\
    {Andorra} & Claude 3.5 & 0.40*** (0.00) & 0.30*** (0.00) & 1.45*** (0.01) & 1.40*** (0.01) & -0.11*** (0.00) & -0.32*** (0.00) \\
     & GPT-3.5 & 0.43*** (0.00) & 0.25*** (0.02) & 0.17*** (0.01) & -0.06*** (0.01) & -0.19*** (0.00) & -0.35*** (0.01) \\
     & GPT-4 Mini & 0.31*** (0.00) & 0.08*** (0.00) & 2.00*** (0.03) & 2.54*** (0.01) & -0.05*** (0.01) & -0.45*** (0.00) \\
     & LLAMA-3 & 0.19*** (0.00) & 0.33*** (0.00) & 1.92*** (0.01) & 2.24*** (0.01) & -0.02* (0.01) & -0.08*** (0.00) \\
     \midrule
    \multicolumn{6}{l}{\emph{Note: *: $P < 0.05$, **: $P < 0.01$, ***: $P < 0.001$}} \\    
    \bottomrule
    \end{tabular}
    \caption{AMEs for the discrete choice models of \cref{tab:real_world_data_environments}.}
    \label{tab:ame}
\end{table*}

%% file: table_5.tex
\begin{table*}[h!]
\centering
\tiny
\begin{tabular}{lccccccccc}
\toprule
 Temp. &        $\hat \theta_{\mathrm{PA}}$ & $\hat \theta_{\mathrm{H}}$ & $\hat \theta_{\mathrm{TC}}$ & LL & AIC & \% Change & \% Change & \% Change & $\Delta Q$ (t-stat) \\
  & & &  & &  & Acc. & $L$ & $C$  &  \\
\midrule
         & \multicolumn{9}{c}{Caltech36 (769 nodes, 33,312 edges)}  \\
\midrule
       0.5 & 0.41*** (0.01) & 1.95*** (0.02) & 0.59*** (0.01) & -1,377.47 & 2,762.94 & 171.8 & -0.008 & -9.94 & 3.45** \\
       1.0 & 0.36*** (0.005) & 1.85*** (0.02) & 0.58*** (0.01) & -1,435.07 & 2,878.13 & 179.6 & -0.18 & -11.08 & 3.49** \\
       1.5 & 0.36*** (0.006) & 1.72*** (0.01) & 0.55*** (0.007) & -1,522.47 & 3,052.94 & 127.6 & -0.06 & -11.46 & 3.37** \\

\midrule
        & \multicolumn{9}{c}{Swarthmore42 (1,659 nodes, 12,2100 edges)} \\
\midrule
         0.5 & 0.18*** (0.003) & 1.62*** (0.006) & 0.65*** (0.002) & -2,838.33 & 5,684.66 & 124.2 & 0.01 & -11.46 & 7.42*** \\
         11.0 & 0.26*** (0.002) & 1.70*** (0.008) & 0.58*** (0.003) & -2,927.99 & 5,863.97 & 91.6 & -0.10 & -4.25 & 1.96* \\
         1.5 & 0.19*** (0.004) & 1.50*** (0.008) & 0.59*** (0.002) & -3,139.42 & 6,286.83 & 87.39 & -0.20 & -4.52 & 4.03*** \\
\midrule
        & \multicolumn{9}{c}{UChicago30 (6,591 nodes, 416,206 edges)} \\
\midrule
         0.5 & 0.23*** (0.001) & 2.00*** (0.005) & 0.41*** (0.002) & -3,444.33 & 6,896.67 & 217.2 & -0.24 & -2.52 & 7.46*** [0.34] \\
         1.0 & 0.23*** (0.002) & 1.98*** (0.004) & 0.38*** (0.001) & -3,578.18 & 7,164.36 & 219.2 & -0.12 & -2.66 & 9.56*** [1.05] \\
         1.5 & 0.22*** (0.004) & 1.78*** (0.008) & 0.41*** (0.002) & -2,033.49 & 4,074.98 & 222.4 & -0.17 & -2.42 & 10.19*** [0.24] \\
\midrule
        \emph{Notes} & \multicolumn{9}{r}{$\hat \theta_{\mathrm{PA}}$ = Coefficient of log degree, $\hat \theta_{\mathrm{H}}$ = Coefficient of log \# of common attributes, $\hat \theta_{\mathrm{TC}}$ = Coefficient of log \# common neighbors} \\ 
        & \multicolumn{9}{r}{LL = Log-likelihood, AIC = Akaike Information Criterion} \\
        & \multicolumn{9}{r}{Acc. = Accuracy, $L$ = Average Path Length, $C$ = Average Clustering Coefficient, $\Delta Q$ (t-stat) = Modularity change t-statistic} \\
         & \multicolumn{9}{r}{$^*: P < 0.05, \; ^{**}: P < 0.01, \; ^{***}: P < 0.001$} \\
\bottomrule
\end{tabular}
\caption{{Multinomial logit coefficients for three networks from the Facebook100 dataset and GPT-4 (gpt-4-1106-preview). The standard errors of the estimates are shown in parentheses. The null hypothesis corresponds to the respective parameter being equal to 0. We report the percent change in accuracy, average path length, and average clustering coefficient compared to the initial network (before the deletion of edges). For the change in modularity, we run the Louvain algorithm ten times and perform a t-test with the resulting modularities. For the UChicago30 dataset, we report the t-statistic value in the subgraph induced by the 2,000 sampled nodes, since the newly added edges would have a very small effect on the change in the community structure if we were to measure it in the whole network. We also report the modularity change (t-statistic) of the whole graph inside brackets.}}
\label{tab:regression}
\end{table*}

%% file: table_6.tex
\begin{table*}[!h]
\tiny
    \centering
    \begin{tabular}{lcccccccccc}
\toprule
Name & Temp & Degrees (KS) & (P-value) & Sizes of CCs (KS) & (P-value) & Spectrum (KS) & (P-value) & Local CC (KS) & (P-value) \\
\midrule
Caltech36 & 0.5 & 0.0481 & 0.336 & 0.125 & 0.999 & 0.0546 & 0.202 & 0.0234 & 0.984 \\
 & 1.0 & 0.0481 & 0.336 & 0.176 & 0.926 & 0.0559 & 0.181 & 0.0325 & 0.811 \\
 & 1.5 & 0.0481 & 0.336 & 0.111 & 1 & 0.0559 & 0.181 & 0.0351 & 0.731 \\
\midrule
Swarthmore42 & 0.5 & 0.0229 & 0.777 & 0.2 & 0.987 & 0.0151 & 0.992 & 0.0127 & 0.999 \\
 & 1.0 & 0.0217 & 0.83 & 0.2 & 0.987 & 0.0133 & 0.999 & 0.0133 & 0.999 \\
 & 1.5 & 0.0211 & 0.854 & 0.2 & 0.987 & 0.0157 & 0.987 & 0.0175 & 0.962 \\
\midrule
UChicago30 & 0.5 & 0.00228 & 1 & 0.81 & 3.4e-08 (***) & 0.00303 & 1 & 0.0188 & 0.194 \\
 & 1.0 & 0.00228 & 1 & 0.805 & 4.62e-08 (***) & 0.00288 & 1 & 0.0184 & 0.217 \\
 & 1.5 & 0.00789 (***) & 0.986 & 0.873 & 3.86e-11 (***) & 0.00819 & 0.98 & 0.0188 & 0.194 \\
\bottomrule
\end{tabular}
\caption{Change in graph statistics for the GPT-4 model (gpt-4-1106-preview) and the Facebook100 data by the metrics outlined in \cite{leskovec2006sampling}. The results for the other datasets, models, and temperatures are similar. (***) denotes $P < 0.001$.}
\label{tab:graph_stats}
\end{table*}

%% file: table_7.tex
\begin{table*}[!h]
    \centering
    \tiny
    \begin{tabular}{lllllll}
    \toprule
    Dataset & Preferential Attachment ($\hat \theta_{\mathrm{PA}}$) & Homophily ($\hat \theta_{\mathrm{H}}$) & Triadic Closure ($\hat \theta_{\mathrm{TC}}$) & Log Likelihood & AIC \\
    \midrule
    \multicolumn{6}{c}{$A = 50$} \\
    \midrule 
    Caltech36 & 0.30*** (0.002) & 2.74*** (0.01) & 0.25*** (0.006) & -2,080.91 & 4,169.83 \\
    Swarthmore42 & 0.17*** (0.002) & 2.23*** (0.004) & 0.40*** (0.002) & -2,708.70 & 5,425.41 \\
    UChicago30 & 0.16*** (0.004) & 2.04*** (0.005) & 0.49*** (0.002) & -2,553.28 & 5,114.57 \\
    Andorra & 0.27*** (0.01) & 6.32*** (0.02) & 0.30*** (0.01) & -1,762.45 & 3,532.89 \\
    MobileD &  0.27** (0.005) & -0.84*** (0.02) & 1.53*** (0.002) & -2,359.88 & 4,727.77 \\
    \midrule
    \multicolumn{6}{c}{$A = 100$} \\
    \midrule 
    Caltech36 & 0.35*** (0.003) & 3.23*** (0.02) & 0.26*** (0.002) & -2,450.46 & 4,908.91 \\
    Swarthmore42 & 0.15** (0.005) & 2.51*** (0.004) & 0.33*** (0.002) & -3,522.14 & 7,052.27 \\
    UChicago30 & 0.16*** (0.003) & 2.66*** (0.004) & 0.40*** (0.003) & -3,100.58 & 6,209.16 \\
    Andorra & 0.23*** (0.006) & 6.72*** (0.04) & 0.56*** (0.002) & -1,838.66 & 3,685.33 \\
    MobileD & 0.32*** (0.004) & -0.23 (0.008) & 1.58*** (0.004) & -2,848.65 & 5,705.30 \\
    \midrule
    \multicolumn{6}{l}{\emph{Note: *: $P < 0.05$, **: $P < 0.01$, ***: $P < 0.001$}} \\    
    \bottomrule
    \end{tabular}
    \caption{Robustness of results to large contexts. Experiments have been performed with gpt-4.1-mini with context windows $A \in \{ 50, 100 \}$. The temperature has been set to 0.5.}
    \label{tab:context_window}
\end{table*}